\documentclass[preprint,trackchanges]{aastex7}

\usepackage{makecell}
\usepackage{booktabs}

\usepackage{graphicx}	
\usepackage{amsmath}	
\usepackage{subfigure}
\usepackage{caption}

\begin{document}
\title{Feedback-Driven Interaction of Expanding Molecular Clouds in the LDN 1204–Sh2-145–Sh2-140 Complex}

\author[0009-0007-6397-3220]{C Rishi$^*$}
\affiliation{Aryabhatta Research Institute of Observational Sciences (ARIES), Manora Peak, Nainital 263002, India}
\affiliation{Department of Applied Physics, Mahatma Jyotiba Phule Rohilkhand University, Bareilly, 243006, U.P., India}
\email[show]{rishictk@gmail.com}  
\author[0000-0002-0151-2361]{Neelam Panwar$^{**}$} 
\affiliation{Aryabhatta Research Institute of Observational Sciences (ARIES), Manora Peak, Nainital 263002, India}
\email[show]{neelam\_1110@yahoo.co.in}
\author[0000-0001-6725-0483]{L. K. Dewangan}
\affiliation{Astronomy \& Astrophysics Division, Physical Research Laboratory, Navrangpura, Ahmedabad 380009, India}
\email[show]{lokeshd@prl.res.in}
\author[0000-0001-7881-7748]{Ajay Kumar Singh}
\affiliation{Department of Applied Physics/Physics, Bareilly College, Bareilly-243001, Uttar Pradesh, India}
\email{}

\date{Accepted XXX. Received YYY; in original form ZZZ}



\begin{abstract}

We present a multi-wavelength study of an extended molecular cloud hosting the LDN~1204, Sh2-145, and Sh2-140 regions to investigate the interplay between stellar feedback, cloud dynamics, and star formation. Infrared images reveal two semi-arc/shell-like structures (extent $\sim$14--18 pc) associated with known massive O-type stars. {\it Herschel} temperature and column density maps identify one hub-filament system (HFS) candidate toward both LDN~1204A and LDN~1204B sub-regions, with peak column densities of $\sim3.1 \times 10^{23}$ and $\sim1.3 \times 10^{23}$ cm$^{-2}$, respectively. These sub-regions are located at the periphery of the Sh2-145 H\,{\sc ii} region. Molecular line data reveal a large arc-/shell-like cloud resolved into two kinematically distinct components spanning $\sim-15$ to $-9.5$ km s$^{-1}$ and $\sim-9$ to $-3$ km s$^{-1}$. Each component exhibits an arc-/shell-like morphology, is centered on one of the massive stars, and displays an inverted C-shaped position-velocity structure, indicating expansion driven by its associated H\,{\sc ii} region. Both LDN sub-regions are located at the projected interface of the two expanding cloud components. Young stellar objects (YSOs), particularly Class~I YSOs, are concentrated toward these sub-regions, while moment~2 maps reveal enhanced velocity dispersion, suggesting increased turbulent motions at the interface. Altogether, the presence of dense filaments, candidate hub–filament systems, and active star formation in LDN 1204 sub-regions demonstrates that feedback-driven shell expansion and interaction compressed the gas at the interface, establishing the necessary conditions for filaments and ongoing star formation. These results highlight the important role of feedback-driven cloud interactions in regulating dense structure formation and star formation in molecular clouds.


\end{abstract}

\keywords{H II regions(694); Low mass stars(2050); Young star clusters(1833); Star-forming regions(1565); Star formation (1569)}



\section{Introduction}

Molecular clouds are subject to ongoing transformations due to a range of processes, including persistent external effects from adjacent massive stars, supernova explosions, star formation within the cloud, and interactions between clouds. Some of these feedback mechanisms influence the star formation rate within the molecular cloud \citep{2007prpl.conf..245A,2015NewAR..68....1D,2019MNRAS.488.4753D,2024MNRAS.52710077H}. In certain cases, these processes can induce star formation within molecular clouds, whereas in other instances, they can cause disruption to the cloud  \citep{2005MNRAS.358..291D,2018ApJ...859..166F}. For example, the ionizing radiation from massive stars can disrupt the outer envelope of young stellar objects (YSOs) that are already in their formation path \citep{2021ApJ...913...95P}. In other cases, the same ionizing radiation can promote star formation by inducing the collapse of the nearby clumps \citep{1989ApJ...346..735B,1994A&A...289..559L,2006A&A...450..625U}.

Over the past decade, wide-field far-infrared surveys (notably with the \textit{Herschel} Space Observatory) have established that cold molecular clouds are ubiquitously structured into complex networks of filaments, where dense cores and stars predominantly form \citep{2010A&A...518L.102A, 2014prpl.conf...27A}. Within this framework, intersections of multiple filaments often form dense, central junctions termed hub--filament systems \citep[HFSs;][]{2009ApJ...700.1609M, 2012A&A...540L..11S, 2022A&A...658A.114K}. These hubs serve as convergence centers where material is channeled along longitudinal filaments to sustain active, clustered star formation. 

Massive stars formed inside the molecular cloud create H{\sc ii} regions, which are composed primarily of ionized hydrogen. 
These regions expand further, influencing nearby molecular cloud \citep{2011MNRAS.414.1747A,2023MNRAS.522.2369S}. 
This expanding H {\sc ii} region expanding uniformly, can create bubbles \citep{2011ApJ...742..105A}. 
Recently, the ``bubble--filament paradigm'' has emerged to bridge the gap between large-scale feedback and the origin of these filamentary structures \citep{2015A&A...580A..49I, 2020A&A...638A...7Z}. In this scenario, expanding ionization fronts sweep up the surrounding medium into dense shells, which subsequently fragment into elongated filaments. Furthermore, when multiple expanding bubbles or feedback-driven flows converge, the enhanced compression at the stagnation interface preferentially facilitates the formation of magnetized filaments and candidate hub--filament configurations \citep{2015ApJ...799...64D, 2017A&A...599A..94N, 2021MNRAS.505..459F}.

\begin{figure*}
    \centering
    \subfigure{
    \includegraphics[width = 0.48\textwidth]{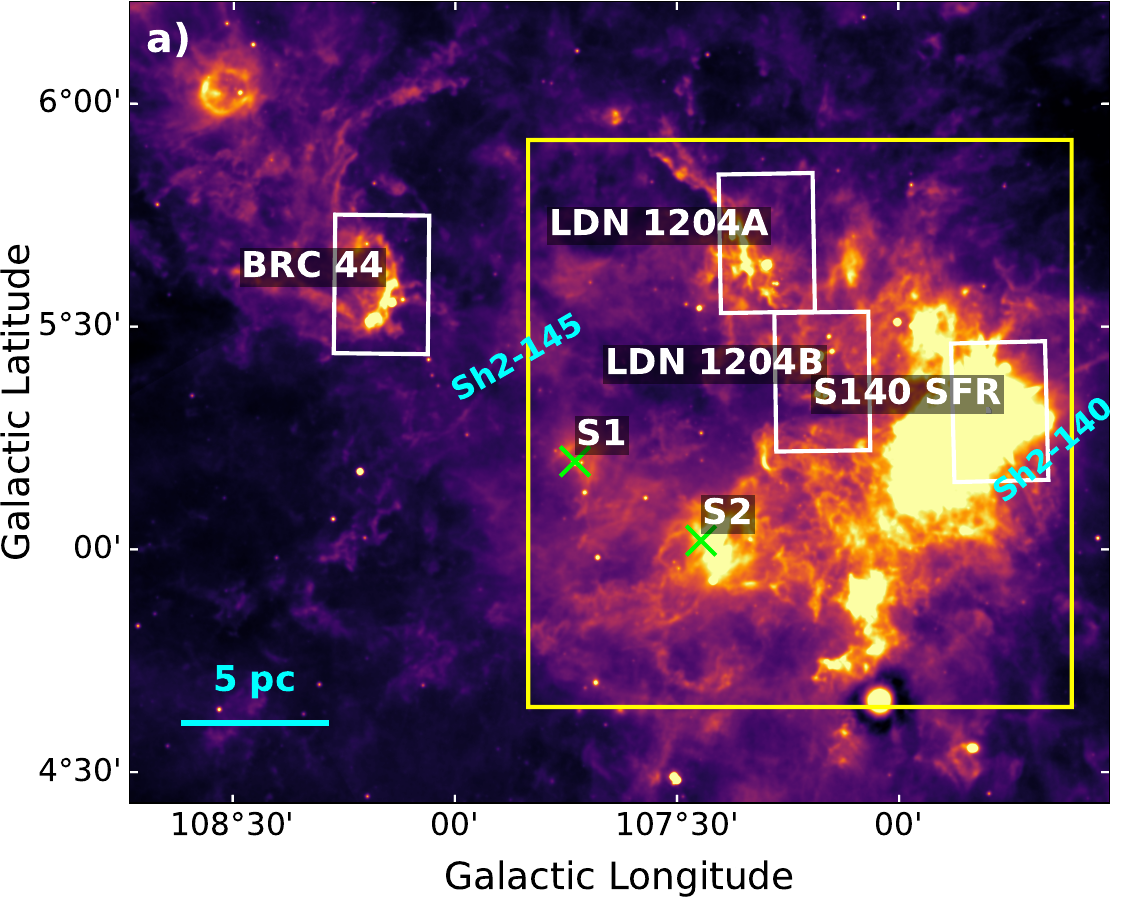}
     \label{fig:wise}
    }
    
    \subfigure{
    \includegraphics[width = 0.48\textwidth]{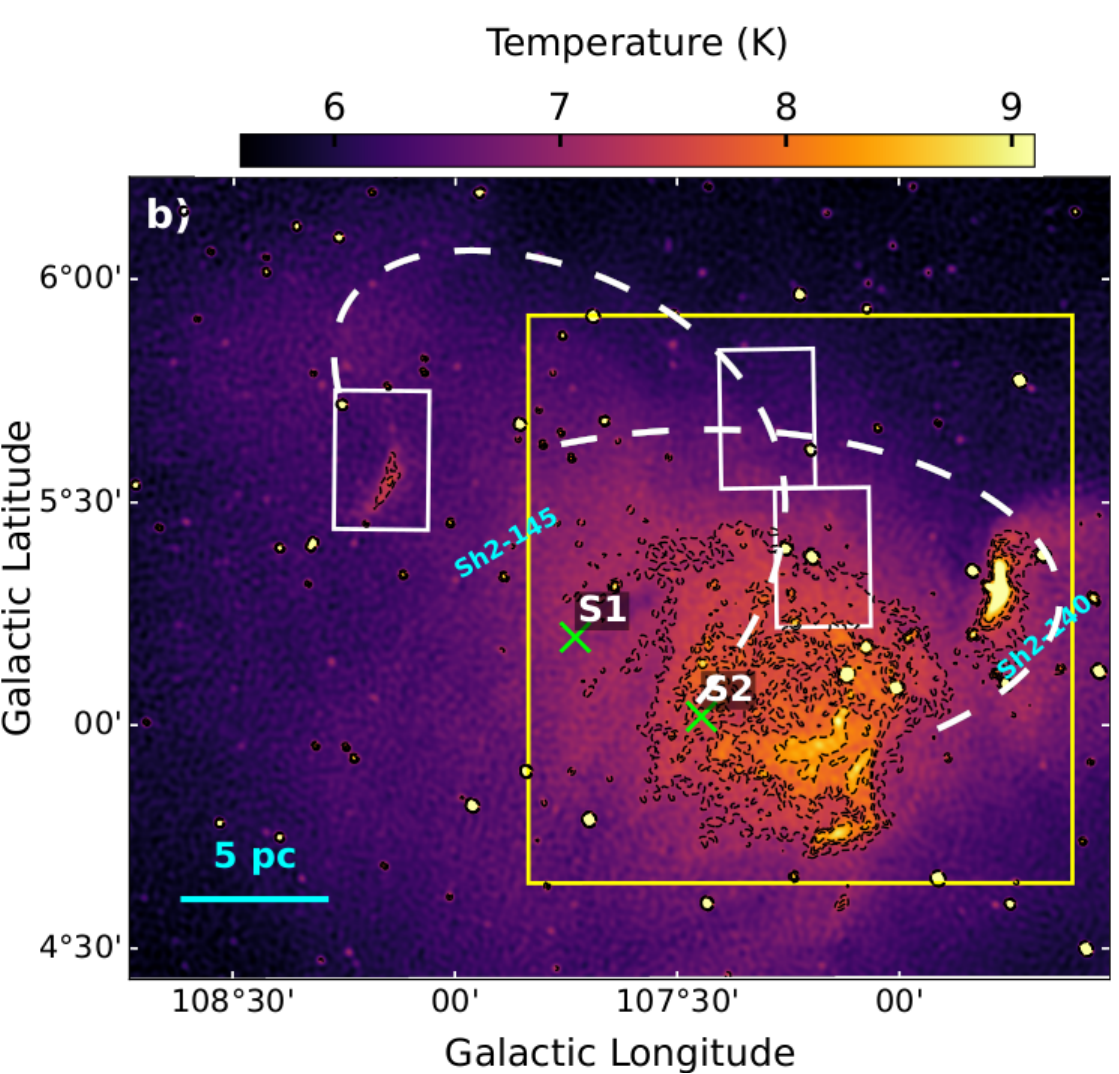}
     \label{fig:cgps}
    }
    \caption{a) WISE 3.6$\mu$m image of a larger region containing both clouds BRC 44 and LDN 1204 clouds. The yellow box is the region that is concentrated in this work with parameters, center longitude : +107$^\circ$13$^{\prime}$21.32$^{\prime \prime}$, center Latitude : +5$^\circ$17$^{\prime}$, width : 1.22$^\circ$ and height : 1.27$^\circ$. b) CGPS 1420 MHz map of the same region. The arc-like features is shown as the dashed arc. Scale length is constructed by taking the distance to the region as 900 pc. Massive stars are marked as `X' in both the images.}
    \label{fig:sandwich}
\end{figure*}

In this study, we investigate the region designated LDN 1204 (Figure \ref{fig:sandwich}), located in the periphery of H {\sc ii} region Sh2-145 (distance $\sim$ 930 pc \citep{2015AJ....150..147F}). Another  H{\sc ii} region Sh2-140, west of Sh2-145 is ionized by the B0.5 V star HD 211880 \citep{1996A&A...315L.281T} with a distance of $\sim$ 921 pc, while Sh2-145 is ionized by the massive stars HD 213023 (O9 V) and BD 62 2078 (O7.5 V) \citep{1994A&A...283..963C},both at a distance of $\sim$ 1000 pc. All three stars have similar proper motion direction. The distance and proper motion is extracted from \citet{2020yCat.1350....0G}.

An active star-forming region is located in the south-western part of LDN~1204, identified as S140 SFR in Figure~\ref{fig:sandwich}, where a couple of early B-type stars are currently forming \citep{2001ApJ...561..282K}. The LDN~1204 complex comprises of multiple dark clouds of which we concentrate on two clouds, LDN~1204A and LDN~1204B (hereafter `A' and `B').
Detailed studies of dense cores suggest that the observed velocity gradients are influenced by outflow activity.
\citet{1993ApJ...403..175T} studied the entire Sh2-140 region and had detected multiple clouds of different velocities. The dense cores along the region were studied in detail, which explained the velocity shifts due to outflows. They detected three clouds in the region. Cloud 1 encompasses substructures `A' and `B' and is characterized by a radial velocity range of -12 km s$^{-1}$ to -10 km s$^{-1}$. Cloud 2 is identified within the velocity interval -10 km s$^{-1}$ to -8 km s$^{-1}$ while Cloud 3, characterized by strong infrared emission and harboring candidate massive stars, spans a radial velocity range of -8 km s$^{-1}$ to -6 km s$^{-1}$.   
In region `A', a filament-shaped cloud seems to harbor an intermediate mass star \citet{2010ApJ...721L.107S}. High-resolution observations by \citet{2013ApJ...762..120P} indicate minimal fragmentation down to 0.1 M$\odot$, with millimeter emission dominated by a single source. In contrast, cloud `B' exhibits double-peaked profiles in $^{12}$CO observations, which have been interpreted as evidence for a potential CCC scenario \citep{2012RAA....12.1499L}.

Motivated by these findings, the present work aims to investigate the effects of dynamical expansion of the H{\sc ii} region. We analyze molecular line data along with complementary archival datasets to examine the dynamical state of the clouds and assess evidence for ongoing interaction. Additionally, we study the YSO population in the region to evaluate their properties, evolutionary stages and spatial distribution along the molecular cloud. The structure of the paper is as follows. Section 2 describes the datasets used in this study. The results and analysis are presented in Section 3. In Section 4, we discuss the possible scenarios for cloud dynamics and star formation in the `A' and `B' regions. Finally, Section 5 summarizes the main findings and presents the conclusions.

\section{data used}

\subsection{Spitzer-IRAC and MIPS data}
We obtained Spitzer Enhanced Imaging Products (SEIP) catalog data in IRAC and MIPS bands from the $Spitzer$ heritage archive. The Spitzer-IRAC and MIPS magnitudes
were taken at wavelengths of 3.6 $\micron$ (Ch1), 4.5 $\micron$ (Ch2), 5.8 $\micron$ (Ch3), 8.0 $\micron$ (Ch4) and 24 $\micron$ (M1) (resolutions $\sim$ 1.66$^{\prime \prime}$, 1.72$^{\prime \prime}$, 1.88$^{\prime \prime}$, 1.98$^{\prime \prime}$ and 6$^{\prime \prime}$ respectively). We considered sources in the yellow rectangular area as shown in Figure \ref{fig:wise}.

\subsection{Herschel Far-infrared data}
Far infrared (FIR) data were used to study the cold dust emission and to detect filaments in the region. We used level 2.5 products from NASA/IPAC Infrared Science Archive at 100 and 160 $\mu$m taken with the Photoconductor Array Camera
and Spectrometer (PACS; \citet{2010A&A...518L...2P} and at 250, 350,
and 500 $\mu$m taken with the Spectral and Photometric Imaging Receiver
(SPIRE; \citet{2010A&A...518L...3G}).  The scan speed of every image was medium (20"$s^{-1}$), and the resolution of the images was 6.8", 10.7", 17.6", 23.9" and 35.2" for 100, 160, 250, 350 \& 500 $\mu$m, respectively. 
\subsection{CO molecular line data}
We obtained molecular line calibrated data from  PMO (Purple Mountain Observatory) 13.7-m millimetre-wave radio telescope located in Delingha in the $^{12}$CO (J=1–0), $^{13}$CO (J=1–0), and C$^{18}$O (J=1–0) lines along the Galactic Plane.
The $rms$ noise level, which is about 0.5 K for $^{12}$CO (J=1–0) and 0.3 K for $^{13}$CO (J=1–0) and C$^{18}$O (J=1–0), is uniform across the whole covered area. The beam width is 50$^{\prime \prime}$. 
\section{Results}

\subsection{Morphology of the whole complex as seen through WISE and 1420 MHz CGPS emission}
Figure \ref{fig:sandwich} represents the images of the whole complex encompassing a bright rimmed cloud, BRC 44 \citep{2025ApJ...994..222C}, the LDN 1204 molecular clouds with sub-regions LDN 1204A (`A') and LDN 1204B (`B'), the S140 SFR star forming region South-west of the LDN 1204 clouds, and the H {\sc ii} regions Sh2-145 and Sh2-140. The massive stars associated with Sh2-145 is also shown as `S1' and `S2' as mentioned in the introduction. In the WISE 22 $\mu$m image as shown in Figure \ref{fig:wise}, we see a bubble or shell like structure formed by the H {\sc ii} region Sh2-145, surrounded by BRC 44, and the molecular cloud sub-regions `A' and `B'. 

For further analysis in the upcoming sections, we concentrate around the region inside the yellow rectangular box in the figure, to study the the cloud dynamics that may be altered due to the expansion of the H {\sc ii} regions which caused the prominent shell structure in the infrared image. 

\subsection{Distribution of cold dust from Herschel}
\begin{figure*}

    \subfigure{
    \includegraphics[width =0.48\textwidth]{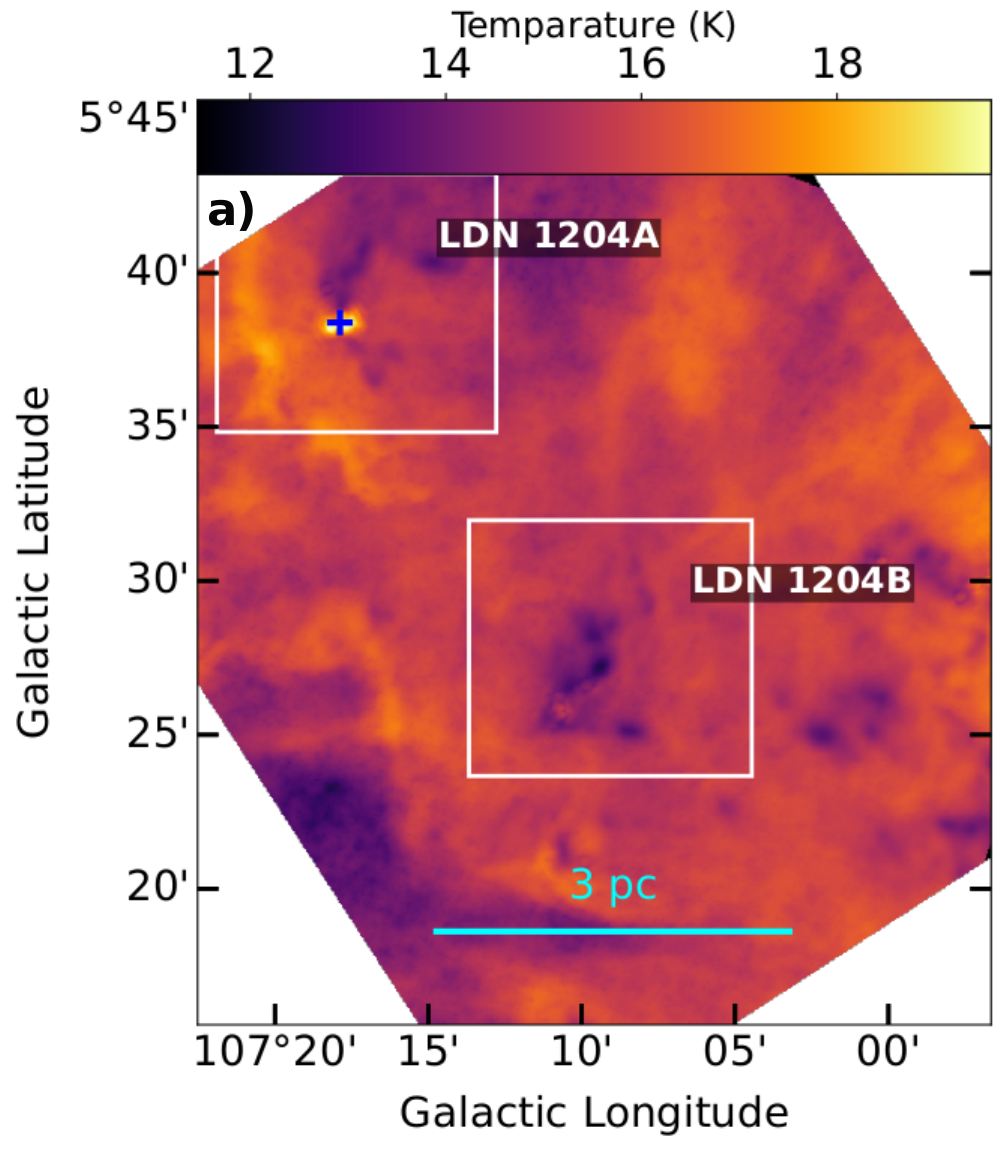} 
    \label{fig:temp}
    }
    \subfigure{
    \includegraphics[width =0.48\textwidth]{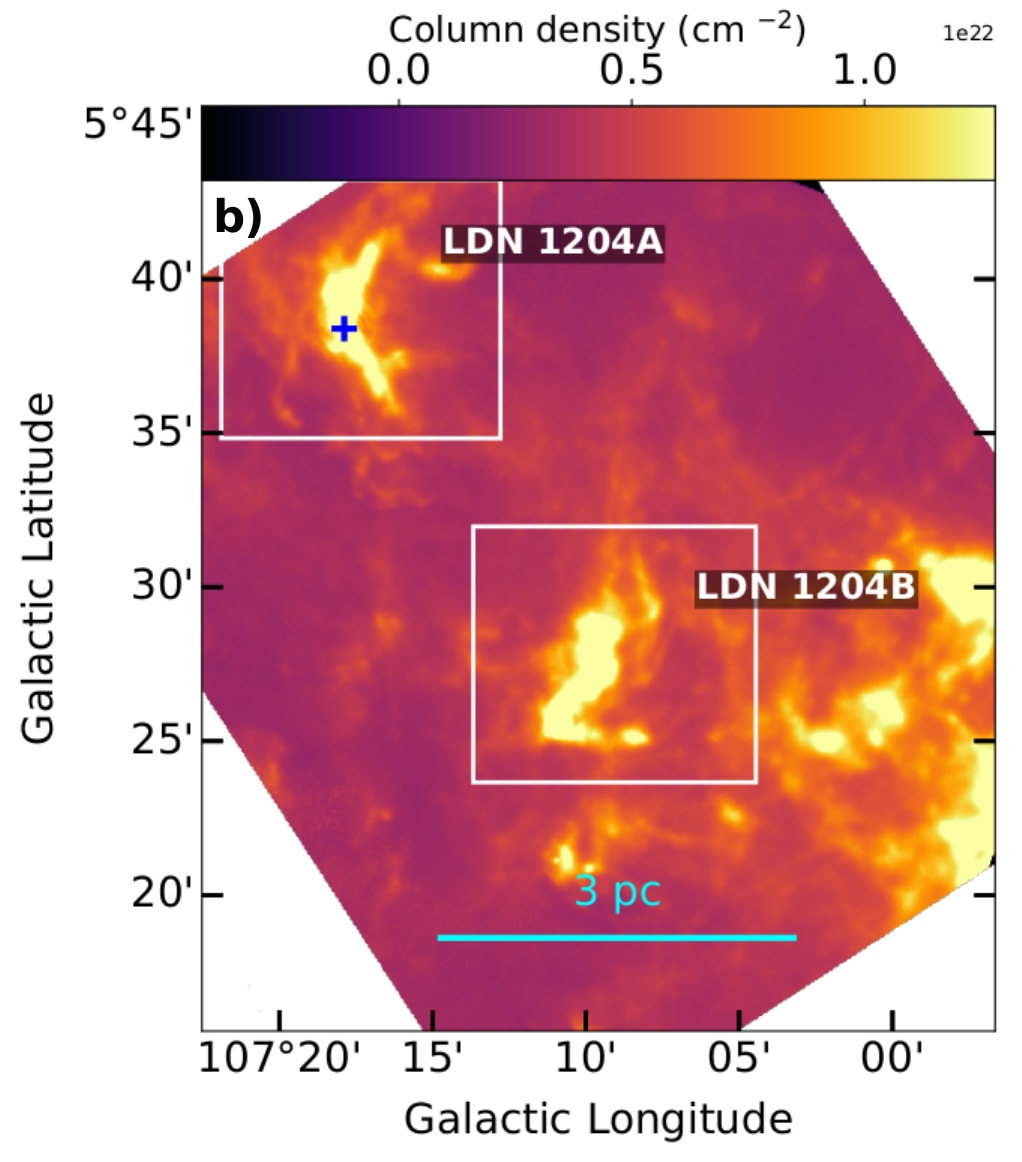}  \label{fig:col}
    }
    \subfigure{
    \includegraphics[width =0.48\textwidth]{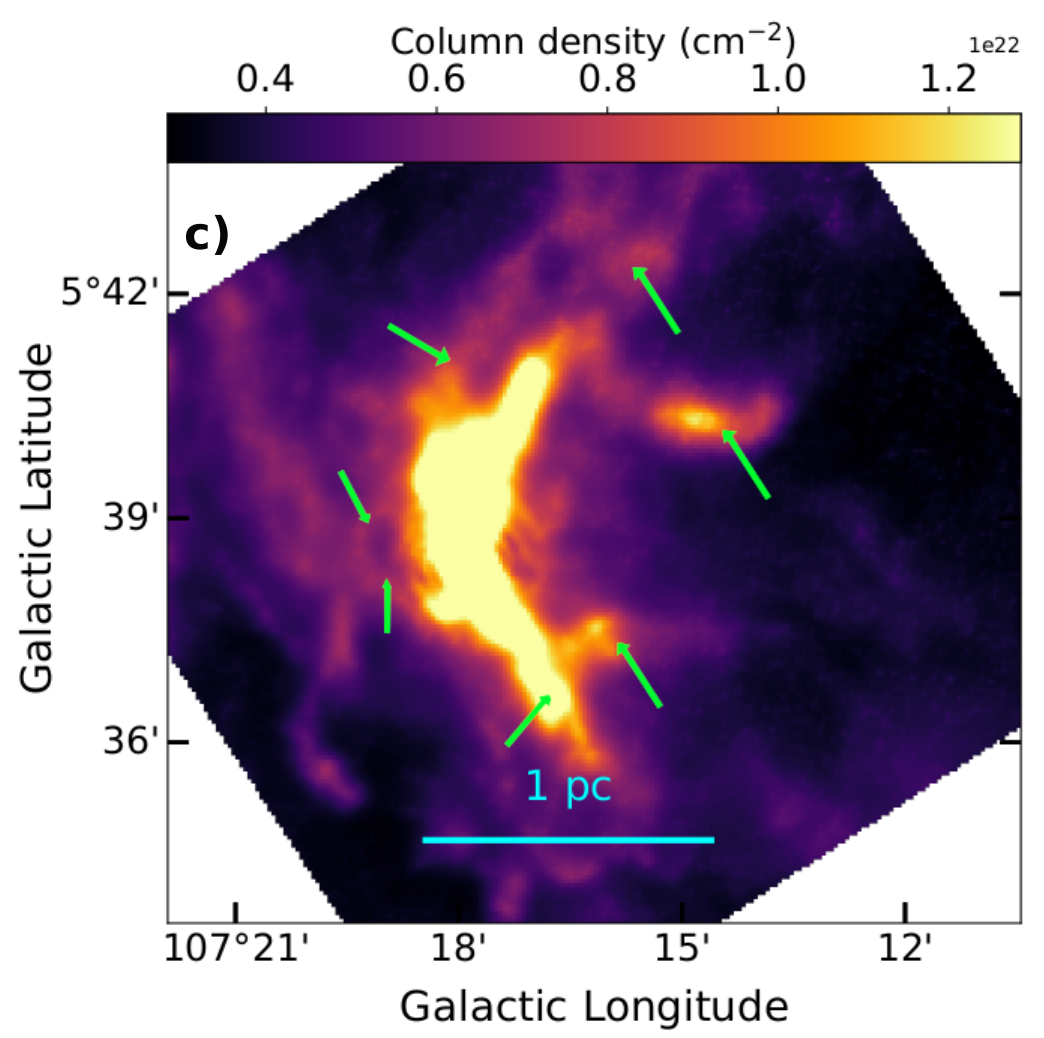}  \label{fig:fil_l1204a}
    }
    \subfigure{
    \includegraphics[width =0.48
    \textwidth]{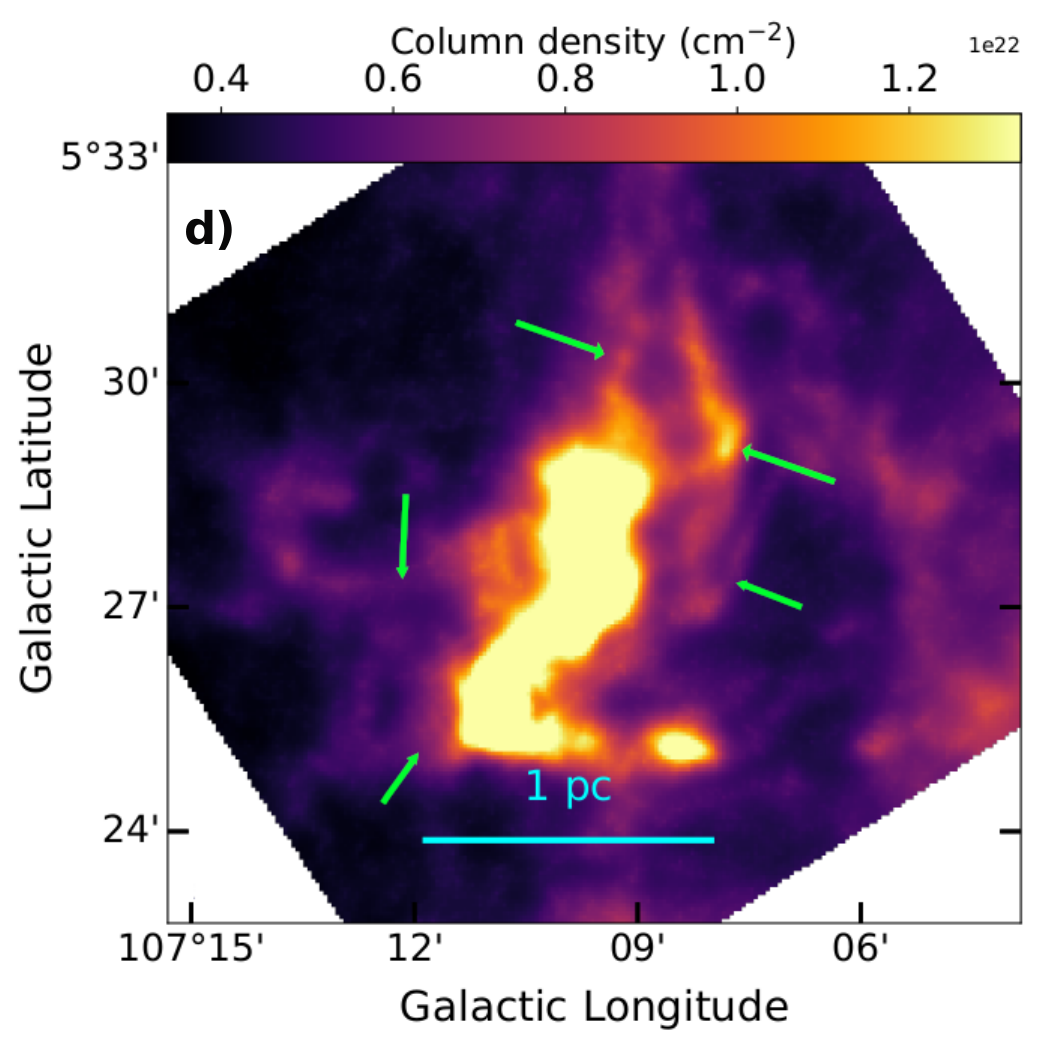}  \label{fig:fil_l1204b}
    }
    \caption{a)The temperature map of clouds derived from $Herschel$ using $getsf$. The intermediate mass star present in `A' is situated in the high temperature region as shown in blue cross. b)Similarly, the column density map, c)sub filaments identified visually in the column density map of the region LDN 1204A, d)similarly, filaments identified in LDN 1204B. All the images have resolution of 10.7 $^{\prime \prime}$.}
    \label{fig:temp_coldensemap}
   
\end{figure*}

The far-infrared emission from dust provides temperature information of the dust, which can be used to study if any heated or extreme cold dust is present in the region. Here, we studied the cold dust emission in the region using $Herschel$ images $\geq$ 160$\mu$m. We used the $getsf$ software suite to produce the temperature and column density map of the region. For this purpose, we first re-projected the Herschel images at 100, 160, 250, 350, and 500 $\mu$m to a common grid, ensuring the same pixel scale, image size, and angular resolution (beam width). Then a pixel-to-pixel fit was performed using a modified
blackbody function for optically thin emission using the $hires$ package in the $getsf$. To account for the excess hot dust emission in 160 $\mu \mathrm{m}$, we had to give offset values for each wavelength image to make the temperature distribution uniform. The $100\ \mu \mathrm{m}$ was neglected because it may contain an excess of hot dust emission. The final images is shown in Figure \ref{fig:temp} and \ref{fig:col}. Here we concentrate only on the region around `A' and `B'. The resolution of the temperature and the column density map corresponds to that of 160 $\mu$m (10.7 $^{\prime \prime}$).
 
The colder region in temperature map seen as thread-like features in `A' and `B' are filaments, which are also roughly traced by C$^{18}$O, discussed previously. The temperature of these structures is found to be relatively lower compared to the surrounding diffuse gas \citep{2011A&A...529L...6A, 2013MNRAS.428.3425H}. The temperature ranges from $\sim$ 12 - 21 K along the filaments seen here. In the filament in `A', we see a region of high temperature (marked as a blue cross). This region, as discussed in the introduction section, exhibits active star formation, and there is also a well-studied intermediate-mass star. The higher temperature may be due to the excess heating of the dust by this star. Along the filament, the temperature ranges from $\sim$ 13.5 - 20.5 K, compared to a mean temperature of $\sim$ 15 K away from the filament. The maximum temperature around the star-forming site is $\sim$ 20.5 K and the maximum column density calculated in `A' is $\sim$ $3.10\ \times \ 10^{23} \ \mathrm{cm^{-3}}$.

In region `B', along the filament, the temperature ranges from $\sim$ 12.5 - 16.3 K and the calculated maximum column density is $\sim$ $1.3\ \times \ 10^{23} \ \mathrm{cm^{-3}}$, both of which are lesser compared to `A'. This suggests that the dust embedded in the filaments does not get heated much. For the entire region outside dense structures, the average column density is in the order of $10^{21}$ $\mathrm{cm^{-3}}$.

Further, we also visually identified multiple smaller filaments attached to the main filaments as shown in Figure \ref{fig:fil_l1204a} and \ref{fig:fil_l1204b} for both `A' and `B' regions. This intertwined filamentary structures may be an indication of a Hub-filament system, where multiple filaments are interconnected and the longitudinal flow along these interconnected filaments feeds the central hub. These hubs are found to be the site of potential massive star formation or clusters of low-mass stars \citep{2025AJ....169...56M, 2026AJ....171...69D}.

\subsection{Structure of the molecular cloud}
 Motivated by the shell structure, we analyzed molecular line data ($^{12}$CO, $^{13}$CO, and C$^{18}$O) from PMO, to infer the cloud morphology in millimeter wavelengths within the yellow box in Figure \ref{fig:sandwich}. The three isotopologues of CO trace different physical components of the molecular cloud due to their varying optical depths. $^{12}$CO emission becomes optically thick and therefore traces the more extended outer layers of the cloud. $^{13}$CO, with a lower abundance, probes the intermediate-density gas, while the optically thin C$^{18}$O traces the optically thin part of the cloud. 


\begin{figure*}
    \subfigure{
        \includegraphics[width=0.48\textwidth]{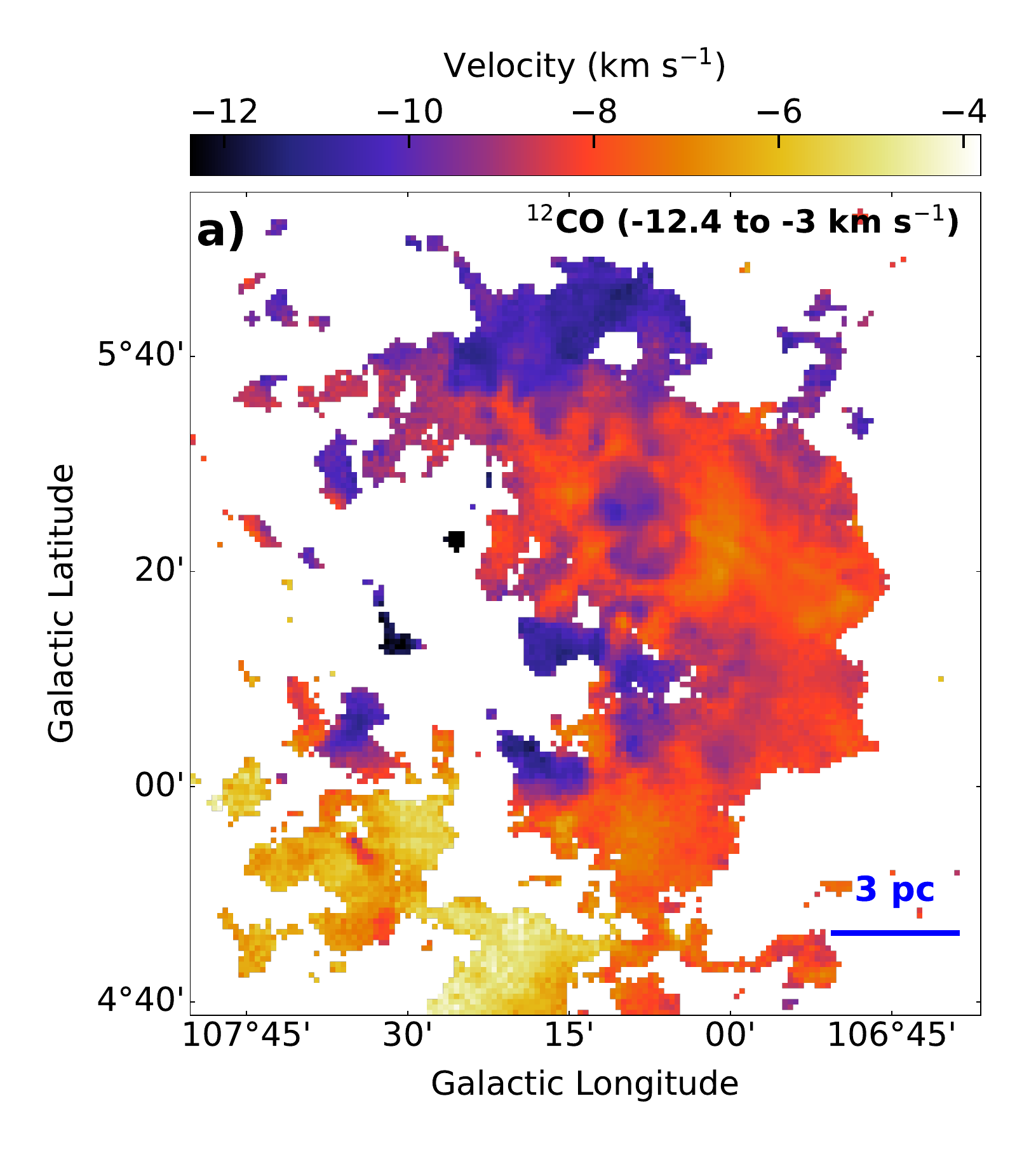}  
        \label{fig:12co_moment1}
    }
    \subfigure{
        \includegraphics[width=0.48\textwidth]{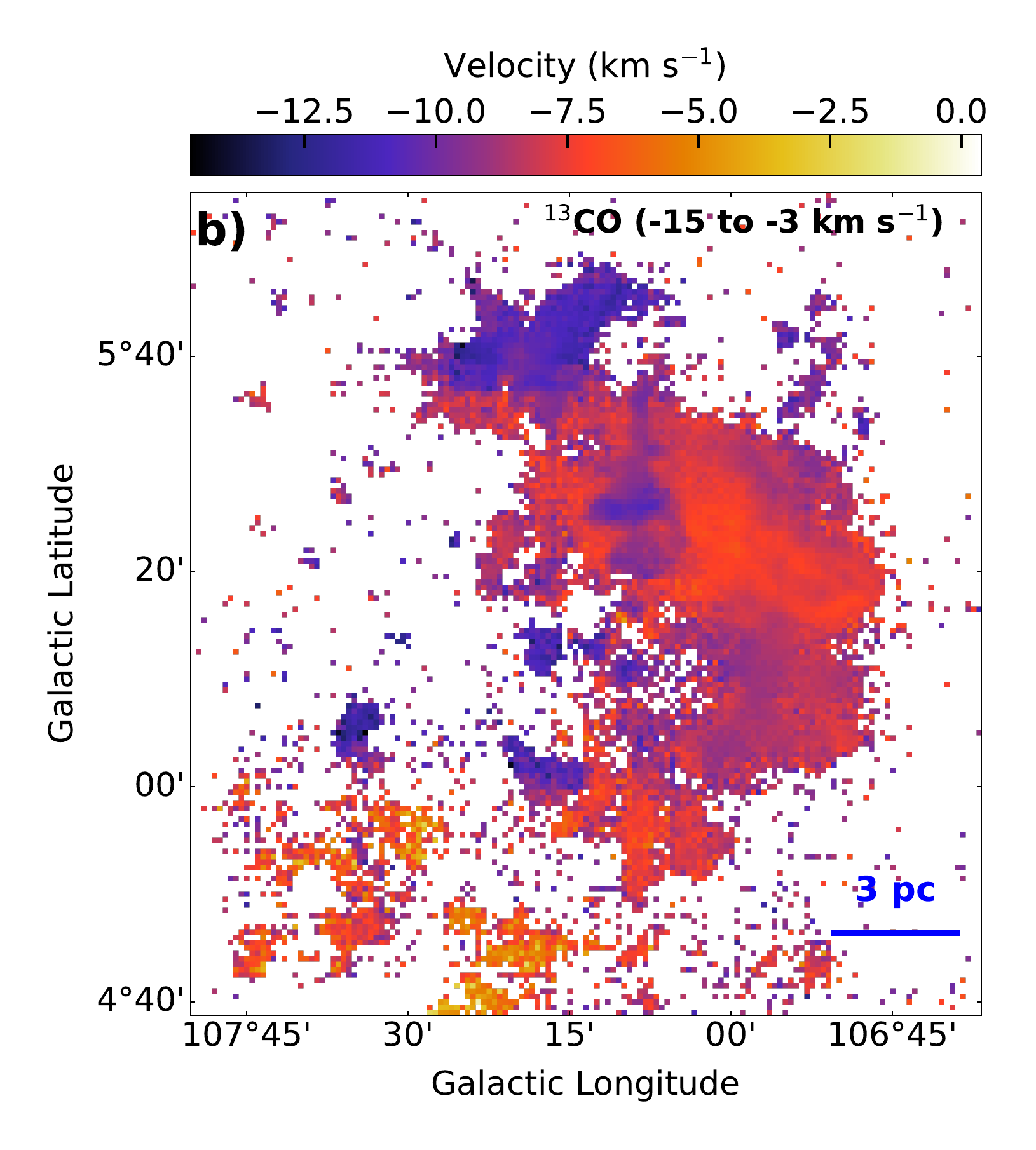}
        \label{fig:13co_moment1}
    }
     \subfigure{
        \includegraphics[width=0.48\textwidth]{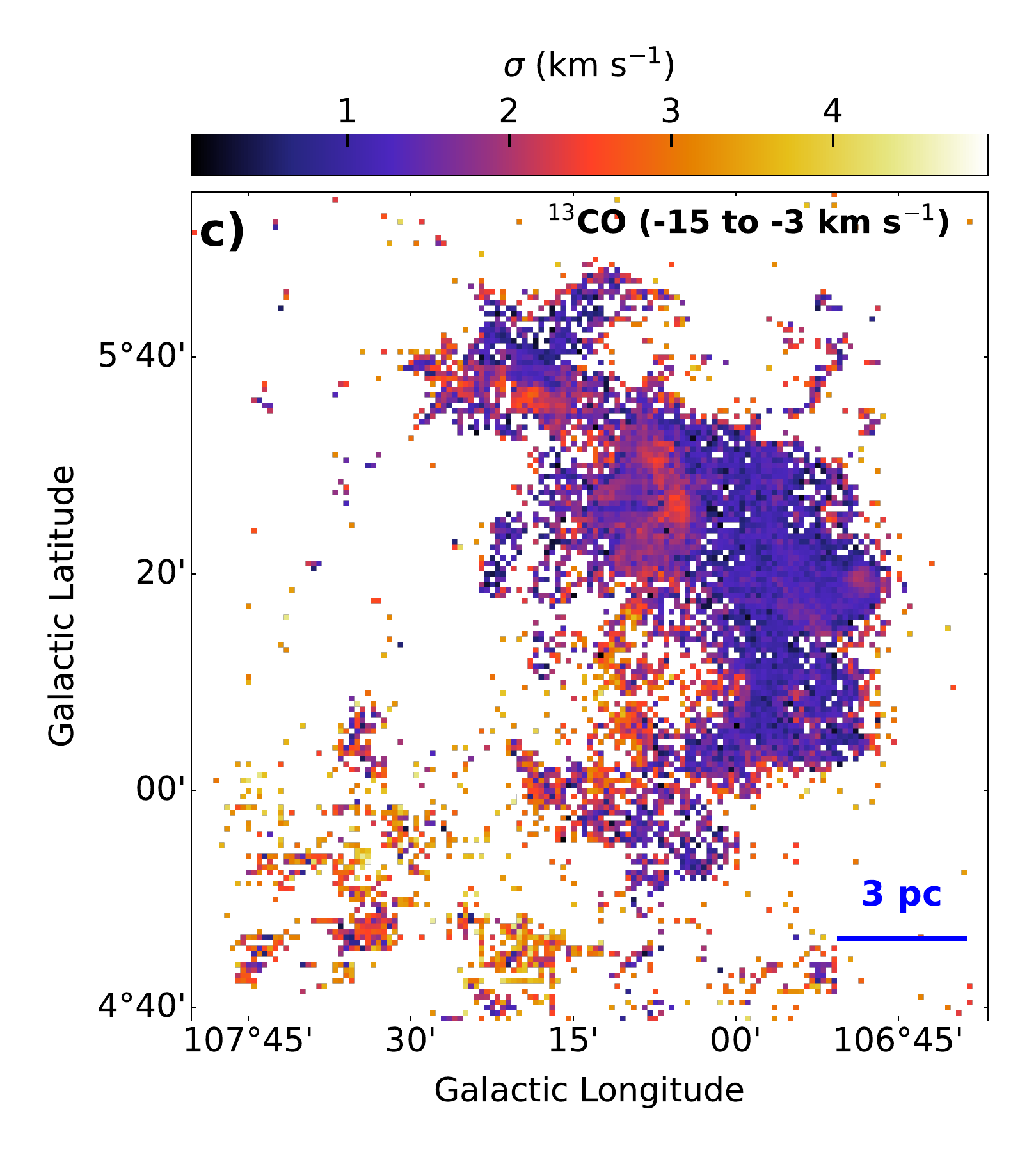}
        \label{fig:13co_moment2}
    }
    \caption{a)$^{12}$CO moment 1 map, b)$^{13}$CO velocity moment 1 map.
    c)$^{13}$CO velocity dispersion or moment 2 map}
    \label{fig:moments_part2}
\end{figure*}

Using molecular data, we created the integrated intensity maps or moment 0 maps, moment 1 maps and $^{13}$CO moment 2 map of the entire region, as shown in Figure \ref{fig:moments_part1} and \ref{fig:moments_part2}. The velocity range is taken between -15 km s$^{-1}$ to 0 km s$^{-1}$ for all the molecules. Although the resolution of the observation is not high enough to detect individual cores, a filamentary shape is easily discernible in region `A' as traced by C$^{18}$O (Figure \ref{fig:moments_c18o}). However, filamentary features are not clearly seen in `B' region. In Figure \ref{fig:moments_part1} images, we can see S140 SFR region, where a highly intense emission peak is clearly visible in all isotopologues. This strong peak highlights the dense and active molecular material associated with S140 SFR. On the $^{12}$CO moment 1 map in the figure, we see a distinct velocity component different from the velocity range of the rest of the cloud in the south (yellow cloud). This cloud may not be part of the main cloud and hence we are ignoring it for rest of the study.

\begin{figure*}
    \centering

    \includegraphics[width=\textwidth]{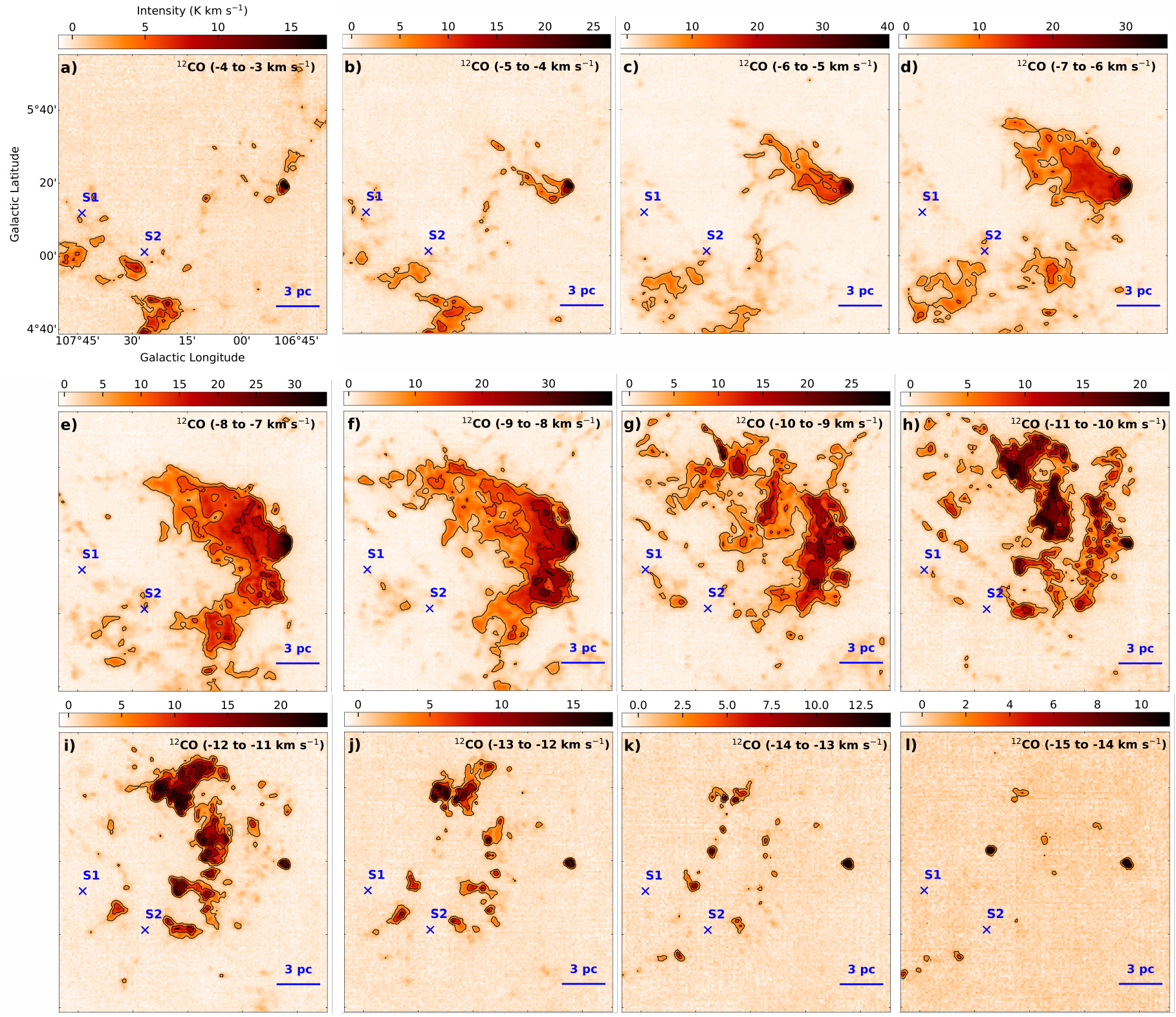}
        
    \caption{a) $^{12}$CO channel maps of the region starting from -3 km s$^{-1}$ (top left) to -15 km s$^{-1}$ (bottom right). Each figure represents channel maps with a velocity width of 1 km s$^{-1}$. The contour levels are defined as seven linearly spaced values between the minimum and maximum of the moment-0 map.}
\label{fig:channelmap}
\end{figure*}

In addition to the moment 0 and moment 1 maps, we also constructed channel maps of regions with a width of 1 km s$^{-1}$ and a velocity of -15 to -3 km s$^{-1}$ as shown in Figure \ref{fig:channelmap}. Detailed analysis of the map is provided in the discussion section.

\subsection{Molecular gas around `A' and `B'}
\label{sec:results_twoclouds}
\begin{figure*}
    \subfigure{
        \includegraphics[width=0.48\textwidth]{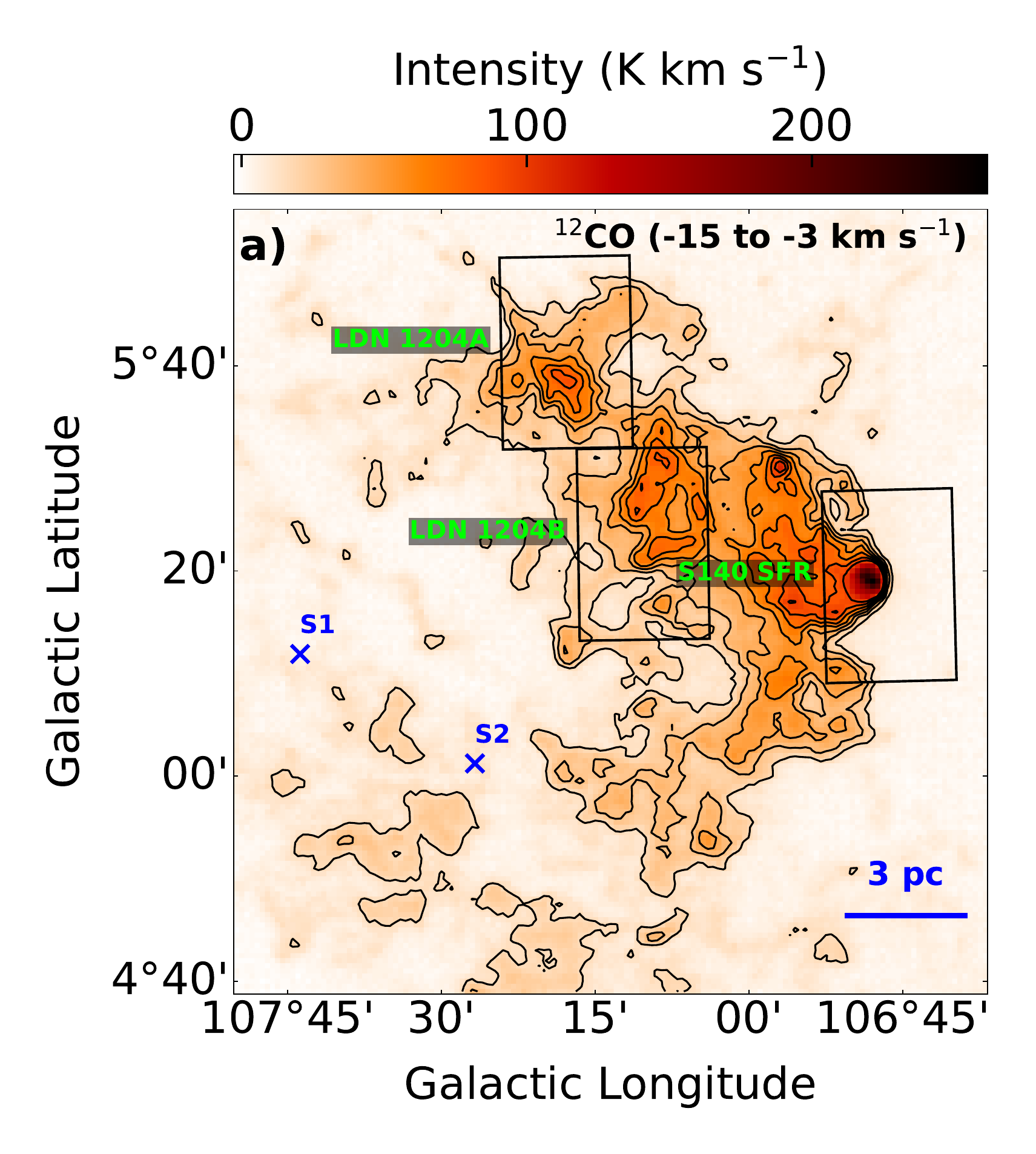}
        \label{fig:moments_12co}
    }
    \subfigure{
        \includegraphics[width=0.48\textwidth]{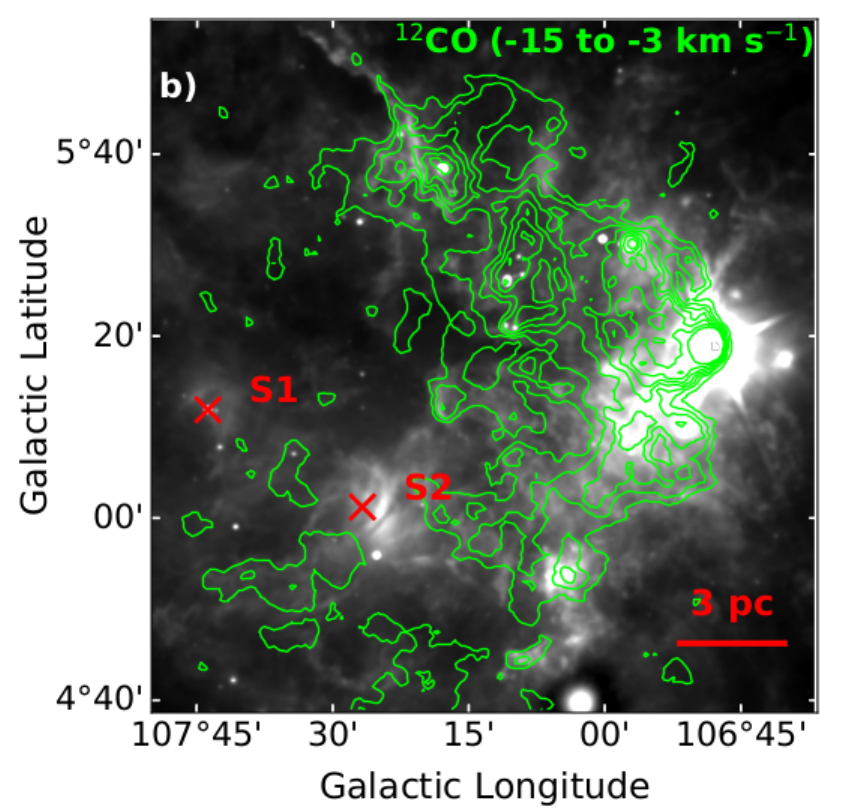}
        \label{fig:moments_22}
    }
    \subfigure{
    
        \includegraphics[width=0.48\textwidth]{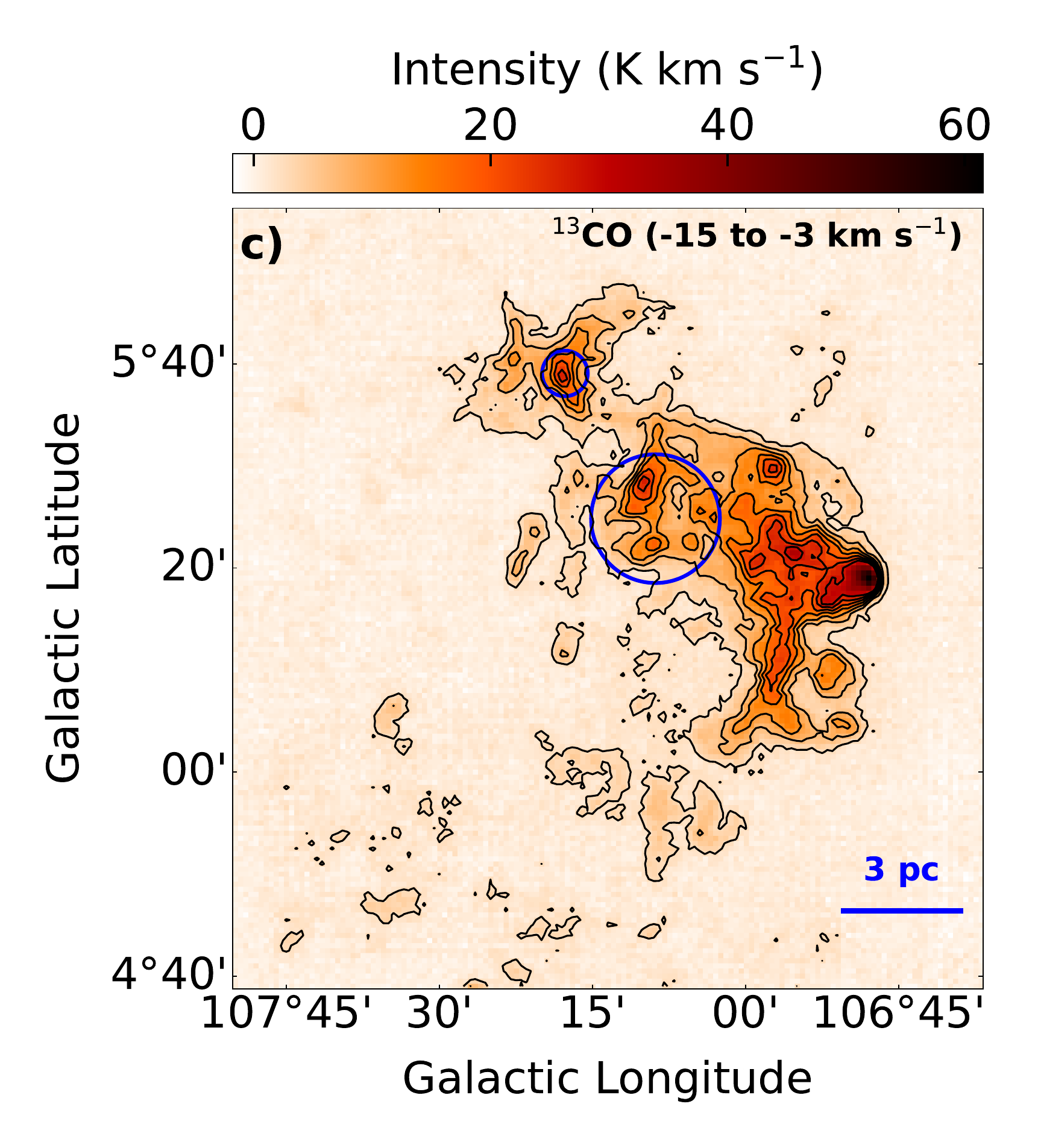}
        \label{fig:moments_13co}
    }
    \subfigure{
        \includegraphics[width=0.48\textwidth]{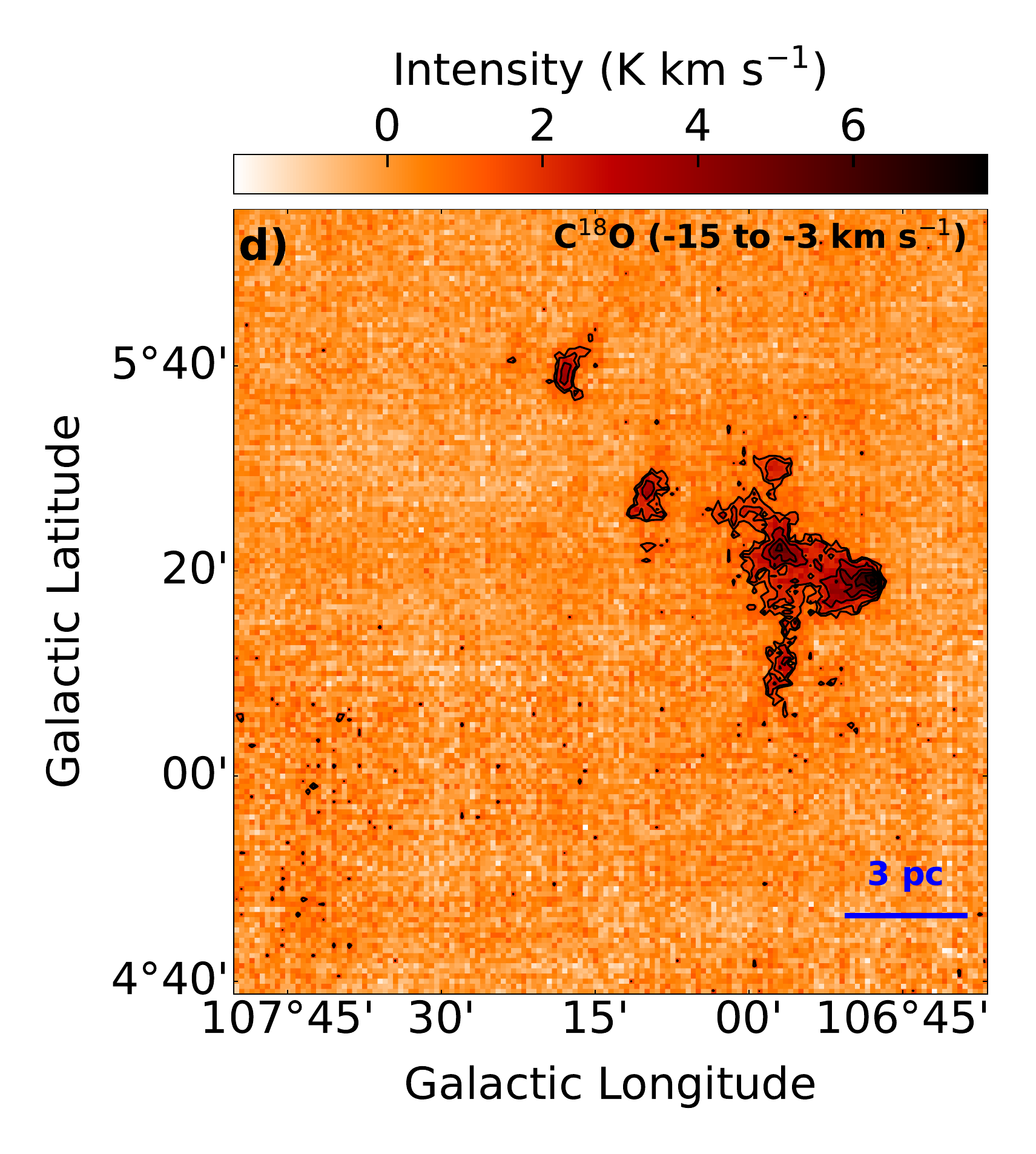}
        \label{fig:moments_c18o}
    }
    \caption{$^{12}$CO moment 0 map with contour intervals 15.45, 29.13, 42.80, 56.48, 70.15,
       83.83 and 97.51 K km s$^{-1}$, b) $^{12}$CO moment 0 contours overplotted on 22 $\mu$m WISE image, c) $^{13}$CO moment 0 map with contour intervals 2.46,  6.66, 10.87, 15.07, 19.28,
       23.48 and 27.69 K km s$^{-1}$. The circular region is the area from which the spectrum is produced, d) C$^{18}$O moment 0 map  with contour intervals 1.40, 1.90, 3.01, 4.13, 5.24,
       6.36 and 7.47 K km s$^{-1}$.}
    \label{fig:moments_part1}
\end{figure*}

\begin{figure*}
    \subfigure{
    \includegraphics[width = 0.48\textwidth]{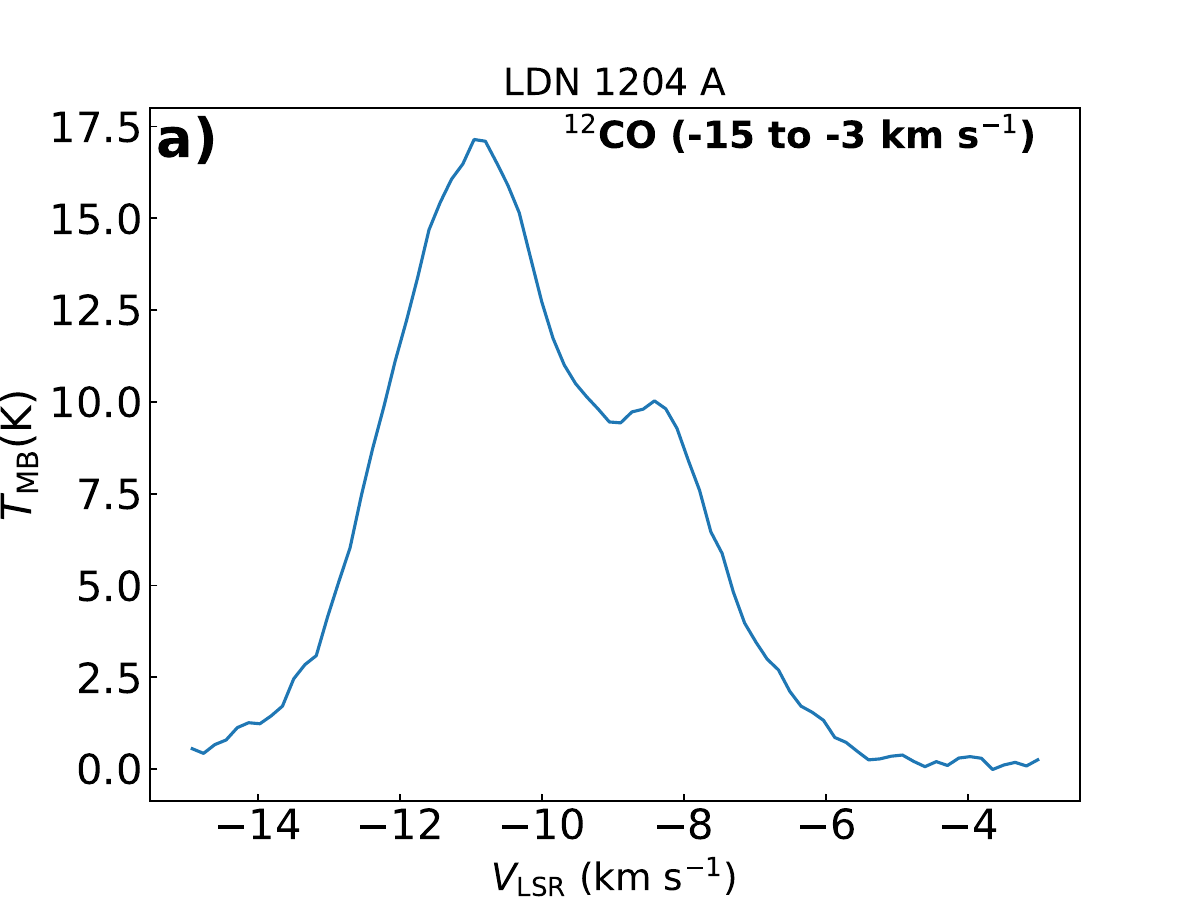}}
    \subfigure{
        \includegraphics[width=0.48\textwidth]{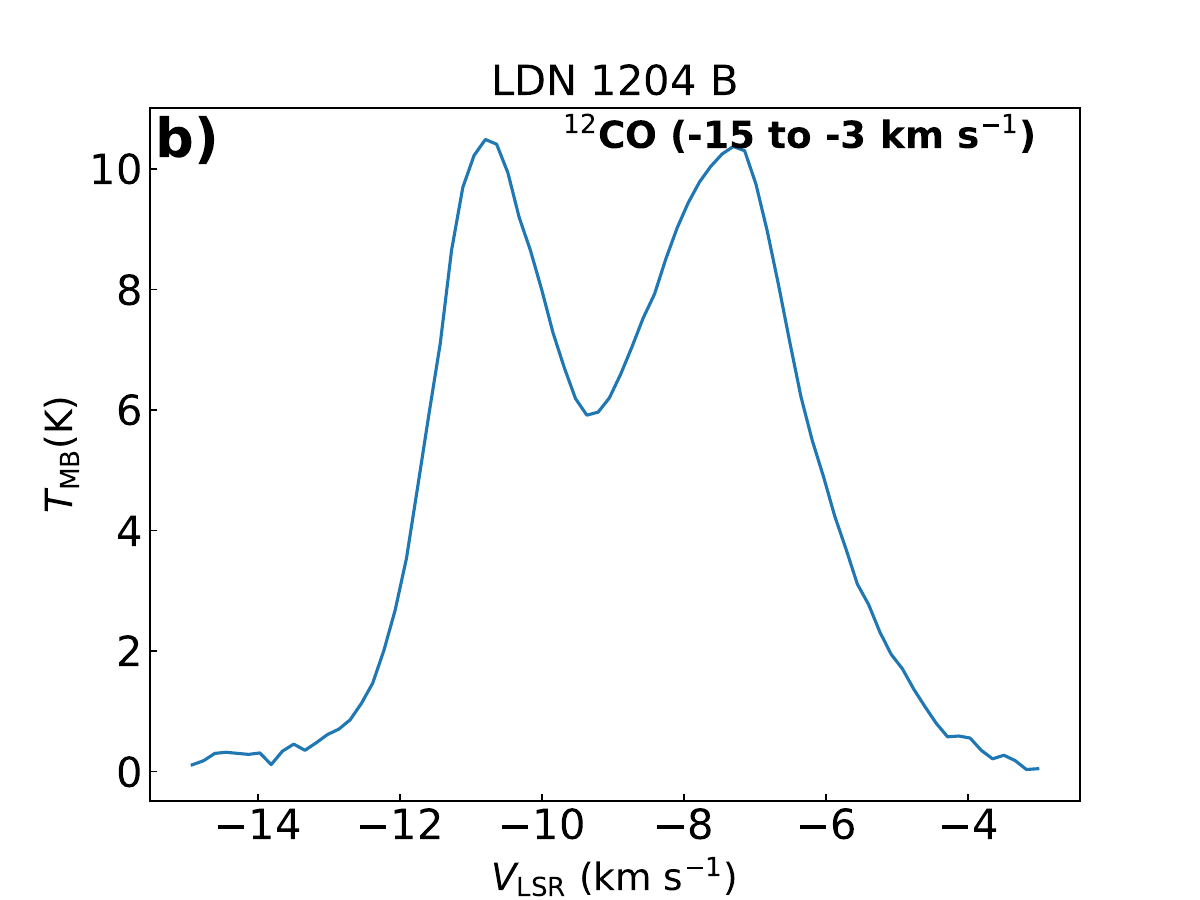}
    }
    \subfigure{
        \includegraphics[width=0.48\textwidth]{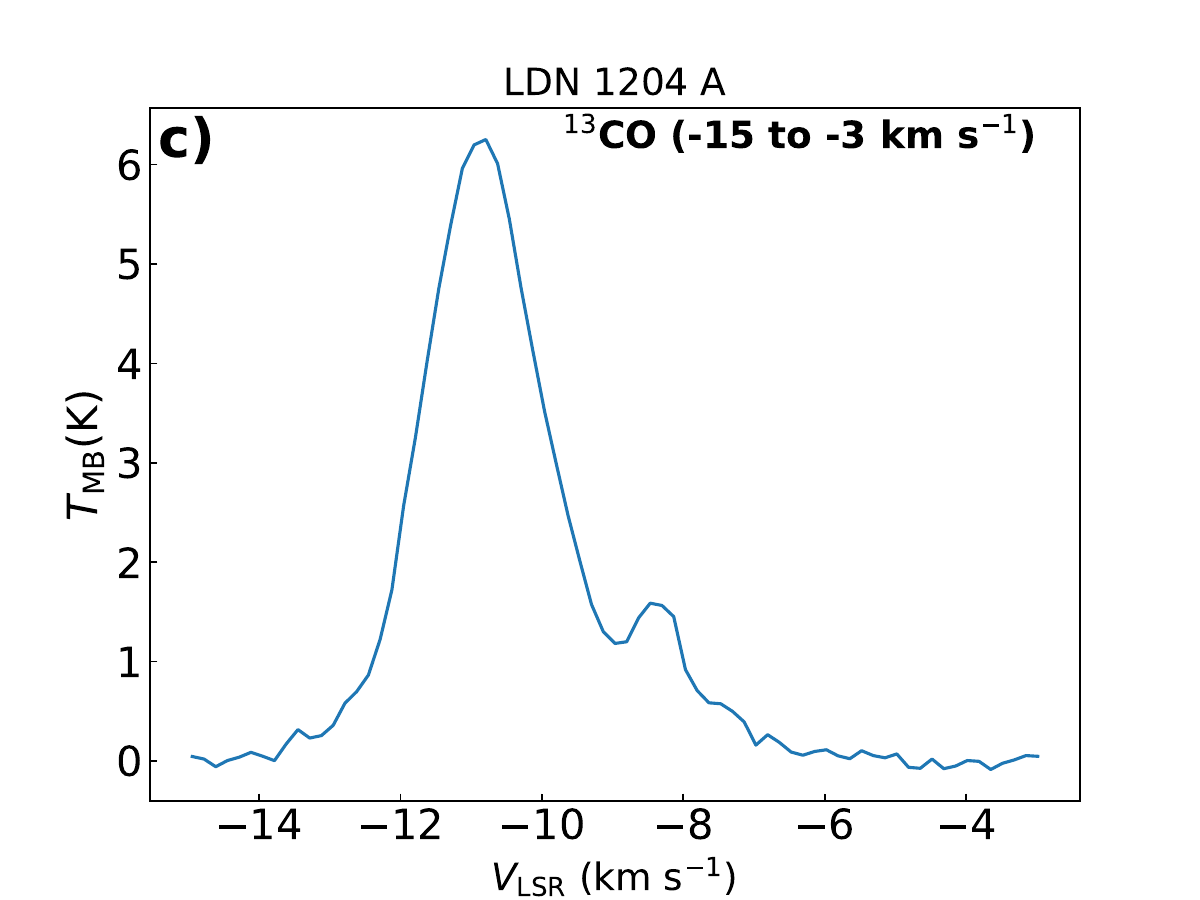}
    }
    \subfigure{
        \includegraphics[width=0.48\textwidth]{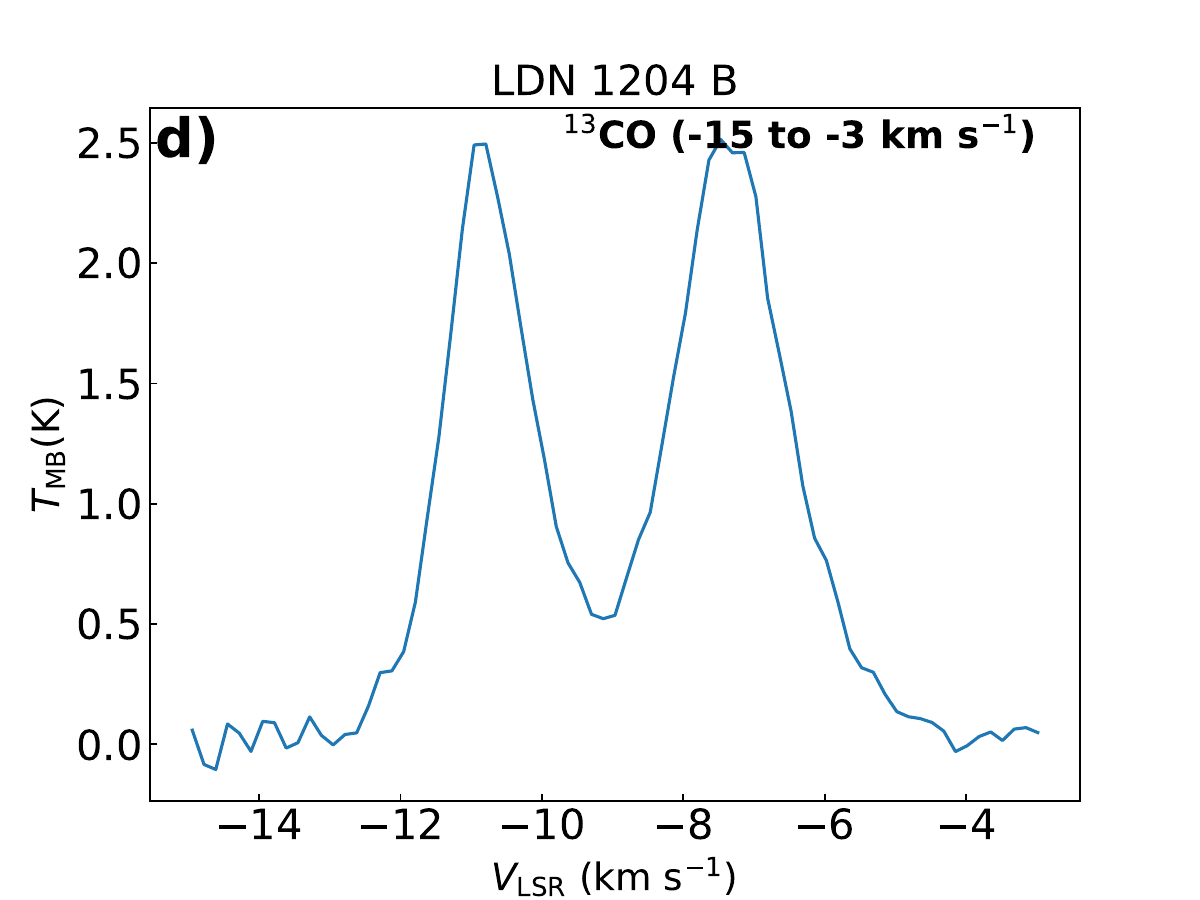}
    }
    \caption{a) and b)$^{12}$CO(1-0) average molecular line spectra for LDN 1204 A and B, c) and d) similarly,$^{13}$CO(1-0) average molecular line spectra for both the regions. The spectrums have velocity range from -15 km s$^{-1}$ to -3 km s$^{-1}$.}
    \label{fig:spectrum}
\end{figure*}

The molecular data cube obtained from PMO has a velocity resolution of 0.158 km s$^{-1}$, which is sufficient to resolve the interesting features in the region containing `A' and `B'. We see a distinct double peak in the molecular line spectrum around `A' and `B' which motivated us to plot the spectrum around
a circular aperture of radius $134^{\prime \prime}$ around `A' and $338^{\prime \prime}$ around `B' (Figure \ref{fig:moments_13co}). These numbers are selected to obtain the clear and distinct double peaks in the two regions. Here, we use $^{12}$CO and $^{13}$CO to plot the mean molecular spectra as shown in Figure~\ref{fig:spectrum}. The parameters obtained from Gaussian fitting are shown in the Table. \ref{tab:cloudvel}. 

\begin{figure*}
    \subfigure{
        \includegraphics[width=0.48\textwidth]{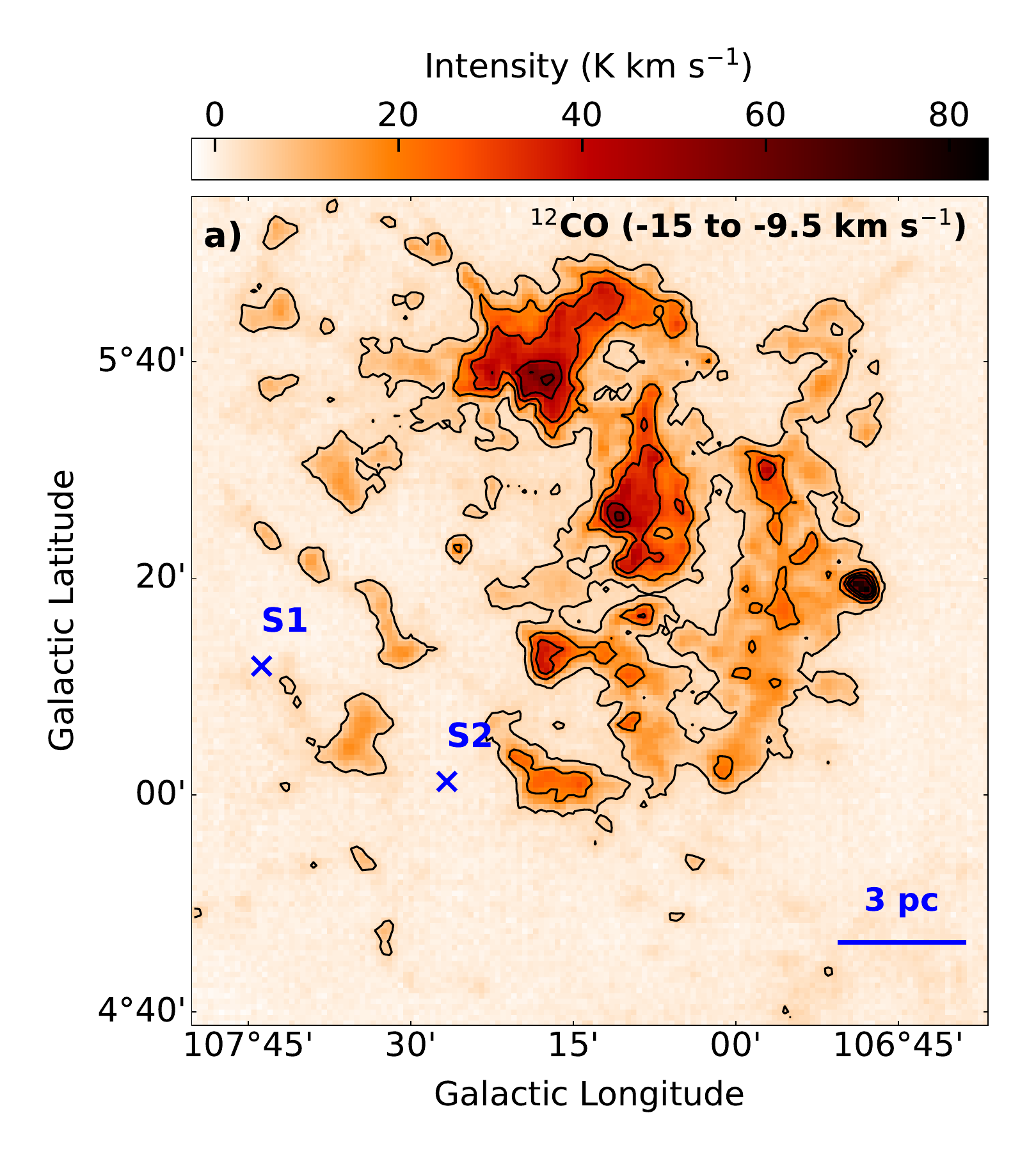}
        \label{fig:twoclouds12co_1}
    }
    \subfigure{
        \includegraphics[width=0.48\textwidth]{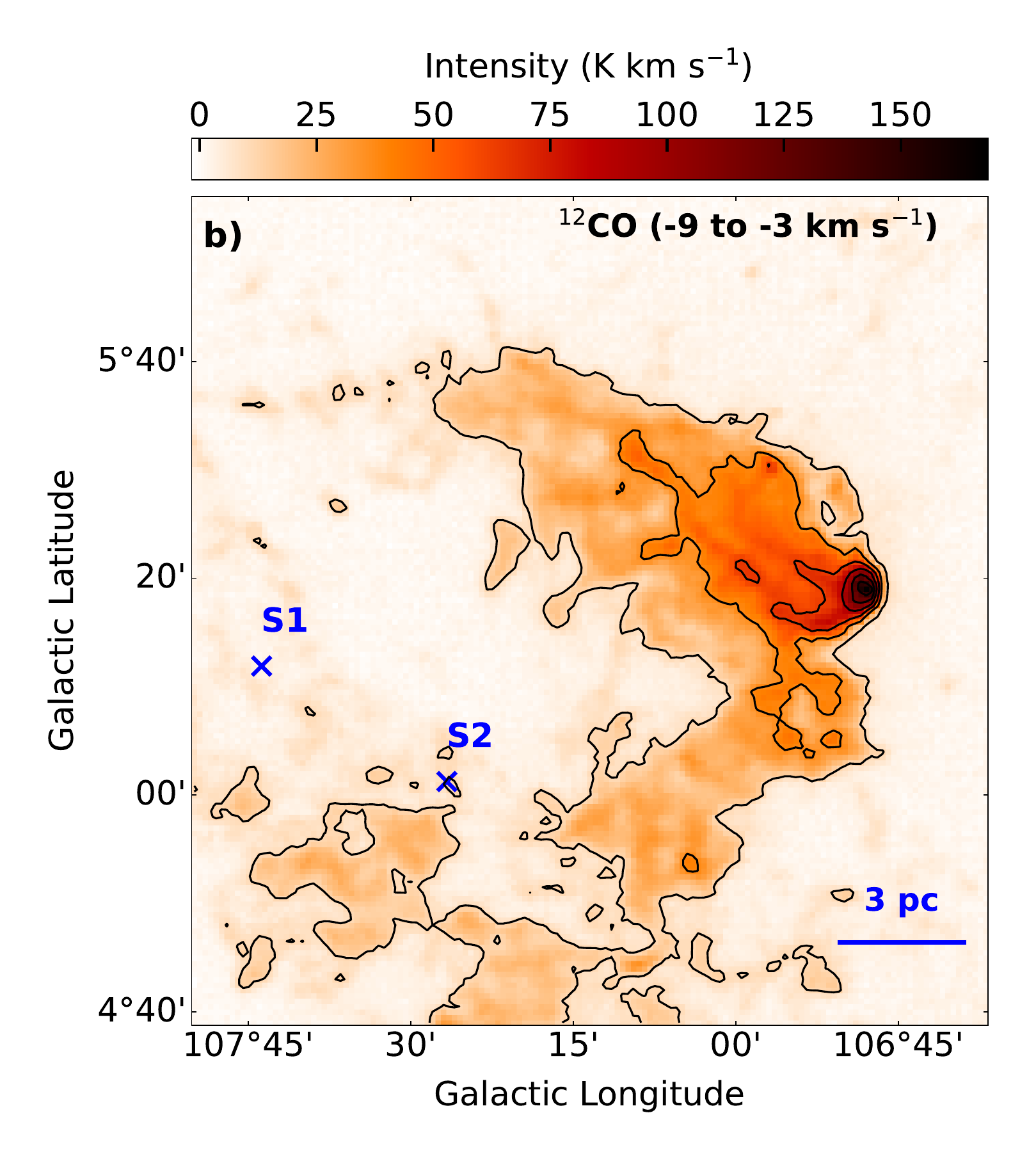}
        \label{fig:twoclouds12co_2}
    }
    \subfigure{
        \includegraphics[width=0.48\textwidth]{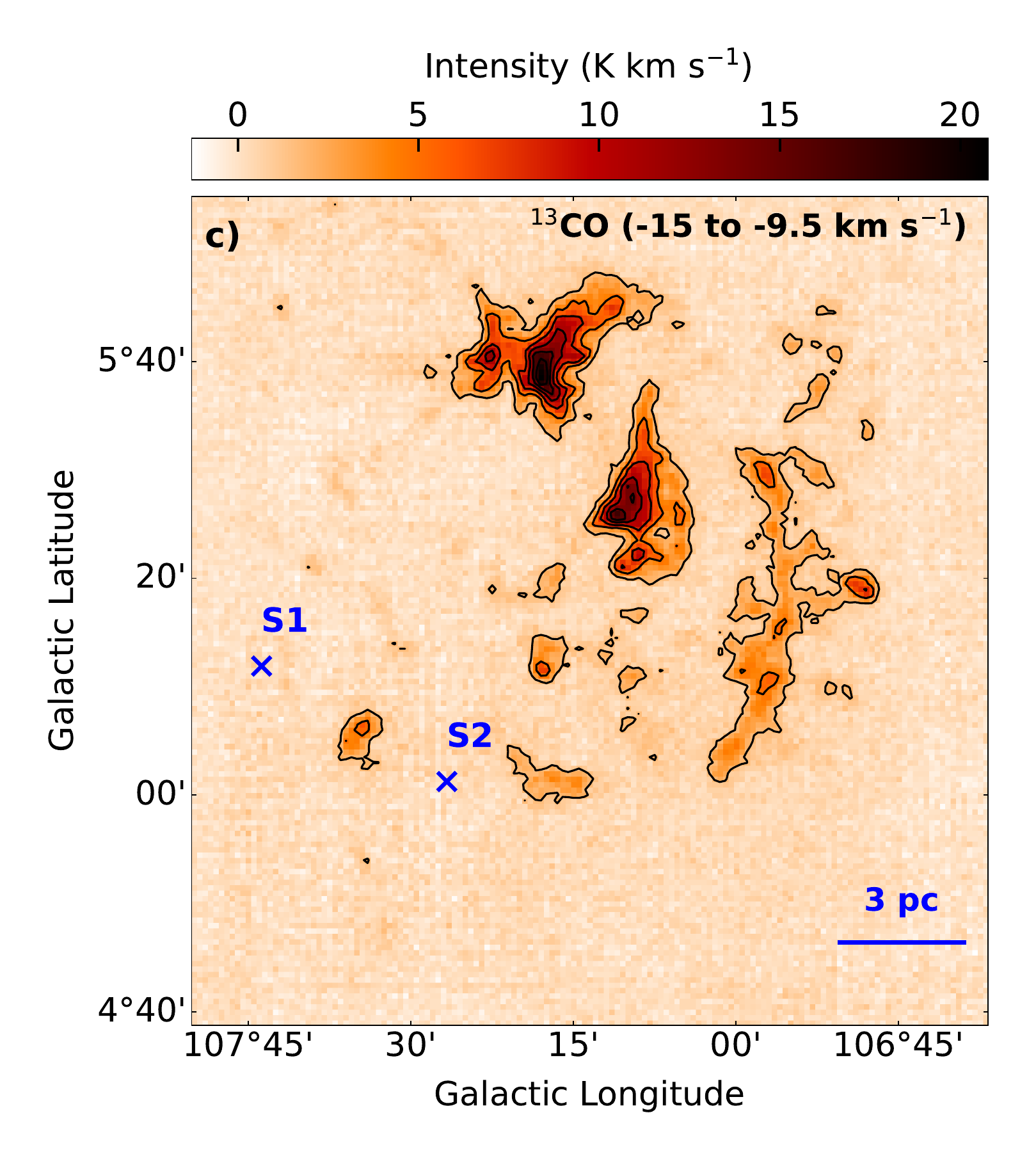}
        \label{fig:twoclouds13co_1}
    }
    \subfigure{
        \includegraphics[width=0.48\textwidth]{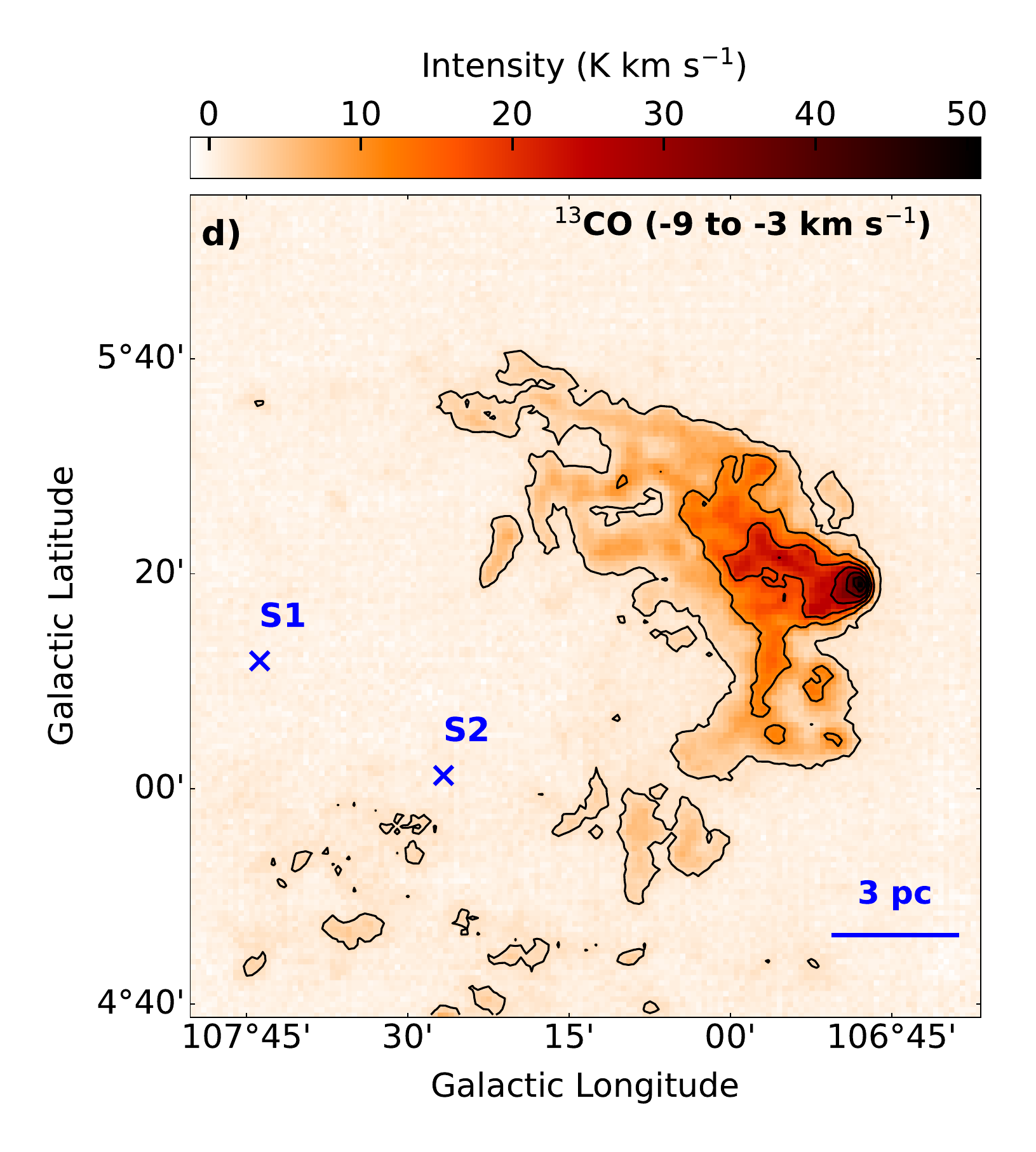}
        \label{fig:twoclouds13co_2}
    }
    
    \caption{Moment 0 map ($^{12}$CO) of Cloud 1 (velocity range: -15 km $s^{-1}$ - -9.5 km $s^{-1}$) with contour levels 4.95, 18.16, 31.37, 44.58, 57.79,
       71.00 and 84.21K km s$^{-1}$, b) moment 0 map ($^{12}$CO) of Cloud 2 (velocity range: -9 km $s^{-1}$ - -3 km $s^{-1}$) with contours 11.05,  37.32, 63.58, 89.85,
       116.12, 142.38 and 168.65 K km s$^{-1}$ ,c) moment 0 map ($^{13}$CO) of Cloud 1 (velocity range: -15 km $s^{-1}$ - -9.5 km $s^{-1}$) with contour levels 1.63,  4.82,  8.01, 11.20, 14.39,
       17.57, and 20.76 K km s$^{-1}$, d) moment 0 map ($^{13}$CO) of Cloud 2 (velocity range -9 km $s^{-1}$ - -3 km $s^{-1}$) with contours  2.09, 10.22, 18.36, 26.49, 34.62,
       42.76 and 50.89 K km s$^{-1}$.}
    \label{fig:twoclouds}
\end{figure*}

From the spectra, it is evident that there are two clouds with distinct velocity ranges. We categorize the clouds according to their velocity range from $^{13}$CO spectra as Cloud 1 (velocity range $\sim$ -15 km $s^{-1}$ to -9.5 km $s^{-1}$) and Cloud 2 (velocity range $\sim$ -9 km $s^{-1}$ to -3 km $s^{-1}$). There is a small overlap in velocity between the two clouds as their velocity difference is very small. The integrated intensity map of both clouds in the given velocity range is shown in Figure \ref{fig:twoclouds}. Since the S140 SFR region, which is seen to be part of Cloud 2 is interacting with Sh2-140, we set the distance to Cloud 2 as 906 pc \citep{2023MNRAS.520.1390S}, similarly we set the distance to Cloud 1, which is part of Sh2-145 as 930 pc. For simplicity, we consider both clouds to be at a distance of 900 pc. From the spectra, the velocity difference between two clouds in `A' is $\sim$ 2.3 km $s^{-1}$ and in the region `B' is $\sim$ 3.4 km $s^{-1}$. 

Further, we measured the masses of both clouds 1 and 2 using optically thick $^{12}$CO and optically thin $^{13}$CO. Assuming local thermodynamic equilibrium (LTE), we can find the column density of $^{13}$CO \citep{2008ApJ...679..481P} and then estimate the molecular hydrogen column density using,
\begin{equation}
\frac{N(\mathrm{^{13}CO})}{N(\mathrm{H_2})} = 1.4 \times 10^{-6},
\quad\quad \text{\citep{1982ApJ...262..590F}}.
\end{equation}

Then the masses of individual clouds are measured, from

\begin{equation}
M = \mu_{\mathrm{H_2}}m_H\int N(\mathrm{H_2})dA ,
\end{equation}
where $\mu_{H_{2}} = 2.8$, represents the mean molecular weight per (H$_{2}$) molecule, including the contribution from helium, $m_H$ is the
mass of the hydrogen atom, and A is the area of the cloud.
We calculated the cloud masses only that encompass the regions shown in Figure \ref{fig:moments_part1}, taking special care to remove the outlier yellow cloud in the south (visible in $^{12}$CO moment 1 map). The calculated masses are; for Cloud 1 $\sim$ 3700 M$_\odot$ and for Cloud 2 $\sim$ 11000 M$_\odot$. Cloud 2 is more massive likely due to the presence of the S140 SFR.
\begin{table}
	\centering
	\caption{Parameters obtained from Gaussian fitting of $^{13}$CO spectra for the both embedded region.}
	\begin{tabular}{lccr} 
		\hline
		Cloud & $V_{LSR}$   & $V_{LSR}$ \\
        & ($\Delta$V)  & ($\Delta$V) &\\
        &of first peak (km/s) & of second peak (km/s)\\
		\hline
		L1204 A & -10.85 & -8.52 \\
        &0.82 & 0.75 \\
	      L1204 B & -10.76 & -7.36 \\
          &0.66 & 0.87 \\

		\hline
	\end{tabular}
        \label{tab:cloudvel}
\end{table}





\subsection{Spatial distribution of young stars identified using Spitzer}
We investigate whether young stars are located close to the filaments in `A' and `B' by analyzing the mid-IR data. 
Using $Spitzer$ IRAC/ MIPS data, we extracted sources covering the entire region which also covers both `A' and `B'. Then, we identify the YSOs and classify them into class {\sc i} and {\sc ii} using the criterion of \citep{2009ApJS..184...18G}. It is plotted in Figure \ref{fig:overlap_yso}.

\section{Discussion}
\subsection{Overall morphology of the whole region and star formation}
The two H{\sc ii} regions Sh2-140 and Sh2-145 are located in close proximity to each other (see Figure \ref{fig:sandwich}). The H{\sc ii} region Sh2-145 contains the massive stars HD 213023 of spectral type O9 V (`S1') and BD 62 2078 of spectral type O7.5 V (`S2'), whereas Sh2-140 is associated with the star HD 211880 of spectral type B0.5 V. In Figure \ref{fig:moments_22}, at the positions of `S1' and `S2', a small arc like feature possibly formed due to heating by respective massive stars can also be seen. In the north-east periphery of the H{\sc ii} region, a bright-rimmed cloud (BRC) 44 is located, whereas the dark clouds labeled `LDN~1204A' and `LDN~1204B' (`A' and `B' in our work) are seen projected between the two H{\sc ii} regions. 
In Figure \ref{fig:cgps}, the CGPS 1420 MHz ionized emission shows the two H {\sc ii} regions. 

The BRC 44 in Figure \ref{fig:sandwich} hosts a significant number of YSOs.In \citet{2025ApJ...994..222C}, we presented a detailed analysis of BRC 44, demonstrating that it is subject to the radiative and mechanical influence of Sh2-145. On the basis of the spatial alignment, morphology, and age distribution of the embedded sources, we argued that at least a group of young stars residing within the bright rim may have formed as a consequence of triggered star formation driven by the massive stars in Sh2-145. Nonetheless, robustly inferring a causal connection between stellar feedback and subsequent star formation remains non-trivial. As shown by \citet{2015MNRAS.450.1199D}, observational diagnostics frequently taken to indicate triggered star formation can also arise through spontaneous star formation occurring in the absence of feedback. Therefore, distinguishing between feedback-driven and spontaneous star-formation scenarios requires additional observational constraints beyond spatial coincidence and age gradients. On the other hand, given the morphology and relative location of the dark clouds `A' and `B', it is highly plausible that these clouds, together with adjacent molecular clouds, are likewise affected by the feedback from Sh2-145.
\begin{figure*}
    \subfigure{
    \includegraphics[width =
    0.48\textwidth]{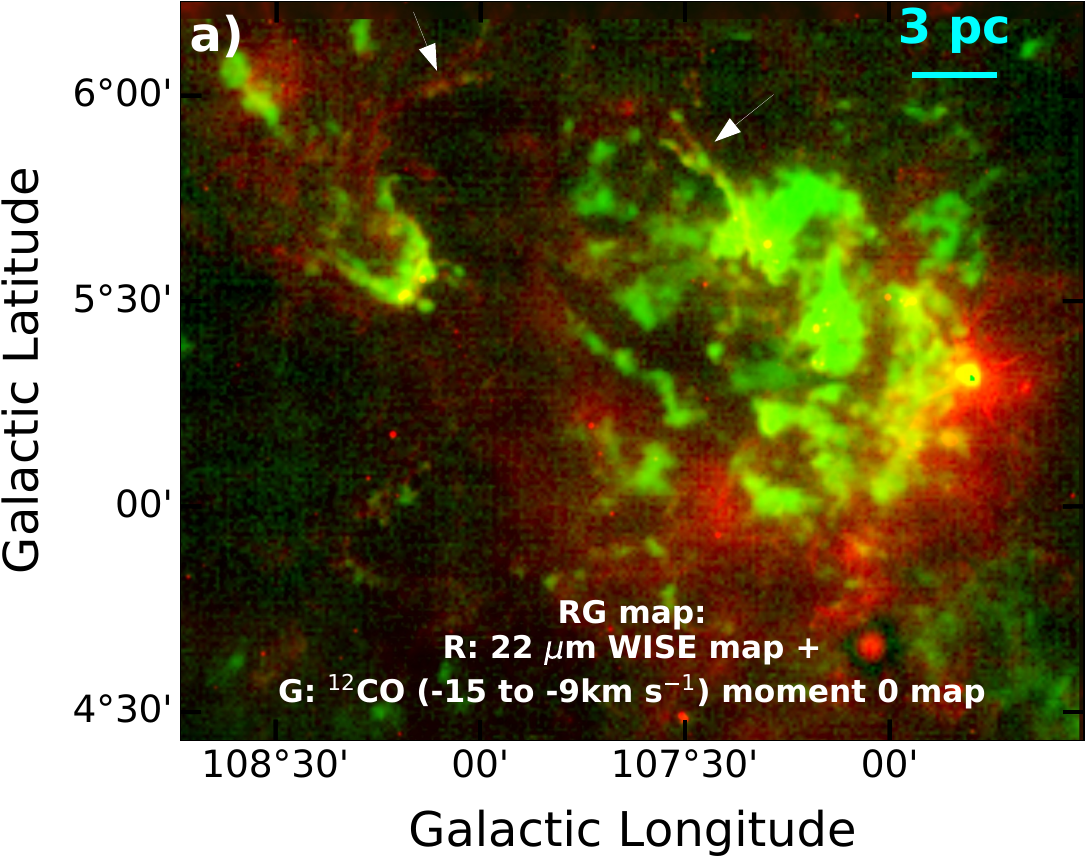}  
    \label{fig:arc1}
    
    }
    \subfigure{
    \includegraphics[width = 
    0.48\textwidth]{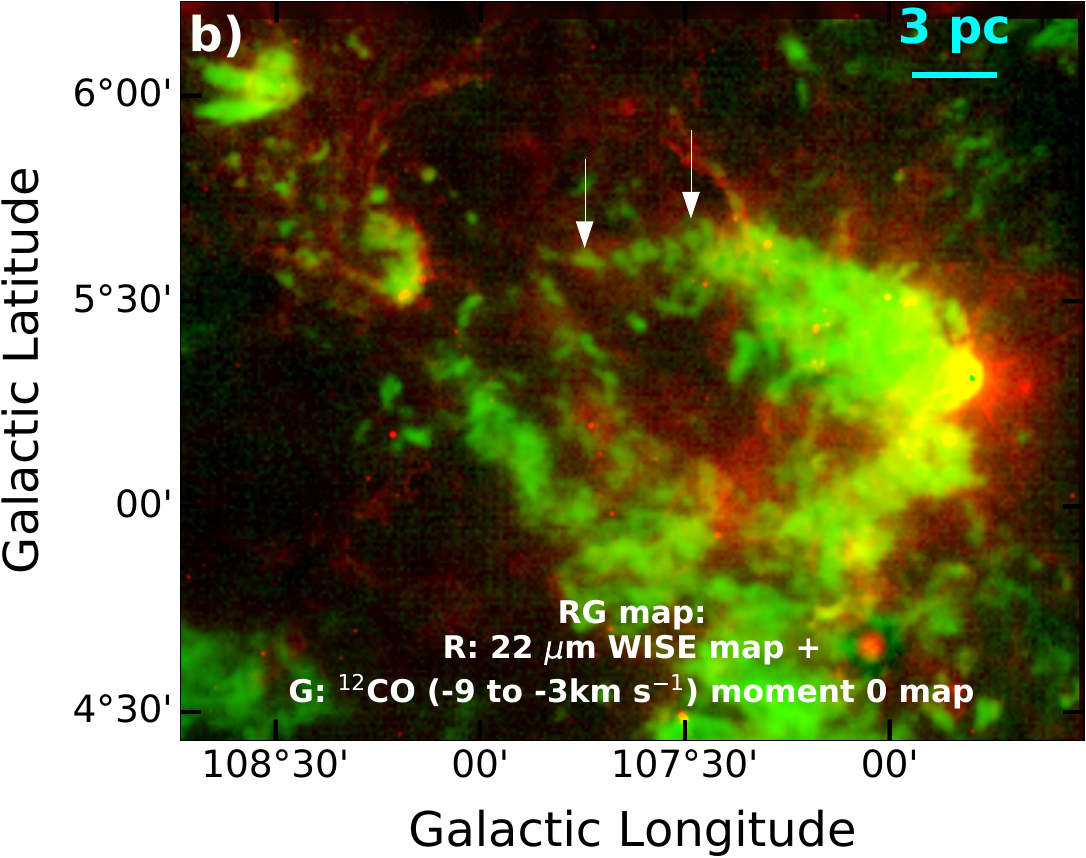}  
    \label{fig:arc2}
    
    }

    \caption{Color composite map of the larger region, a) for Cloud 1, and b)for Cloud 2, green is $^{12}$CO moment 0 map and red is WISE 22 $\mu$m image. Arc-like feature is indicated by the arrows, .}
    \label{fig:arcs}
\end{figure*}

A clearly prominent half-bubble (`arc-like') structure is visible in the mid infrared WISE image in  Figure \ref{fig:wise}, likely formed by the Sh2-145 H {\sc ii} region. Apart from the prominent arc, we also separate another arc, which is less evident in the image. Both arcs are overplotted in CGPS image in Figure \ref{fig:cgps}. Moving to the molecular line data, we note that in all the moment 0 maps, (Figure \ref{fig:moments_part1}) there is a single arc-like feature, opening towards the south-east direction approximately towards the positions of massive stars (`S1' and `S2'). This is the prominent arc-like feature seen in the mid infrared image. In the 22 $\mu$m WISE image shown in Figure \ref{fig:moments_22}, we show that the $^{12}$CO emission closely overlaps with the 22 $\mu$m WISE emission, where the molecular emission also closely traces the arc-like morphology of the WISE image. In Figure \ref{fig:arcs}, this situation is further delineated using the the two clouds separately (Cloud 1 and 2 : mentioned in the section \ref{sec:results_twoclouds}). Here, we see that both expanding clouds also have an arc-like footprint in the 22 $\mu$m WISE image, as indicated by the arrows.

From the two clouds identified, with velocity ranges -15 to -9.5 km s$^{-1}$ (Cloud 1: Figure \ref{fig:twoclouds12co_1} and \ref{fig:twoclouds13co_1}) and -9 to -3 km s$^{-1}$ (Cloud 2: Figure \ref{fig:twoclouds12co_2} and \ref{fig:twoclouds13co_2}), the velocity map is constructed as shown in Figure \ref{fig:moments_part2}, where it clearly distinguishes the two clouds, in the above mentioned velocity ranges. The Cloud 1 is depicted as a darker color, while Cloud 2 is represented as a lighter color. 
The morphological structure of both the clouds in different velocities is clearly visible in the channel maps prepared from $^{12}$CO moment 0 maps. The channel map spanning radial velocities from -15 km s$^{-1}$ to -3 km s$^{-1}$, with a channel width of 1 km s$^{-1}$, is shown in Figure \ref{fig:channelmap}. The molecular clouds become particularly prominent in the velocity range -8 km s$^{-1}$ to -7 km s$^{-1}$ for Cloud 2 and -11 km s$^{-1}$ to -10 km s$^{-1}$ for Cloud 1. 
The wing component of the Gaussian in the spectra (Figure \ref{fig:spectrum}) is traced by the remaining channel maps. It is noteworthy that the S140 SFR region is detected in every channel map, but it is particularly prominent in Figure \ref{fig:channelmap} (c). In the other channels, S140 SFR predominantly exhibits a broad Gaussian wing profile.

Furthermore, to study the star formation scenario in these regions, we identified YSOs using $spitzer$ data in the LDN 1204 clouds which are shown in Figure \ref{fig:overlap_yso}. 
We also classified the detected YSOs into Class {\sc i} and Class {\sc ii}. Most of them show Class {\sc ii} signature, with class {\sc i} sources present inside the cloud around `A' and `B' and S140 SFR, and most Class {\sc ii} outside the cloud where gas is almost dispersed. 

Now, since we see that at `A' and `B' the double peak was detected, this suggests that Cloud 1 and 2 overlap in those regions.
Figure \ref{fig:overlap} shows the overlap regions, where a color composite image is constructed from both clouds. With enough evidence suggesting H {\sc ii} region shaping the molecular cloud, and the presence of two separate clouds with different but overlapping velocities, we move on to study if the H {\sc ii} region is expanding and so is the cloud too expanding with it.

We also note an asymmetry in the blue profiles in  the spectrum in Figure \ref{fig:spectrum} for $^{12}$CO/ $^{13}$CO, with the blue shifted emission being relatively stronger than the red shifted emission, which may suggest a collapse scenario \citep{2012ApJ...750...64S}. However, in the present case, the line profile is complicated by the superposition of two molecular components, Cloud 1 (-15 to -9.5 km s$^{-1}$) and Cloud 2 (-9 to -3 km s$^{-1}$), with distinct line of sight velocities. Moreover, the $^{12}$CO emission traces the outer molecular boundary, which appears to be influenced by the interaction of the two expanding H {\sc ii} regions. Therefore, the observed blue asymmetry cannot be uniquely attributed to gravitational collapse. Although active star formation is present in the region, higher angular resolution observations combining optically thick and optically thin molecular tracers are required to establish whether infall is occurring toward `A' and `B'. 

\subsection{Expansion of the clouds}
\label{sec:exp}
To investigate cloud dynamics, which may be modified by the influence of the H{\sc ii} region produced by massive stars in the vicinity, we carried out an analysis of the molecular line emission. In this study, we construct position – velocity ($p$ –$v$) diagrams to examine the characteristic `C'-shaped morphology that arises when a molecular cloud undergoes expansion in the proximity of massive stars \citep[e.g.,][]{2011ApJ...742..105A}. \citet{2011ApJ...742..105A} analyzed the behavior of molecular clouds in and around bubbles driven by massive stars by means of $p$ – $v$ diagrams. Such diagrams, in conjunction with integrated intensity maps, serve as powerful diagnostics for probing the kinematics of molecular gas.

\begin{figure*}
    \subfigure{
    \includegraphics[width =
    0.5\textwidth]{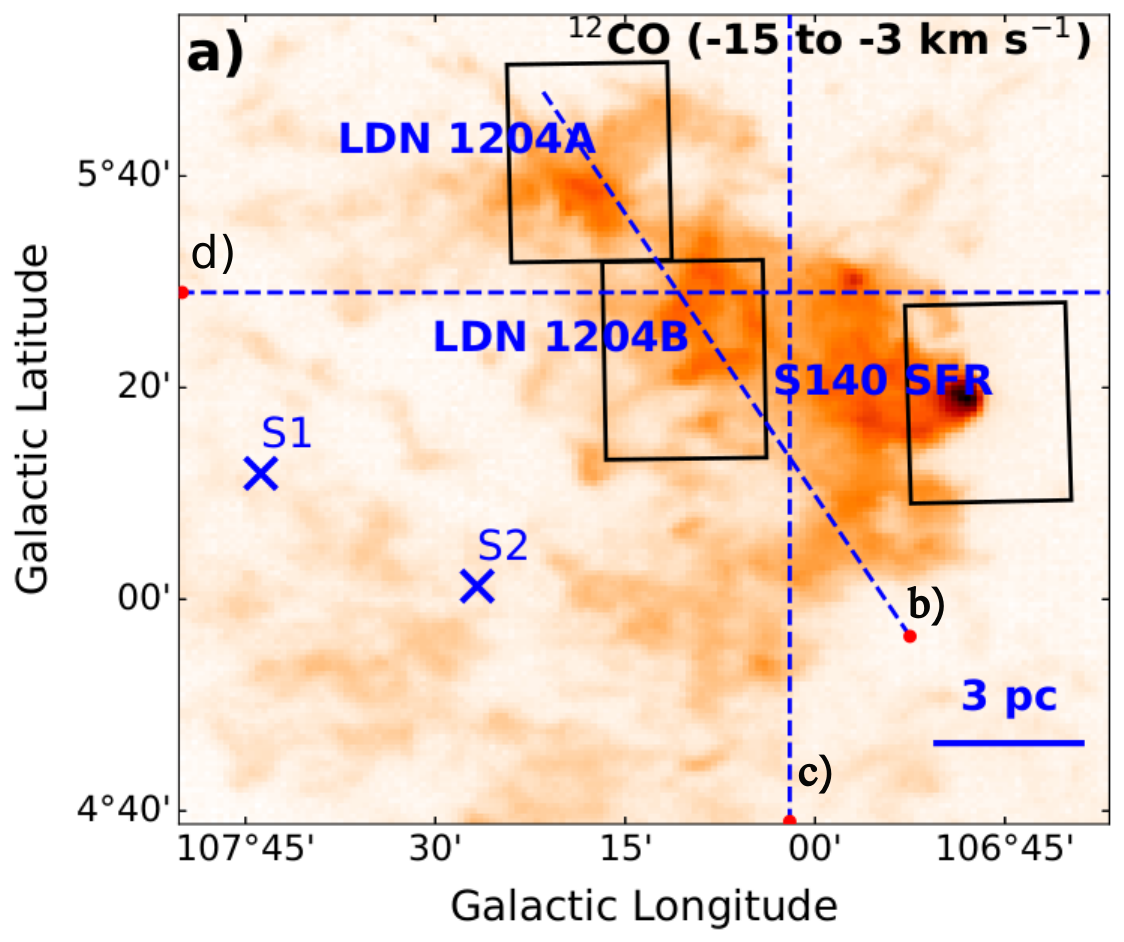}  
    \label{fig:pv_moment}
    
    }
    \subfigure{
    \includegraphics[width = 
    0.5\textwidth]{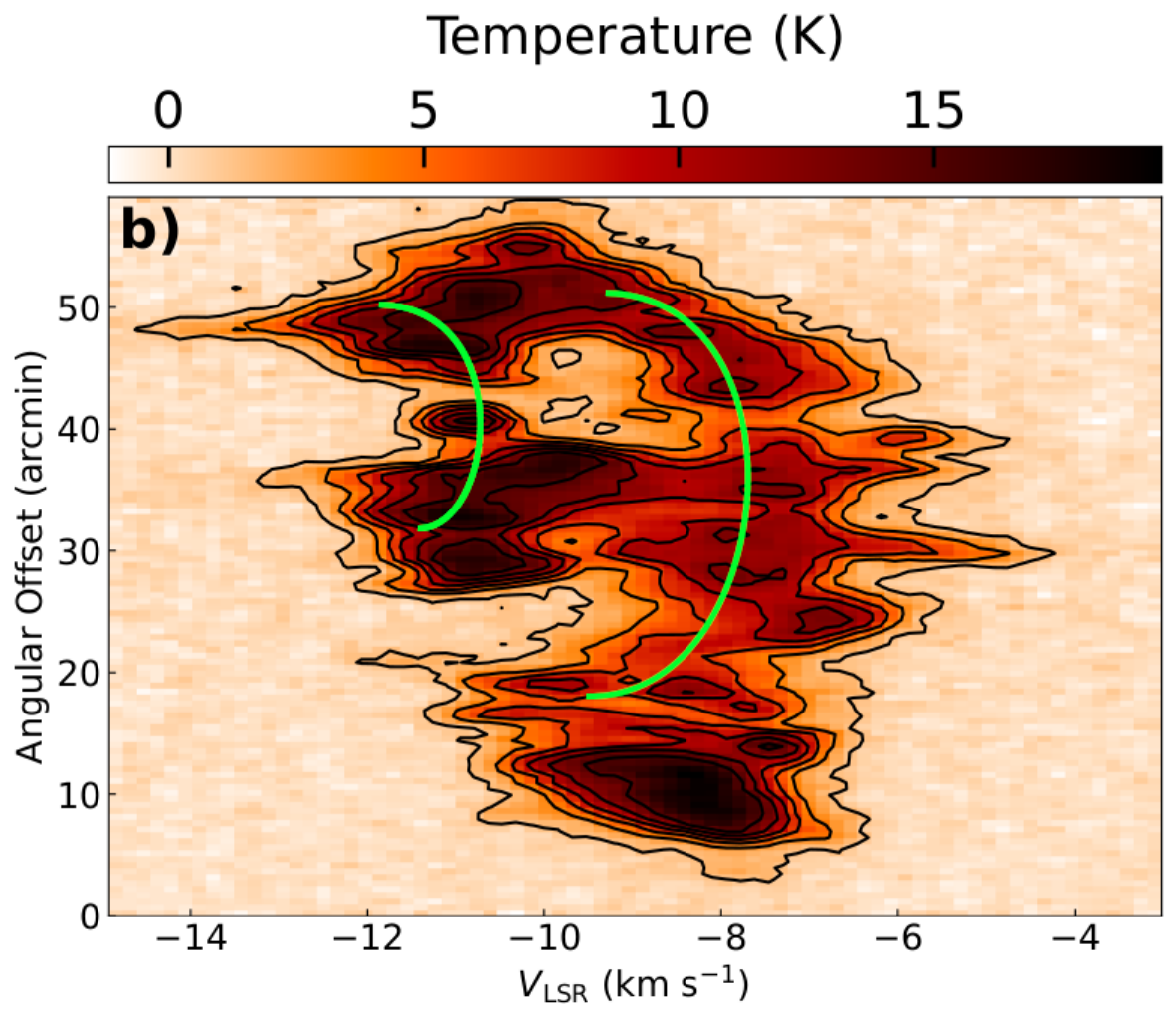}  
    \label{fig:pv}
    
    }
        \subfigure{
    \includegraphics[width = 
    0.5\textwidth]{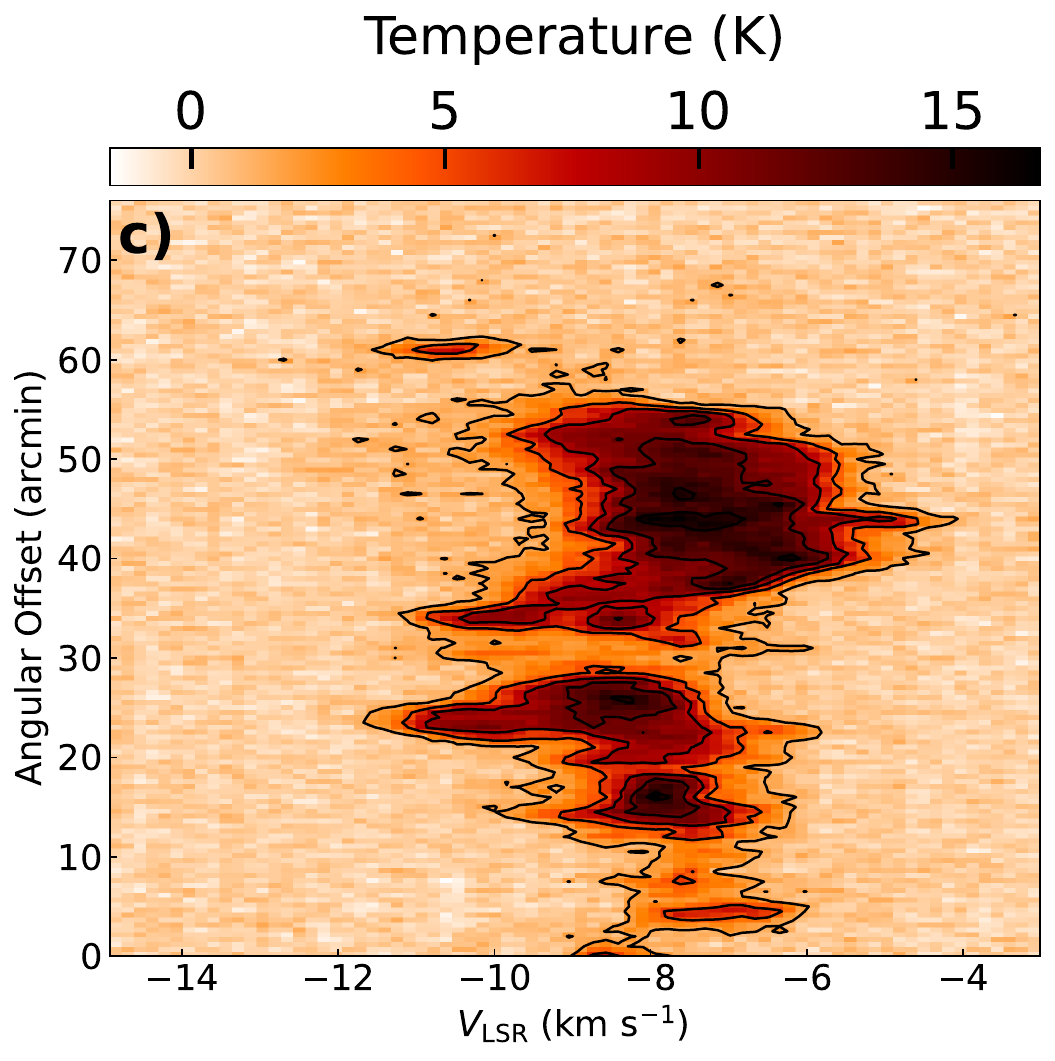}  
    \label{fig:pv_lon}
    
     }
        \subfigure{
    \includegraphics[width = 
    0.5\textwidth]{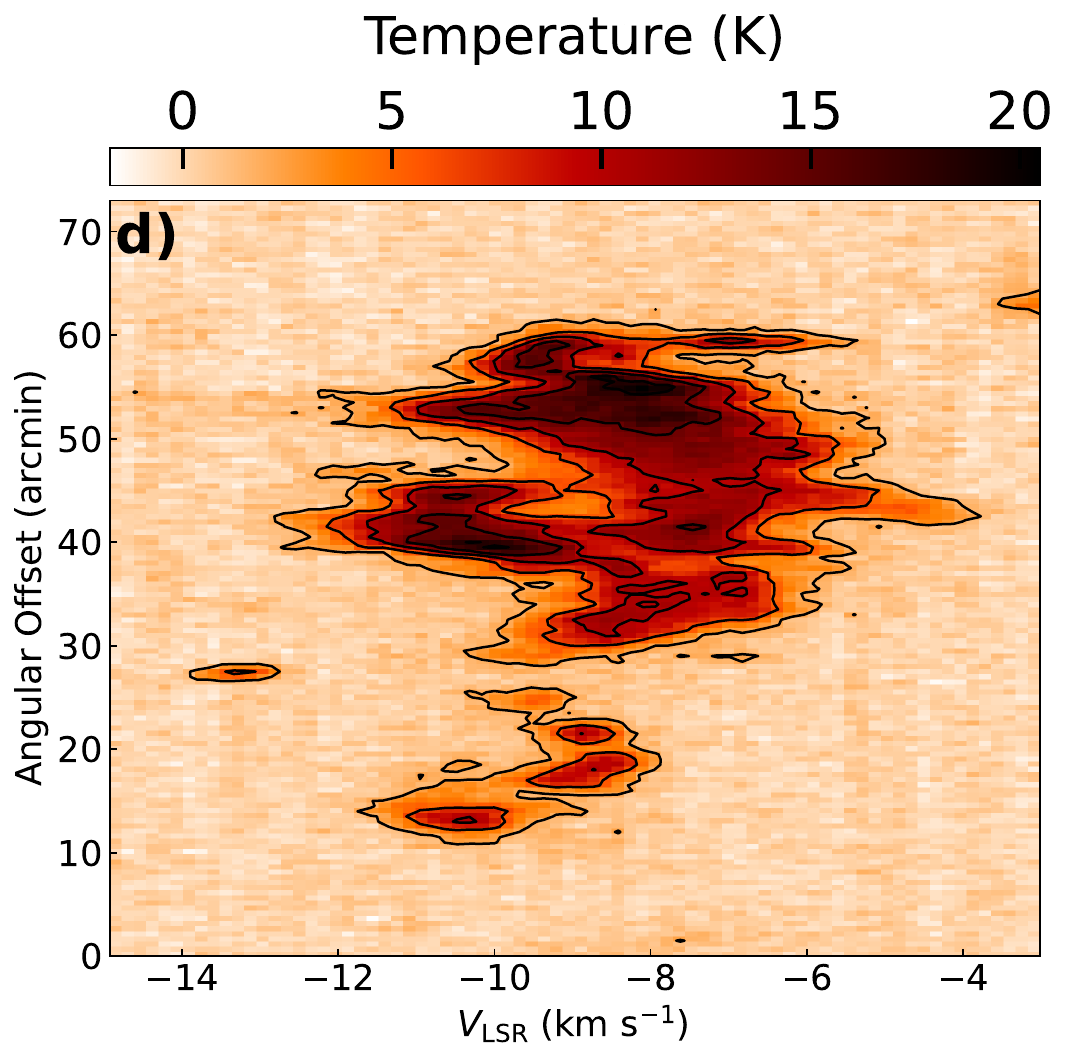}  
    \label{fig:pv_lat}
    
    }

    \caption{a)$^{12}$CO moment-0 map of the region, along with line along which the $p$ - $v$ diagrams are constructed. b)$p$ - $v$ diagram along the diagonal cut taken perpendicular to the direction of the positions of the massive stars. The mirror `C' morphology is marked in the $p$ - $v$ diagram, c)$p$ - $v$ diagram along constant longitude cut, d)$p$ - $v$ diagram along constant latitude cut.
     The origin of angular offset in $p$ - $v$ diagram is marked as red dot. The cross marks indicate the position of massive stars.}
\end{figure*}

In Figure \ref{fig:pv}, we have shown the $p$ – $v$ diagram. The `C'-shaped morphology is clearly visible as a consequence of a diagonal $p$ – $v$ cut taken along the arc-like structure of the cloud (Figure \ref{fig:pv_moment}). This diagonal cut is oriented approximately perpendicular to the line connecting the positions of the massive star(s), allowing the expansion signature to be effectively captured. The observed inverted `C' pattern indicates red-shifted expansion of the molecular gas, consistent with feedback-driven dynamics. The inverted `C' is pattern is also seen in the $p$ – $v$ diagrams across a constant longitude and latitude cut as shown in Figures \ref{fig:pv_lon} and \ref{fig:pv_lat}. The inverted `C' pattern is also discernible in $p$ – $v$ constructed individually for both clouds as given in \ref{appendix}.

\begin{figure*}
    \subfigure{
    \includegraphics[width =
    0.5\textwidth]{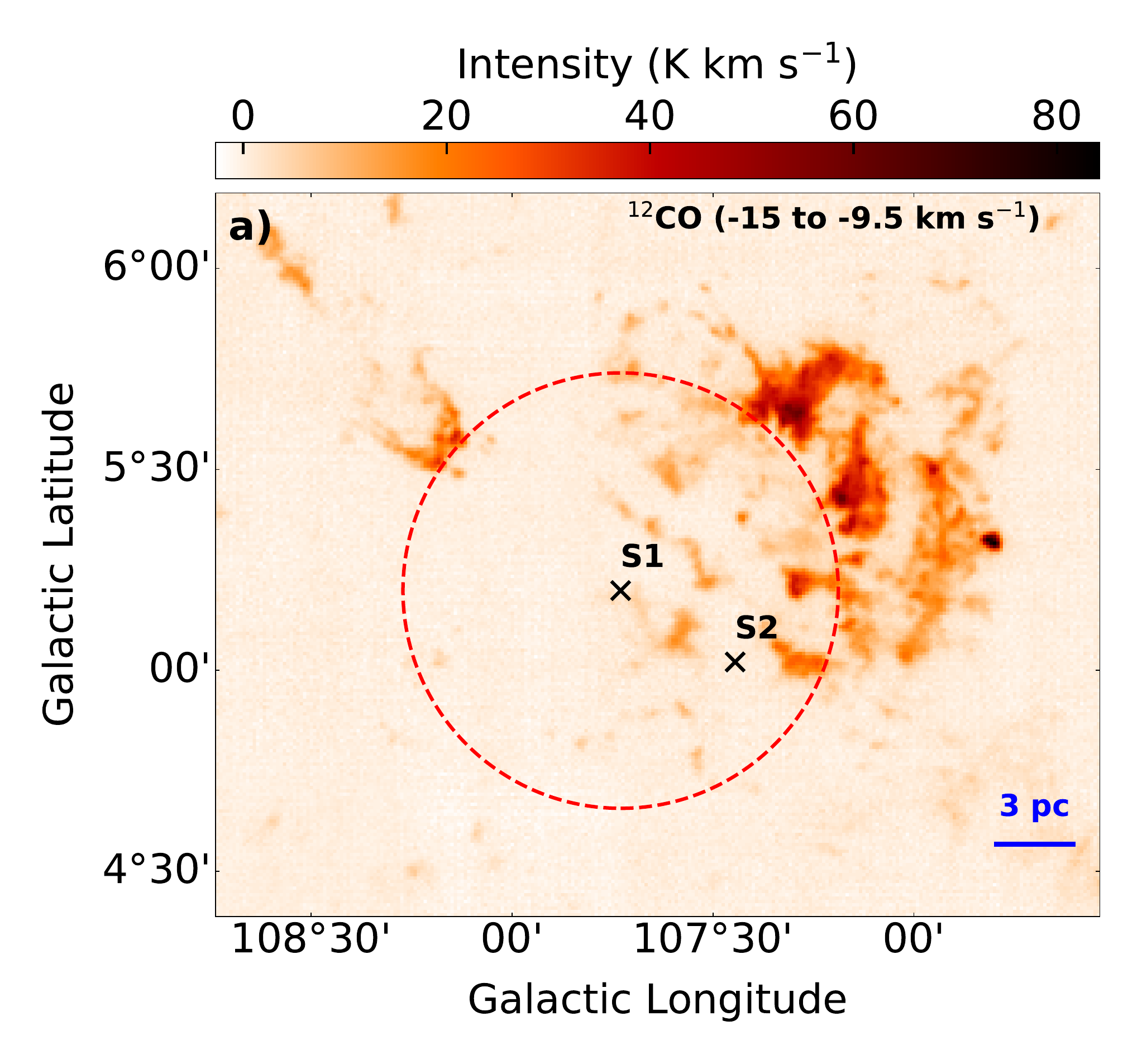}
    \label{fig:hiiregion1}
    
    }
    \subfigure{
    \includegraphics[width = 
    0.5\textwidth]{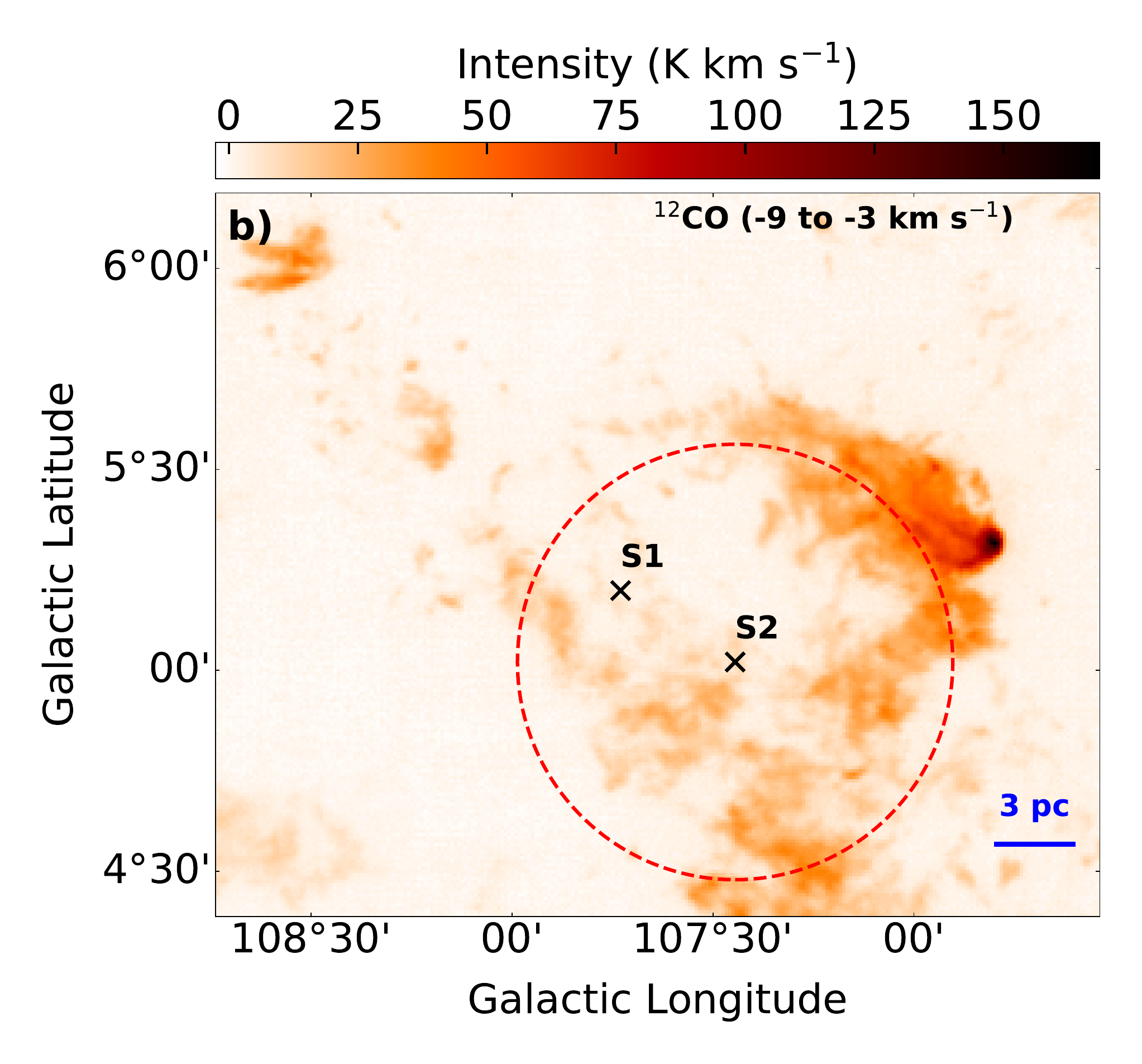} 
    \label{fig:hiiregion2}
    
    }
    \subfigure{
    \includegraphics[width = 
    0.5\textwidth]{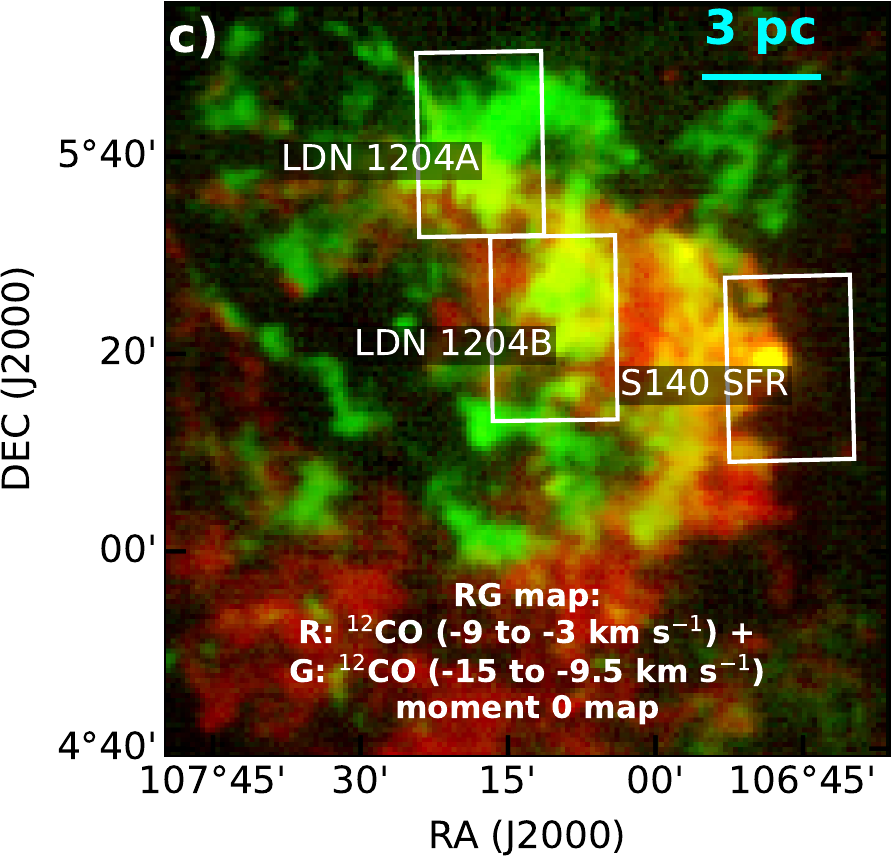} 
    \label{fig:overlap}
    
    }
        \subfigure{
    \includegraphics[width = 
    0.5\textwidth]{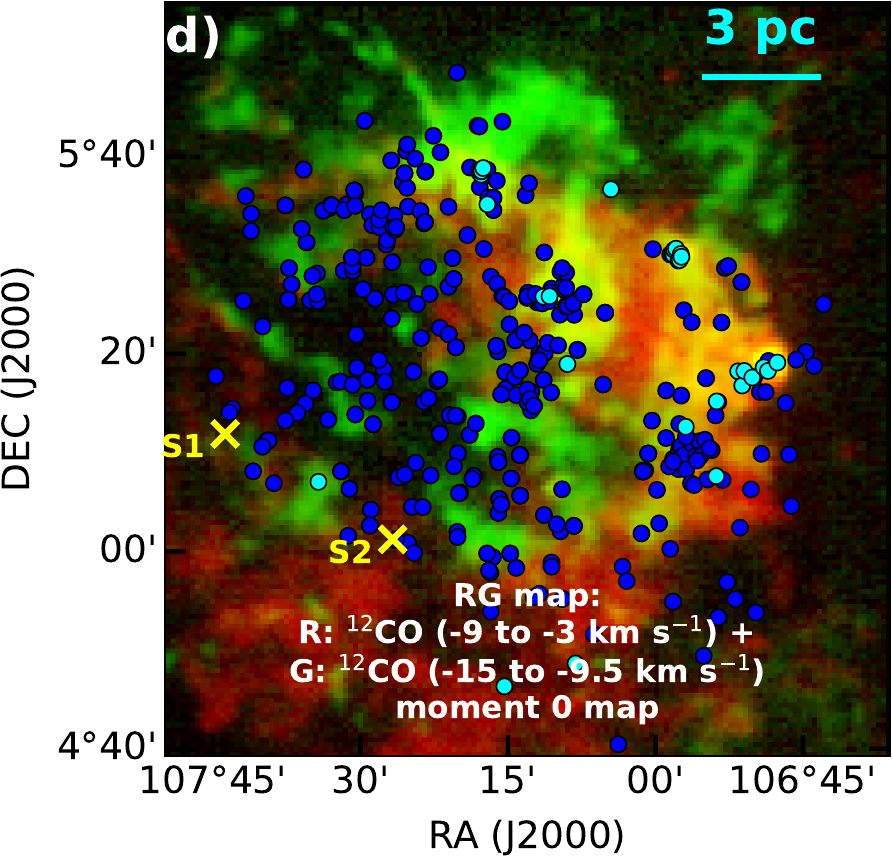} 
    \label{fig:overlap_yso}
    
    }

    \caption{a)H {\sc ii} region (dashed circles) centered around `S1', associated with Cloud 1 ($^{12}$CO moment 0 : -15 km s$^{-1}$ to -9.5 km s$^{-1}$, b)H {\sc ii} region centered around `S2' associated with Cloud 2 ($^{12}$CO moment 0 : -9 km s$^{-1}$ to -3 km s$^{-1}$. c)$^{12}$CO color composite image of Cloud 2 (red) and Cloud 1 (green) showing the overlap regions. d) same figure as (c), with Class {\sc i} (cyan) and Class {\sc ii}(blue) overplotted.}
     \label{fig:hiiregion}
\end{figure*}

Furthermore, we identify clouds that exhibit expansion signatures using channel maps (see Figure \ref{fig:channelmap}). As shown in Figure \ref{fig:hiiregion}, the morphology of the H{\sc ii} region associated with both massive stars is clearly delineated. It appears that the star `S1' predominantly ionizes the region associated with Cloud 1 (Figure \ref{fig:hiiregion1}), while `S2' is responsible for ionizing Cloud 2(Figure \ref{fig:hiiregion2}). However, given the evolved nature of the H{\sc ii} region, the classical `C'-shaped morphology and clear ionization-driven structures are less pronounced. Furthermore, we can also conclude that Sh2-145 is likely a combined H {\sc ii} region formed out of the two ionizing sources.


\subsection{The Bubble-filament paradigm : Feedback-driven formation of filamentary structures and ongoing star
formation}
We now discuss the possible origin and evolutionary state of the filamentary structures identified
toward regions A and B. The \textit{Herschel} temperature and column density maps (Figures \ref{fig:temp} and \ref{fig:col}) provide a first indication of hub--filament-like configurations in both regions, where the
filamentary structures are embedded within relatively dense molecular material. The presence of
YSOs, including the S140 SFR (Figure \ref{fig:overlap_yso}), further demonstrates that these
structures are associated with ongoing star-formation activity. The molecular gas is also
kinematically disturbed, as indicated by the enhanced velocity dispersion toward and around A and
B in the moment-2 map (Figure \ref{fig:13co_moment2}). Although enhanced line widths can result from turbulence,
unresolved velocity components, compression, or stellar feedback, their spatial coincidence with the
dense filamentary structures suggests that the gas in these regions is dynamically active rather than
quiescent.
The large-scale morphology of the complex provides a possible explanation for this dynamical
state. The H II regions associated with the massive stars `S1' and `S2' appear to be expanding into their
surrounding molecular material. The spatial association of `S1' with Cloud 1 and `S2' with Cloud 2
suggests that these clouds constitute the respective parent molecular environments of the two
massive stars. Importantly, the two molecular components overlap toward A and B, where double peaked molecular line profiles reveal distinct velocity components along the same lines of sight.

This spatial and kinematic correspondence suggests that the two clouds may be interacting in these
regions. If the expanding H {\sc ii} regions are simultaneously driving their respective parent clouds
toward this interface, the combined effect of feedback-driven compression and interaction between
the molecular components could naturally produce the enhanced density and complex filamentary morphology observed toward `A' and `B'.
Such a feedback-driven origin for dense filamentary structures is supported by both observations
and simulations. \citet{2020A&A...638A...7Z} showed that expansion of the H {\sc ii} region RCW 120 can
compress the surrounding molecular gas into a dense shell and promote the formation of
filamentary structures associated with subsequent star formation. \citet{2015A&A...580A..49I} proposed a
related evolutionary picture in which expanding H II regions generate compressed atomic and
molecular layers that can subsequently fragment into filaments and dense star-forming structures.
Numerical simulations by \citet{2017A&A...599A..94N} further demonstrated that interactions between
expanding shells can preferentially produce filamentary structures, particularly in magnetized
molecular gas. These results provide a physical framework in which stellar feedback can transform
initially more extended molecular material into dense, elongated structures.
The morphology observed in LDN 1204 is particularly interesting in this context because the
candidate hub--filament structures are located close to the apparent interface between the two
molecular components. A similar configuration has been proposed for GMC G288.5+1.5, where
\citet{2015ApJ...799...64D} suggested that the cloud may have formed at the stagnation point between two
expanding H {\sc i} supershells. In such a scenario, converging feedback-driven flows can enhance the
gas density at their interface and promote the formation of molecular and filamentary structures.
The coincidence in LDN 1204 of the molecular-component overlap, enhanced velocity dispersion, dense filaments, and ongoing star formation is therefore consistent with a picture in which the
interaction of the two feedback-affected molecular environments has contributed to the assembly of the filamentary structures in A and B.

The resulting configuration may subsequently evolve through the accumulation and redistribution
of gas along the interconnected filaments. In particular, the presence of sub-filamentary structures
connected to the main filaments suggests that the filament network may provide pathways through
which material can be transported toward the denser central regions. However, the present
observations do not yet establish whether such longitudinal flows are occurring or whether they are
sustaining the star formation within the candidate hubs. Demonstrating this connection would
require direct kinematic evidence for coherent gas motions along the filaments and, ideally, an
estimate of the associated mass-transport rate.

Taken together, our observations favor a scenario in which stellar feedback and large-scale
molecular gas interaction have jointly shaped the present day structure of the LDN 1204 complex.
The expansion of the `S1' and `S2' H {\sc ii} regions provides a mechanism for compressing and
reorganizing their respective parent clouds, while the spatial and kinematic overlap of the two
molecular components toward `A' and `B' may introduce an additional converging or interacting flow.
The resulting dense filamentary structures are associated with enhanced velocity dispersion and
ongoing star formation, suggesting that the interaction is occurring in a dynamically active
environment. Thus, LDN 1204 may represent an example of how feedback-driven flows can
contribute to the assembly of filamentary structures and create favorable conditions for subsequent
star formation, although the detailed evolutionary connection between cloud interaction, filament
formation, and gas accretion onto the central hubs remains to be established.

\section{Summary and Conclusions}
To investigate the ongoing physical processes in the extended molecular cloud hosting the LDN~1204, Sh2-145, and Sh2-140 regions, we employed a multi-wavelength observational approach. We analyzed the cloud morphology, gas kinematics, and star formation activity using infrared data from {\it Spitzer} and WISE, far-infrared and sub-millimeter observations from the {\it Herschel} Space Observatory, and molecular line data from the PMO telescope. The main results are summarized as follows.

\begin{itemize}

\item Mid-infrared images reveal at least two semi-arc-/shell-like structures (extent $\sim$14--18 pc) toward the target region. The previously identified massive stars `S1' and `S2' appear to be physically associated with these infrared structures, both of which contain ionized emission.

\item {\it Herschel} temperature and column density maps (resolution $\sim$10.7 arcsec) reveal two filamentary systems associated with LDN~1204A and LDN~1204B, each containing at least four filamentary structures. Their morphology suggests the presence of a hub--filament system (HFS) candidate in each LDN region. Both LDN~1204A and LDN~1204B are located at the periphery of the H\,{\sc ii} region associated with Sh2-145.

\item The {\it Herschel} temperature map shows dust temperatures of 13.5--20.5~K and 12.5--16.3~K toward LDN~1204A and LDN~1204B, respectively. The corresponding column density map yields peak column densities of approximately $3.1 \times 10^{23}$ cm$^{-2}$ and $1.3 \times 10^{23}$ cm$^{-2}$ toward LDN~1204A and LDN~1204B, respectively.

\item Integrated intensity (moment~0) maps of the $^{12}$CO(1--0), $^{13}$CO(1--0), and C$^{18}$O(1--0) emission, integrated over the velocity range $-15$ to $-3$ km s$^{-1}$, reveal a single extended arc-/shell-like molecular cloud hosting the LDN~1204 and Sh2-140 regions. A similar morphology is also evident in the WISE 22~$\mu$m image, with the arc opening toward the massive stars `S1' and `S2'.

\item The velocity profiles and channel maps reveal two kinematically distinct molecular cloud components: cloud~1 ($\sim -15$ to $-9.5$ km s$^{-1}$) and cloud~2 ($\sim -9$ to $-3$ km s$^{-1}$). The integrated intensity map of each component exhibits an arc-/shell-like morphology. The massive star `S1' is located near the center of cloud~1, whereas `S2' lies close to the center of cloud~2, indicating that each star is physically associated with its respective molecular cloud. Each massive star appears to ionize its local environment, producing a distinct H{\sc ii} region.

\item Position-velocity diagrams of the two molecular cloud components exhibit inverted C-shaped velocity structures, a characteristic signature of expanding shells. Each cloud component appears to be expanding under the influence of its respective massive star. The LDN~1204A and LDN~1204B sub-regions are situated at the projected interface between the two molecular cloud components. The CO moment~2 map also reveals enhanced velocity dispersion toward both sub-regions, suggesting increased turbulent gas motions at the interface of the two cloud components.

\item Most Class~{\sc i} sources are concentrated within LDN 1204 (especially around LDN~1204A and LDN~1204B), whereas the majority of Class~{\sc ii} sources lie outside the cloud boundaries. Only a small fraction of the YSO population is associated with cloud~1, while a substantial concentration of YSOs in cloud~2 is associated with the S140 SFR.

\item The spatial and kinematic correspondence of LDN~1204A and LDN~1204B at the interface of the two expanding H\,{\sc ii} shells supports a feedback-driven bubble--filament paradigm. In this scenario, the combined effects of stellar feedback and interacting molecular clouds have compressed the gas, facilitating the assembly of dense filamentary networks and candidate hub--filament systems that foster and sustain the observed ongoing star formation.

\end{itemize}

Our results support a scenario in which the expansion and interaction of two neighboring H{\sc ii} regions  has likely compressed the molecular gas at their interface, promoting the formation of dense filamentary structures and sustaining the ongoing star formation activity observed toward LDN~1204A and LDN~1204B.\\\\

\section*{Acknowledgements}
We thank the referee for their constructive comments and suggestions, which greatly improved the scientific structure of the paper. We thank Alexander Mensh’chikov for his help in using the getsf code to construct the temperature and column density maps. NP and RC acknowledge the financial support received through the SERB CRG/2021/005876 grant. The research work at the Physical Research Laboratory is funded by the Department of Space, Government of India. This publication makes use of data products obtained with the Spitzer Space Telescope (operated by the Jet Propulsion
Laboratory, California Institute of Technology, under contract
with NASA). The research work at the Physical Research Laboratory is funded by the Department of Space, Government of India.

This research made use of the data from the Milky Way Imaging Scroll Painting (MWISP) project, which is a multi-line survey in $^{12}\mathrm{CO}$/$^{13}\mathrm{CO}$/C$^{18}$O along the northern galactic plane with PMO-13.7m telescope. We are grateful to all the members of the MWISP working group, particularly the staff members at PMO-13.7m telescope, for their long-term support. MWISP was sponsored by National Key R\&D Program of
China with grants 2023YFA1608000 \& 2017YFA0402701 and by CAS Key Research Program of Frontier Sciences with grant QYZDJ-SSW-SLH047.






\appendix \label{appendix}
\restartappendixnumbering

\section{$p$ - $v$ diagrams of individual clouds}
\begin{figure*}

    \subfigure{
    \includegraphics[width=0.48\textwidth]{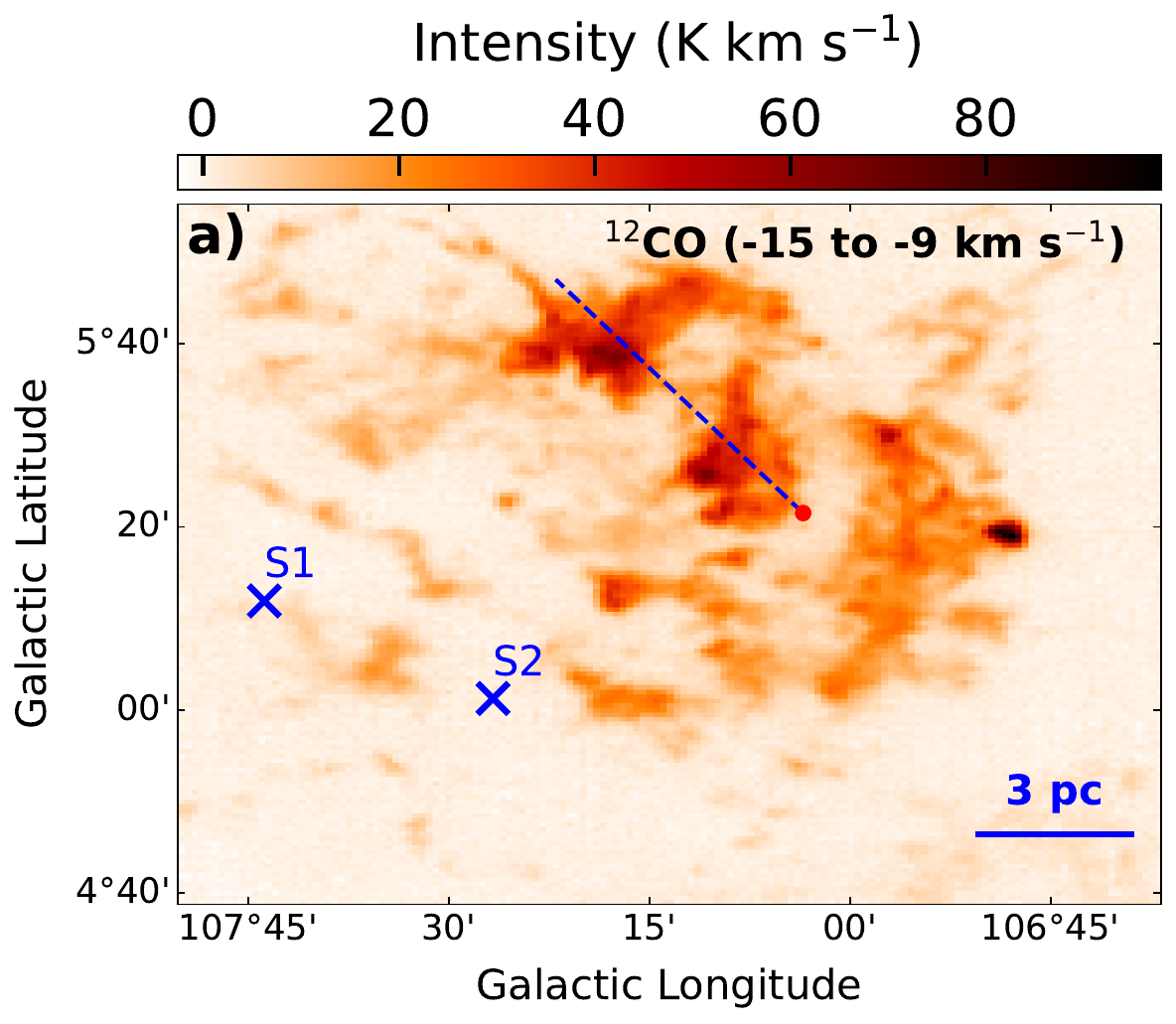}
    \label{fig:12co_moment0_cloud1_pvcut}
    }
    \subfigure{
    \includegraphics[width=0.48\textwidth]{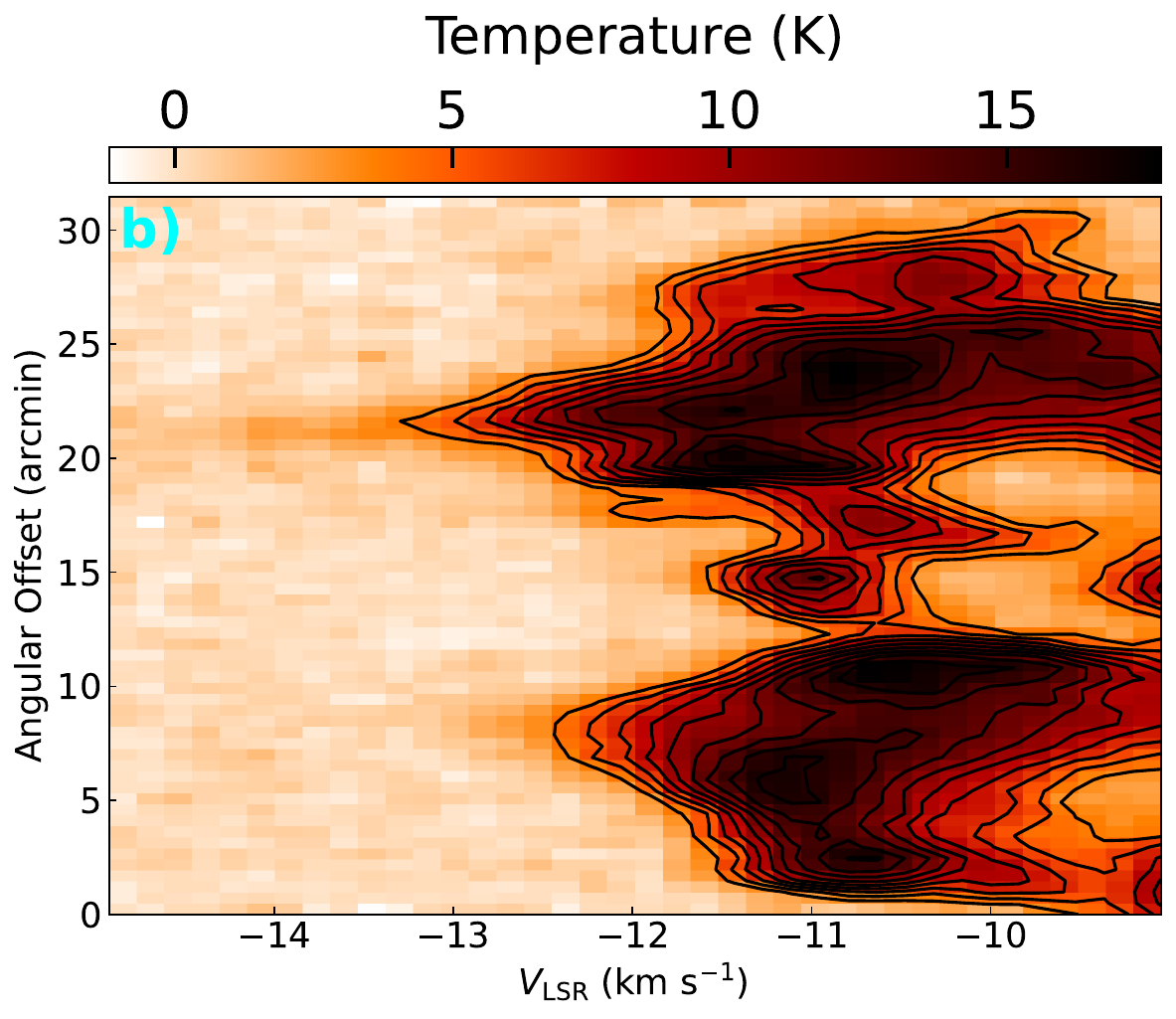}
    \label{fig:pv_diagram_cloud1_diagonal}
    }

    \subfigure{
    \includegraphics[width=0.48\textwidth]{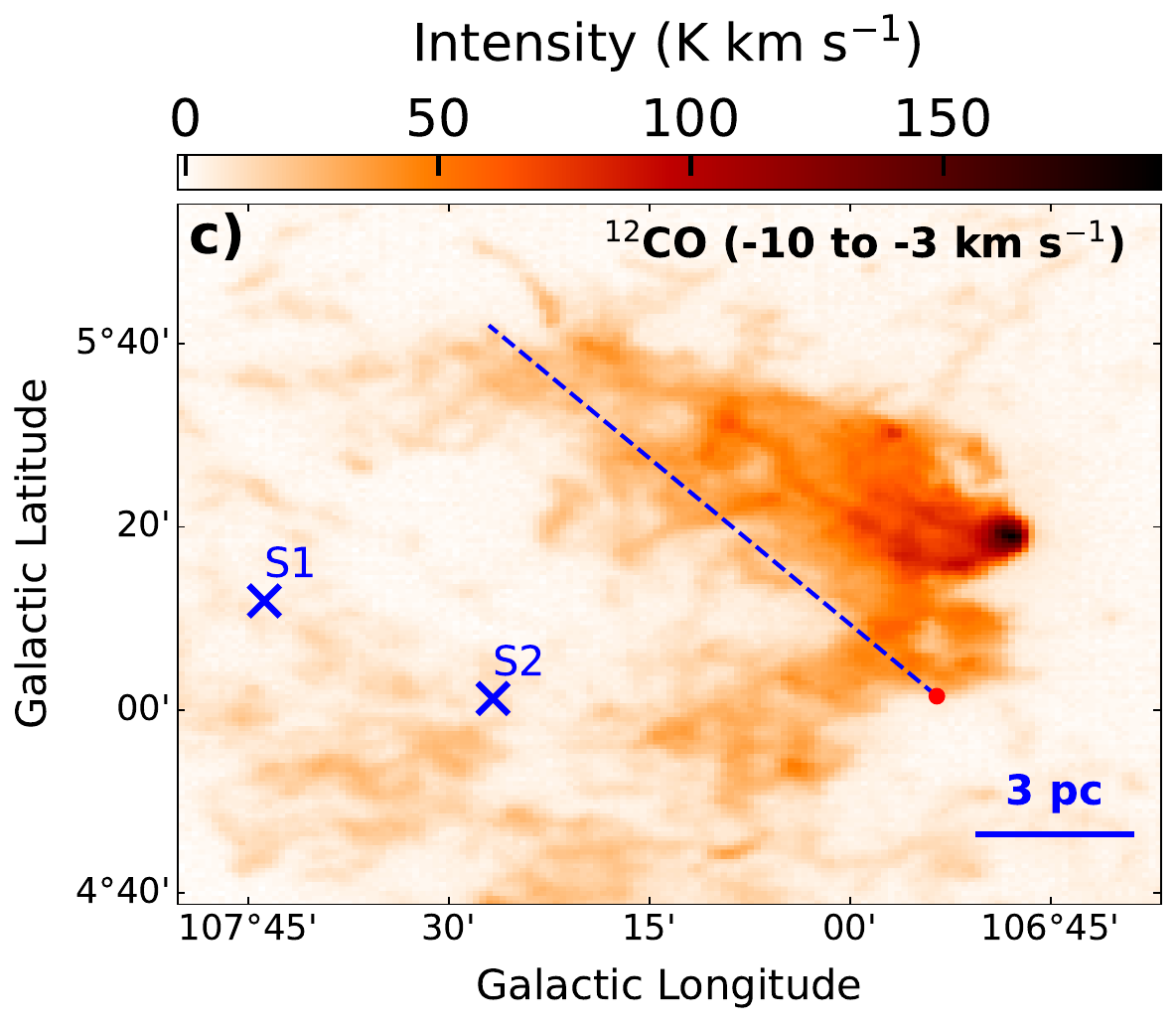}
    \label{fig:12co_moment0_cloud2_pvcut}
    }
    \subfigure{
    \includegraphics[width=0.48\textwidth]{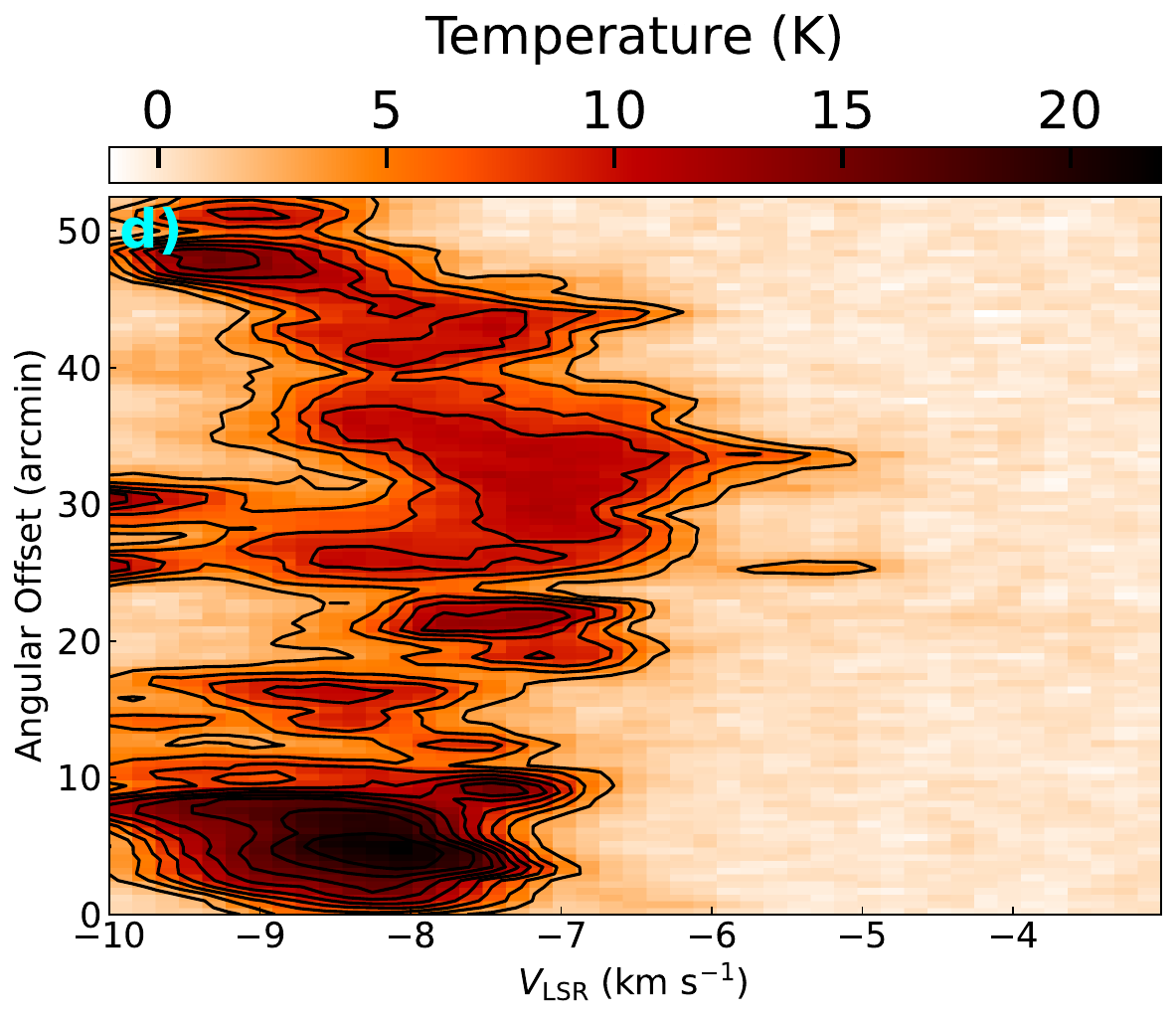}
    \label{fig:pv_diagram_cloud2_diagonal}
    }

    \caption{a) $^{12}$CO moment-0 map of the region in velocity range $-15$ km s$^{-1}$ to $-9$ km s$^{-1}$ (velocity range of Cloud 1 is within this range), b) $p$--$v$ diagram along the diagonal cut shown in a), c) $^{12}$CO moment-0 map of the region in velocity range $-10$ km s$^{-1}$ to $-3$ km s$^{-1}$ (velocity range of Cloud 2 is within this range), d) $p$--$v$ diagram along the diagonal cut shown in c). The origin of angular offset in the $p$--$v$ diagram is marked as a red dot. The cross marks indicate the position of massive stars.}

    \label{fig:pv_diagram_individualclouds}

\end{figure*}

Here, we construct the $p$ - $v$ diagrams considering individual clouds separately (Figure \ref{fig:pv_diagram_individualclouds}). The moment 0 map is constructed for both clouds 1 and 2 in velocity range $\sim -15$ to $-9.5$ km s$^{-1}$ and $\sim -9$ to $-3$ km s$^{-1}$ respectively, and the $p$ - $v$ diagram is constructed along the diagonal. As can be seen, the inverted `C' morphology is visible in both $p$ - $v$ diagrams, which suggests that both clouds are dynamically expanding along with the H {\sc ii} region.

An expanding spherical shell produces an approximately elliptical locus in $p$ - $v$ space, with the velocity extent about the systemic velocity providing an estimate of the expansion velocity (e.g., \citet{2011ApJ...742..105A,2018ApJ...854..106D}).

We therefore constructed simple synthetic $p$ - $v$ models assuming a spherical shell characterized by a radius $(R_{\rm sh})$, systemic velocity $(V_{\rm sys})$, and expansion velocity $(V_{\rm exp})$, such that

$$ \left(\frac{p-p_0}{R_{\rm sh}}\right)^2+ \left(\frac{V_{\rm LSR}-V_{\rm sys}}{V_{\rm exp}}\right)^2=1, $$

where \(p\) is the position along the $p$ - $v$ cut and \(p_0\) is the projected shell center. The model emission was convolved with a Gaussian beam of FWHM \(50^{\prime\prime}\), appropriate for the \(^{12}\)CO observations, and Gaussian noise with an rms of \(0.5\) K was added to reproduce the observational sensitivity.

For the two identified molecular structures, the observed $p$ - $v$ arcs are approximately reproduced with \(V_{\rm sys}\simeq-11.8\) km s\(^{-1}\), \(V_{\rm exp}\simeq2.0\) km s\(^{-1}\), and \(R_{\rm sh}\simeq14^{\prime}\) for the first shell, while the second shell is characterized by \(V_{\rm sys}\simeq-7.0\) km s\(^{-1}\), \(V_{\rm exp}\simeq3.0\) km s\(^{-1}\), and \(R_{\rm sh}\simeq24^{\prime}\). These values should be regarded as characteristic parameters obtained from the simple shell representation, since the observed structures trace incomplete and asymmetric arcs rather than complete ellipses.

Using these values, to assess the causal connection between the expanding H\,{\sc ii} regions and the ongoing star formation toward `A' and `B', we compare the dynamical timescale of the H\,{\sc ii} region with the characteristic ages of the embedded YSO population. 
Assuming an average shell expansion velocity of $v_{\mathrm{exp}} \sim 2\text{--}3\ \mathrm{km\ s^{-1}}$ and a shell radius of $R \sim 7\text{--}9\ \mathrm{pc}$, the dynamical expansion age of the H\,{\sc ii} region is estimated to be $t_{\mathrm{dyn}} \sim R / v_{\mathrm{exp}} \sim 2.3\text{--}4.4\ \mathrm{Myr}$. 
In comparison, the typical lifetimes of Class~{\sc i} and Class~{\sc ii} YSOs are $\sim 0.5\ \mathrm{Myr}$ and $\sim 1\text{--}3\ \mathrm{Myr}$, respectively \citep{2009ApJS..181..321E}. 
The fact that the dynamical timescale of the expanding H\,{\sc ii} bubble exceeds the evolutionary ages of the embedded young stars ($t_{\mathrm{dyn}} > t_{\mathrm{YSO}}$) indicates that 
molecular gas was swept up to the current configuration by the H{\sc ii} region prior to the onset of the current star-forming event.

\section{Mach number map of the clouds}
\begin{figure*}

    \subfigure{
    \includegraphics[width =0.48\textwidth]{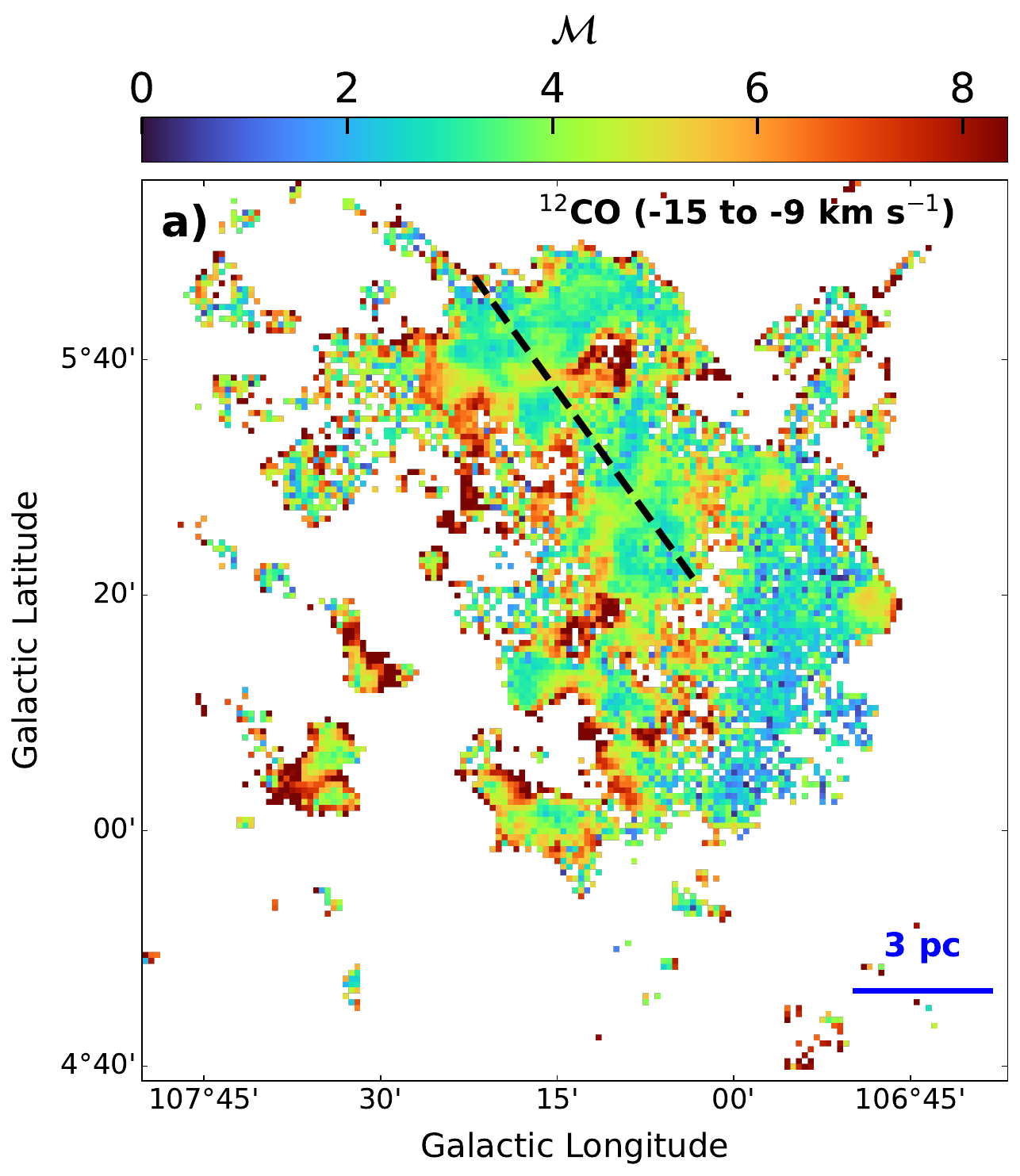} 
    \label{fig:12co_moment2_cloud1_referee}
    }
    \subfigure{
    \includegraphics[width =0.48\textwidth]{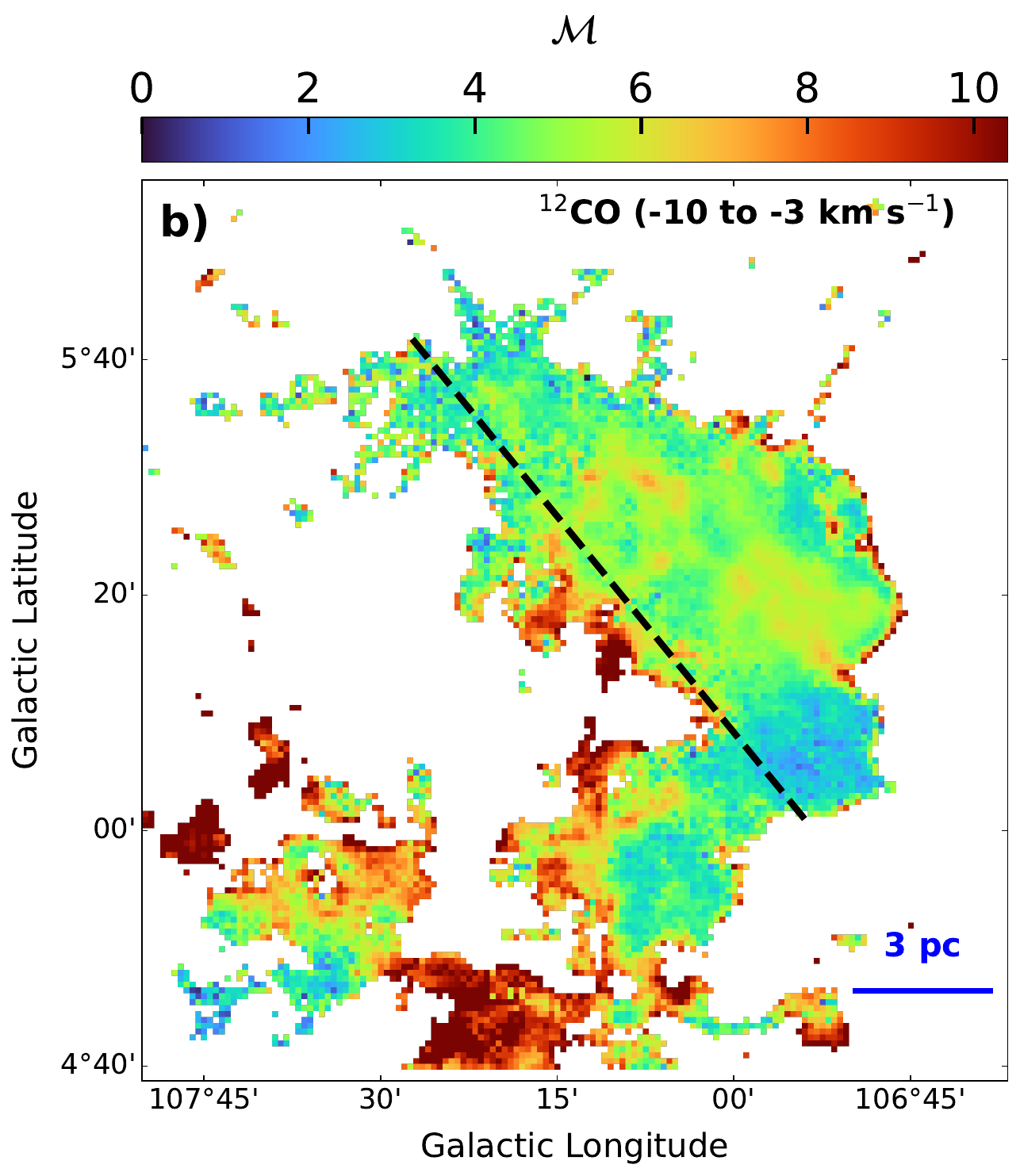}  \label{fig:12co_moment2_cloud2_referee}
    }

    \caption{a)The Mach number map of cloud in velocity range -15 km s$^{-1}$ to -9 km s$^{-1}$ b) and in velocity range -10 km s$^{-1}$ to -3 km s$^{-1}$. The dashed line indicates the line along which the $p$ - $v$ diagram was constructed in the Figure \ref{fig:pv_diagram_individualclouds}, and along which we calculated the mean $\mathcal{M}$.}
 
    \label{fig:12co_moment2_referee}
   
\end{figure*}

To assess the contribution of turbulence to the observed velocity structures, we estimated the Mach number along the diagonal $p$ - $v$ cut for the individual clouds in velocity ranges similar to that taken in the previous section. The $p$ - $v$ cut is also taken as in Figure \ref{fig:pv_diagram_individualclouds}. The excitation temperature map derived from the $^{12}$CO emission was used to calculate the thermal velocity dispersion map,
\begin{equation}
\sigma_{\rm th} = \sqrt{\frac{k_{\rm B}T_{\rm ex}}{m_{\rm CO}}},
\end{equation}
while the non-thermal component was estimated as
\begin{equation}
\sigma_{\rm NT} = \sqrt{\sigma_{\rm obs}^{2}-\sigma_{\rm th}^{2}}.
\end{equation}
$\sigma_{\rm obs}^{2}$ is the moment 2 map of $^{12}$CO.
The corresponding sound speed of the molecular gas was calculated as
\begin{equation}
c_{\rm s} =
\sqrt{\frac{k_{\rm B}T_{\rm ex}}{\mu m_{\rm H}}},
\end{equation}
where $\mu=2.33$ is the mean molecular weight per free particle in molecular gas. The non thermal Mach number was then estimated from
\begin{equation}
\mathcal{M} =
\frac{\sigma_{\rm NT}}{c_{\rm s}}.
\end{equation}
The Mach number map of both clouds are shown in Figure \ref{fig:12co_moment2_referee}. The thermal contribution was considerably low compared to the non-thermal contribution. We estimate the mean value of $\mathcal{M}$ along the $p$–$v$ cut. It ranges from 2.5 to 5 for Cloud 1 and from 3 to 7 for Cloud 2. The resulting mean values are 3.5 for Cloud 1 and 4.6 for Cloud 2, as shown in Figures \ref{fig:12co_moment2_cloud1_referee} and \ref{fig:12co_moment2_cloud2_referee}, respectively. The supersonic Mach numbers and thus high turbulence may be caused due to the cloud-cloud interaction.
 

\newpage
\bibliography{cite}{}

@ARTICLE{2012RAA....12.1499L,
       author = {{Li}, Nan and {Wang}, Jun-Jie},
        title = "{A survey of CO and its isotope lines for possible cloud-cloud collision candidates}",
      journal = {Research in Astronomy and Astrophysics},
         year = 2012,
        month = nov,
       volume = {12},
       number = {11},
        pages = {1499-1518},
          doi = {10.1088/1674-4527/12/11/005},
archivePrefix = {arXiv},
       eprint = {1212.0084},
 primaryClass = {astro-ph.SR},
       adsurl = {https://ui.adsabs.harvard.edu/abs/2012RAA....12.1499L}
}

@ARTICLE{2009ApJS..184...18G,
       author = {{Gutermuth}, R.~A. and {Megeath}, S.~T. and {Myers}, P.~C. and {Allen}, L.~E. and {Pipher}, J.~L. and {Fazio}, G.~G.},
        title = "{A Spitzer Survey of Young Stellar Clusters Within One Kiloparsec of the Sun: Cluster Core Extraction and Basic Structural Analysis}",
      journal = {\apjs},
         year = 2009,
        month = sep,
       volume = {184},
       number = {1},
        pages = {18-83},
          doi = {10.1088/0067-0049/184/1/18},
archivePrefix = {arXiv},
       eprint = {0906.0201},
 primaryClass = {astro-ph.SR},
       adsurl = {https://ui.adsabs.harvard.edu/abs/2009ApJS..184...18G}
}

@ARTICLE{1993ApJ...403..175T,
       author = {{Tafalla}, M. and {Bachiller}, R. and {Martin-Pintado}, J.},
        title = "{Dense Cores in L1204/S140: Star Formation and Velocity Shifts}",
      journal = {\apj},
         year = 1993,
        month = jan,
       volume = {403},
        pages = {175},
          doi = {10.1086/172192},
       adsurl = {https://ui.adsabs.harvard.edu/abs/1993ApJ...403..175T}
}

@ARTICLE{2018ApJ...859..166F,
       author = {{Fukui}, Yasuo and {Torii}, Kazufumi and {Hattori}, Yusuke and {Nishimura}, Atsushi and {Ohama}, Akio and {Shimajiri}, Yoshito and {Shima}, Kazuhiro and {Habe}, Asao and {Sano}, Hidetoshi and {Kohno}, Mikito and {Yamamoto}, Hiroaki and {Tachihara}, Kengo and {Onishi}, Toshikazu},
        title = "{A New Look at the Molecular Gas in M42 and M43: Possible Evidence for Cloud-Cloud Collision that Triggered Formation of the OB Stars in the Orion Nebula Cluster}",
      journal = {\apj},
         year = 2018,
        month = jun,
       volume = {859},
       number = {2},
          eid = {166},
        pages = {166},
          doi = {10.3847/1538-4357/aac217},
archivePrefix = {arXiv},
       eprint = {1701.04669},
 primaryClass = {astro-ph.GA},
       adsurl = {https://ui.adsabs.harvard.edu/abs/2018ApJ...859..166F}
}

@ARTICLE{2025AJ....169...56M,
       author = {{Maity}, A.~K. and {Dewangan}, L.~K. and {Bhadari}, N.~K. and {Fukui}, Y. and {Ismail}, A. Haj and {Jadhav}, O.~R. and {Sharma}, Saurabh and {Sano}, H.},
        title = "{G321.93-0.01: A Rare Site of Multiple Hub-filament Systems with Evidence of Collision and Merging of Filaments}",
      journal = {\aj},
         year = 2025,
        month = feb,
       volume = {169},
       number = {2},
          eid = {56},
        pages = {56},
          doi = {10.3847/1538-3881/ad98ff},
archivePrefix = {arXiv},
       eprint = {2411.13870},
 primaryClass = {astro-ph.GA},
       adsurl = {https://ui.adsabs.harvard.edu/abs/2025AJ....169...56M}
}

@ARTICLE{2010ApJ...721L.107S,
       author = {{S{\'a}nchez-Monge}, {\'A}lvaro and {Palau}, Aina and {Estalella}, Robert and {Kurtz}, Stan and {Zhang}, Qizhou and {Di Francesco}, James and {Shepherd}, Debra},
        title = "{IRAS 22198+6336: Discovery of an Intermediate-mass Hot Core}",
      journal = {\apjl},
         year = 2010,
        month = oct,
       volume = {721},
       number = {2},
        pages = {L107-L111},
          doi = {10.1088/2041-8205/721/2/L107},
archivePrefix = {arXiv},
       eprint = {1007.5258},
 primaryClass = {astro-ph.GA},
       adsurl = {https://ui.adsabs.harvard.edu/abs/2010ApJ...721L.107S}
}

@ARTICLE{2009ApJS..181..321E,
       author = {{Evans}, II, Neal J. and {Dunham}, Michael M. and {J{\o}rgensen}, Jes K. and {Enoch}, Melissa L. and {Mer{\'\i}n}, Bruno and {van Dishoeck}, Ewine F. and {Alcal{\'a}}, Juan M. and {Myers}, Philip C. and {Stapelfeldt}, Karl R. and {Huard}, Tracy L. and {Allen}, Lori E. and {Harvey}, Paul M. and {van Kempen}, Tim and {Blake}, Geoffrey A. and {Koerner}, David W. and {Mundy}, Lee G. and {Padgett}, Deborah L. and {Sargent}, Anneila I.},
        title = "{The Spitzer c2d Legacy Results: Star-Formation Rates and Efficiencies; Evolution and Lifetimes}",
      journal = {\apjs},
         year = 2009,
        month = apr,
       volume = {181},
       number = {2},
        pages = {321-350},
          doi = {10.1088/0067-0049/181/2/321},
archivePrefix = {arXiv},
       eprint = {0811.1059},
 primaryClass = {astro-ph},
       adsurl = {https://ui.adsabs.harvard.edu/abs/2009ApJS..181..321E}
}

@ARTICLE{2013MNRAS.428.3425H,
       author = {{Henshaw}, J.~D. and {Caselli}, P. and {Fontani}, F. and {Jim{\'e}nez-Serra}, I. and {Tan}, J.~C. and {Hernandez}, A.~K.},
        title = "{Complex, quiescent kinematics in a highly filamentary infrared dark cloud}",
      journal = {\mnras},
         year = 2013,
        month = feb,
       volume = {428},
       number = {4},
        pages = {3425-3442},
          doi = {10.1093/mnras/sts282},
archivePrefix = {arXiv},
       eprint = {1210.6835},
 primaryClass = {astro-ph.SR},
       adsurl = {https://ui.adsabs.harvard.edu/abs/2013MNRAS.428.3425H}
}

@ARTICLE{2011A&A...529L...6A,
       author = {{Arzoumanian}, D. and {Andr{\'e}}, Ph. and {Didelon}, P. and {K{\"o}nyves}, V. and {Schneider}, N. and {Men'shchikov}, A. and {Sousbie}, T. and {Zavagno}, A. and {Bontemps}, S. and {di Francesco}, J. and {Griffin}, M. and {Hennemann}, M. and {Hill}, T. and {Kirk}, J. and {Martin}, P. and {Minier}, V. and {Molinari}, S. and {Motte}, F. and {Peretto}, N. and {Pezzuto}, S. and {Spinoglio}, L. and {Ward-Thompson}, D. and {White}, G. and {Wilson}, C.~D.},
        title = "{Characterizing interstellar filaments with Herschel in IC 5146}",
      journal = {\aap},
         year = 2011,
        month = may,
       volume = {529},
          eid = {L6},
        pages = {L6},
          doi = {10.1051/0004-6361/201116596},
archivePrefix = {arXiv},
       eprint = {1103.0201},
 primaryClass = {astro-ph.GA},
       adsurl = {https://ui.adsabs.harvard.edu/abs/2011A&A...529L...6A}
}

@ARTICLE{2015NewAR..68....1D,
       author = {{Dale}, James E.},
        title = "{The modelling of feedback in star formation simulations}",
      journal = {\nar},
         year = 2015,
        month = oct,
       volume = {68},
        pages = {1-33},
          doi = {10.1016/j.newar.2015.06.001},
archivePrefix = {arXiv},
       eprint = {1508.06054},
 primaryClass = {astro-ph.GA},
       adsurl = {https://ui.adsabs.harvard.edu/abs/2015NewAR..68....1D}
}

@ARTICLE{2019MNRAS.488.4753D,
       author = {{Dekel}, Avishai and {Sarkar}, Kartick C. and {Jiang}, Fangzhou and {Bournaud}, Frederic and {Krumholz}, Mark R. and {Ceverino}, Daniel and {Primack}, Joel R.},
        title = "{The global star formation law by supernova feedback}",
      journal = {\mnras},
         year = 2019,
        month = oct,
       volume = {488},
       number = {4},
        pages = {4753-4778},
          doi = {10.1093/mnras/stz1919},
archivePrefix = {arXiv},
       eprint = {1903.00962},
 primaryClass = {astro-ph.GA},
       adsurl = {https://ui.adsabs.harvard.edu/abs/2019MNRAS.488.4753D}
}

@INPROCEEDINGS{2007prpl.conf..245A,
       author = {{Arce}, H.~G. and {Shepherd}, D. and {Gueth}, F. and {Lee}, C.-F. and {Bachiller}, R. and {Rosen}, A. and {Beuther}, H.},
        title = "{Molecular Outflows in Low- and High-Mass Star-forming Regions}",
    booktitle = {Protostars and Planets V},
         year = 2007,
       editor = {{Reipurth}, Bo and {Jewitt}, David and {Keil}, Klaus},
        month = jan,
        pages = {245},
          doi = {10.48550/arXiv.astro-ph/0603071},
archivePrefix = {arXiv},
       eprint = {astro-ph/0603071},
 primaryClass = {astro-ph},
       adsurl = {https://ui.adsabs.harvard.edu/abs/2007prpl.conf..245A}
}

@ARTICLE{2024MNRAS.52710077H,
       author = {{Horie}, Shu and {Okamoto}, Takashi and {Habe}, Asao},
        title = "{Cloud-cloud collisions triggering star formation in galaxy simulations}",
      journal = {\mnras},
         year = 2024,
        month = feb,
       volume = {527},
       number = {4},
        pages = {10077-10094},
          doi = {10.1093/mnras/stad3798},
archivePrefix = {arXiv},
       eprint = {2311.04175},
 primaryClass = {astro-ph.GA},
       adsurl = {https://ui.adsabs.harvard.edu/abs/2024MNRAS.52710077H}
}

@ARTICLE{2005MNRAS.358..291D,
       author = {{Dale}, J.~E. and {Bonnell}, I.~A. and {Clarke}, C.~J. and {Bate}, M.~R.},
        title = "{Photoionizing feedback in star cluster formation}",
      journal = {\mnras},
         year = 2005,
        month = mar,
       volume = {358},
       number = {1},
        pages = {291-304},
          doi = {10.1111/j.1365-2966.2005.08806.x},
archivePrefix = {arXiv},
       eprint = {astro-ph/0501160},
 primaryClass = {astro-ph},
       adsurl = {https://ui.adsabs.harvard.edu/abs/2005MNRAS.358..291D}
}

@ARTICLE{2021ApJ...913...95P,
       author = {{Parker}, Richard J. and {Alcock}, Hayley L. and {Nicholson}, Rhana B. and {Pani{\'c}}, Olja and {Goodwin}, Simon P.},
        title = "{External Photoevaporation of Protoplanetary Disks: Does Location Matter?}",
      journal = {\apj},
         year = 2021,
        month = jun,
       volume = {913},
       number = {2},
          eid = {95},
        pages = {95},
          doi = {10.3847/1538-4357/abf4cc},
archivePrefix = {arXiv},
       eprint = {2104.03973},
 primaryClass = {astro-ph.EP},
       adsurl = {https://ui.adsabs.harvard.edu/abs/2021ApJ...913...95P}
}

@ARTICLE{2006A&A...450..625U,
       author = {{Urquhart}, J.~S. and {Thompson}, M.~A. and {Morgan}, L.~K. and {White}, G.~J.},
        title = "{Radio continuum and molecular line observations of four bright-rimmed clouds}",
      journal = {\aap},
         year = 2006,
        month = may,
       volume = {450},
       number = {2},
        pages = {625-643},
          doi = {10.1051/0004-6361:20053417},
archivePrefix = {arXiv},
       eprint = {astro-ph/0601718},
 primaryClass = {astro-ph},
       adsurl = {https://ui.adsabs.harvard.edu/abs/2006A&A...450..625U}
}

@ARTICLE{1996A&A...315L.281T,
       author = {{Timmermann}, R. and {Bertoldi}, F. and {Wright}, C.~M. and {Drapatz}, S. and {Draine}, B.~T. and {Haser}, L. and {Sternberg}, A.},
        title = "{H\_2\_ infrared line emission from S 140: a warm PDR.}",
      journal = {\aap},
         year = 1996,
        month = nov,
       volume = {315},
        pages = {L281-L284},
       adsurl = {https://ui.adsabs.harvard.edu/abs/1996A&A...315L.281T}
}

@ARTICLE{1994A&A...283..963C,
       author = {{Chavarria-K.} and {Moreno-Corral}, M.~A. and {Hernandez-Toledo}, H. and {Terranegra}, L. and {de Lara}, E.},
        title = "{H.alpha interferometric, optical and near IR photometric studies of star forming regions. II. The NGC 7380E/Sh2-142/NGC 7380 complex.}",
      journal = {\aap},
         year = 1994,
        month = mar,
       volume = {283},
        pages = {963-972},
       adsurl = {https://ui.adsabs.harvard.edu/abs/1994A&A...283..963C}
}

@ARTICLE{2001ApJ...561..282K,
       author = {{Kraemer}, Kathleen E. and {Jackson}, James M. and {Deutsch}, Lynne K. and {Kassis}, Marc and {Hora}, Joseph L. and {Fazio}, G.~G. and {Hoffmann}, William F. and {Dayal}, Aditya},
        title = "{Dust Characteristics of Massive Star-forming Sites in the Mid-Infrared}",
      journal = {\apj},
         year = 2001,
        month = nov,
       volume = {561},
       number = {1},
        pages = {282-298},
          doi = {10.1086/323219},
       adsurl = {https://ui.adsabs.harvard.edu/abs/2001ApJ...561..282K}
}

@ARTICLE{2013ApJ...762..120P,
       author = {{Palau}, Aina and {Fuente}, Asunci{\'o}n and {Girart}, Josep M. and {Estalella}, Robert and {Ho}, Paul T.~P. and {S{\'a}nchez-Monge}, {\'A}lvaro and {Fontani}, Francesco and {Busquet}, Gemma and {Commer{\c{c}}on}, Benoit and {Hennebelle}, Patrick and {Boissier}, J{\'e}r{\'e}mie and {Zhang}, Qizhou and {Cesaroni}, Riccardo and {Zapata}, Luis A.},
        title = "{Early Stages of Cluster Formation: Fragmentation of Massive Dense Cores down to <\raisebox{-0.5ex}\textasciitilde 1000 AU}",
      journal = {\apj},
         year = 2013,
        month = jan,
       volume = {762},
       number = {2},
          eid = {120},
        pages = {120},
          doi = {10.1088/0004-637X/762/2/120},
archivePrefix = {arXiv},
       eprint = {1211.2666},
 primaryClass = {astro-ph.GA},
       adsurl = {https://ui.adsabs.harvard.edu/abs/2013ApJ...762..120P}
}

@ARTICLE{2010A&A...518L...2P,
       author = {{Poglitsch}, A. and {Waelkens}, C. and {Geis}, N. and {Feuchtgruber}, H. and {Vandenbussche}, B. and {Rodriguez}, L. and {Krause}, O. and {Renotte}, E. and {van Hoof}, C. and {Saraceno}, P. and {Cepa}, J. and {Kerschbaum}, F. and {Agn{\`e}se}, P. and {Ali}, B. and {Altieri}, B. and {Andreani}, P. and {Augueres}, J.-L. and {Balog}, Z. and {Barl}, L. and {Bauer}, O.~H. and {Belbachir}, N. and {Benedettini}, M. and {Billot}, N. and {Boulade}, O. and {Bischof}, H. and {Blommaert}, J. and {Callut}, E. and {Cara}, C. and {Cerulli}, R. and {Cesarsky}, D. and {Contursi}, A. and {Creten}, Y. and {De Meester}, W. and {Doublier}, V. and {Doumayrou}, E. and {Duband}, L. and {Exter}, K. and {Genzel}, R. and {Gillis}, J.-M. and {Gr{\"o}zinger}, U. and {Henning}, T. and {Herreros}, J. and {Huygen}, R. and {Inguscio}, M. and {Jakob}, G. and {Jamar}, C. and {Jean}, C. and {de Jong}, J. and {Katterloher}, R. and {Kiss}, C. and {Klaas}, U. and {Lemke}, D. and {Lutz}, D. and {Madden}, S. and {Marquet}, B. and {Martignac}, J. and {Mazy}, A. and {Merken}, P. and {Montfort}, F. and {Morbidelli}, L. and {M{\"u}ller}, T. and {Nielbock}, M. and {Okumura}, K. and {Orfei}, R. and {Ottensamer}, R. and {Pezzuto}, S. and {Popesso}, P. and {Putzeys}, J. and {Regibo}, S. and {Reveret}, V. and {Royer}, P. and {Sauvage}, M. and {Schreiber}, J. and {Stegmaier}, J. and {Schmitt}, D. and {Schubert}, J. and {Sturm}, E. and {Thiel}, M. and {Tofani}, G. and {Vavrek}, R. and {Wetzstein}, M. and {Wieprecht}, E. and {Wiezorrek}, E.},
        title = "{The Photodetector Array Camera and Spectrometer (PACS) on the Herschel Space Observatory}",
      journal = {\aap},
         year = 2010,
        month = jul,
       volume = {518},
          eid = {L2},
        pages = {L2},
          doi = {10.1051/0004-6361/201014535},
archivePrefix = {arXiv},
       eprint = {1005.1487},
 primaryClass = {astro-ph.IM},
       adsurl = {https://ui.adsabs.harvard.edu/abs/2010A&A...518L...2P}
}

@ARTICLE{2010A&A...518L...3G,
       author = {{Griffin}, M.~J. and {Abergel}, A. and {Abreu}, A. and {Ade}, P.~A.~R. and {Andr{\'e}}, P. and {Augueres}, J.-L. and {Babbedge}, T. and {Bae}, Y. and {Baillie}, T. and {Baluteau}, J.-P. and {Barlow}, M.~J. and {Bendo}, G. and {Benielli}, D. and {Bock}, J.~J. and {Bonhomme}, P. and {Brisbin}, D. and {Brockley-Blatt}, C. and {Caldwell}, M. and {Cara}, C. and {Castro-Rodriguez}, N. and {Cerulli}, R. and {Chanial}, P. and {Chen}, S. and {Clark}, E. and {Clements}, D.~L. and {Clerc}, L. and {Coker}, J. and {Communal}, D. and {Conversi}, L. and {Cox}, P. and {Crumb}, D. and {Cunningham}, C. and {Daly}, F. and {Davis}, G.~R. and {de Antoni}, P. and {Delderfield}, J. and {Devin}, N. and {di Giorgio}, A. and {Didschuns}, I. and {Dohlen}, K. and {Donati}, M. and {Dowell}, A. and {Dowell}, C.~D. and {Duband}, L. and {Dumaye}, L. and {Emery}, R.~J. and {Ferlet}, M. and {Ferrand}, D. and {Fontignie}, J. and {Fox}, M. and {Franceschini}, A. and {Frerking}, M. and {Fulton}, T. and {Garcia}, J. and {Gastaud}, R. and {Gear}, W.~K. and {Glenn}, J. and {Goizel}, A. and {Griffin}, D.~K. and {Grundy}, T. and {Guest}, S. and {Guillemet}, L. and {Hargrave}, P.~C. and {Harwit}, M. and {Hastings}, P. and {Hatziminaoglou}, E. and {Herman}, M. and {Hinde}, B. and {Hristov}, V. and {Huang}, M. and {Imhof}, P. and {Isaak}, K.~J. and {Israelsson}, U. and {Ivison}, R.~J. and {Jennings}, D. and {Kiernan}, B. and {King}, K.~J. and {Lange}, A.~E. and {Latter}, W. and {Laurent}, G. and {Laurent}, P. and {Leeks}, S.~J. and {Lellouch}, E. and {Levenson}, L. and {Li}, B. and {Li}, J. and {Lilienthal}, J. and {Lim}, T. and {Liu}, S.~J. and {Lu}, N. and {Madden}, S. and {Mainetti}, G. and {Marliani}, P. and {McKay}, D. and {Mercier}, K. and {Molinari}, S. and {Morris}, H. and {Moseley}, H. and {Mulder}, J. and {Mur}, M. and {Naylor}, D.~A. and {Nguyen}, H. and {O'Halloran}, B. and {Oliver}, S. and {Olofsson}, G. and {Olofsson}, H.-G. and {Orfei}, R. and {Page}, M.~J. and {Pain}, I. and {Panuzzo}, P. and {Papageorgiou}, A. and {Parks}, G. and {Parr-Burman}, P. and {Pearce}, A. and {Pearson}, C. and {P{\'e}rez-Fournon}, I. and {Pinsard}, F. and {Pisano}, G. and {Podosek}, J. and {Pohlen}, M. and {Polehampton}, E.~T. and {Pouliquen}, D. and {Rigopoulou}, D. and {Rizzo}, D. and {Roseboom}, I.~G. and {Roussel}, H. and {Rowan-Robinson}, M. and {Rownd}, B. and {Saraceno}, P. and {Sauvage}, M. and {Savage}, R. and {Savini}, G. and {Sawyer}, E. and {Scharmberg}, C. and {Schmitt}, D. and {Schneider}, N. and {Schulz}, B. and {Schwartz}, A. and {Shafer}, R. and {Shupe}, D.~L. and {Sibthorpe}, B. and {Sidher}, S. and {Smith}, A. and {Smith}, A.~J. and {Smith}, D. and {Spencer}, L. and {Stobie}, B. and {Sudiwala}, R. and {Sukhatme}, K. and {Surace}, C. and {Stevens}, J.~A. and {Swinyard}, B.~M. and {Trichas}, M. and {Tourette}, T. and {Triou}, H. and {Tseng}, S. and {Tucker}, C. and {Turner}, A. and {Vaccari}, M. and {Valtchanov}, I. and {Vigroux}, L. and {Virique}, E. and {Voellmer}, G. and {Walker}, H. and {Ward}, R. and {Waskett}, T. and {Weilert}, M. and {Wesson}, R. and {White}, G.~J. and {Whitehouse}, N. and {Wilson}, C.~D. and {Winter}, B. and {Woodcraft}, A.~L. and {Wright}, G.~S. and {Xu}, C.~K. and {Zavagno}, A. and {Zemcov}, M. and {Zhang}, L. and {Zonca}, E.},
        title = "{The Herschel-SPIRE instrument and its in-flight performance}",
      journal = {\aap},
         year = 2010,
        month = jul,
       volume = {518},
          eid = {L3},
        pages = {L3},
          doi = {10.1051/0004-6361/201014519},
archivePrefix = {arXiv},
       eprint = {1005.5123},
 primaryClass = {astro-ph.IM},
       adsurl = {https://ui.adsabs.harvard.edu/abs/2010A&A...518L...3G}
}

@ARTICLE{2011ApJ...742..105A,
       author = {{Arce}, H{\'e}ctor G. and {Borkin}, Michelle A. and {Goodman}, Alyssa A. and {Pineda}, Jaime E. and {Beaumont}, Christopher N.},
        title = "{A Bubbling Nearby Molecular Cloud: COMPLETE Shells in Perseus}",
      journal = {\apj},
         year = 2011,
        month = dec,
       volume = {742},
       number = {2},
          eid = {105},
        pages = {105},
          doi = {10.1088/0004-637X/742/2/105},
archivePrefix = {arXiv},
       eprint = {1109.3368},
 primaryClass = {astro-ph.SR},
       adsurl = {https://ui.adsabs.harvard.edu/abs/2011ApJ...742..105A}
}

@ARTICLE{2011MNRAS.414.1747A,
       author = {{Arthur}, S.~J. and {Henney}, W.~J. and {Mellema}, G. and {de Colle}, F. and {V{\'a}zquez-Semadeni}, E.},
        title = "{Radiation-magnetohydrodynamic simulations of H II regions and their associated PDRs in turbulent molecular clouds}",
      journal = {\mnras},
         year = 2011,
        month = jun,
       volume = {414},
       number = {2},
        pages = {1747-1768},
          doi = {10.1111/j.1365-2966.2011.18507.x},
archivePrefix = {arXiv},
       eprint = {1101.5510},
 primaryClass = {astro-ph.GA},
       adsurl = {https://ui.adsabs.harvard.edu/abs/2011MNRAS.414.1747A}
}

@ARTICLE{2023MNRAS.522.2369S,
       author = {{Scheuermann}, Fabian and {Kreckel}, Kathryn and {Barnes}, Ashley T. and {Belfiore}, Francesco and {Groves}, Brent and {Hannon}, Stephen and {Lee}, Janice C. and {Minsley}, Rebecca and {Rosolowsky}, Erik and {Bigiel}, Frank and {Blanc}, Guillermo A. and {Boquien}, M{\'e}d{\'e}ric and {Dale}, Daniel A. and {Deger}, Sinan and {Egorov}, Oleg V. and {Emsellem}, Eric and {Glover}, Simon C.~O. and {Grasha}, Kathryn and {Hassani}, Hamid and {Jeffreson}, Sarah M.~R. and {Klessen}, Ralf S. and {Kruijssen}, J.~M. Diederik and {Larson}, Kirsten L. and {Leroy}, Adam K. and {Lopez}, Laura A. and {Pan}, Hsi-An and {S{\'a}nchez-Bl{\'a}zquez}, Patricia and {Santoro}, Francesco and {Schinnerer}, Eva and {Thilker}, David A. and {Whitmore}, Bradley C. and {Watkins}, Elizabeth J. and {Williams}, Thomas G.},
        title = "{Stellar associations powering H II regions - I. Defining an evolutionary sequence}",
      journal = {\mnras},
         year = 2023,
        month = jun,
       volume = {522},
       number = {2},
        pages = {2369-2383},
          doi = {10.1093/mnras/stad878},
archivePrefix = {arXiv},
       eprint = {2303.12101},
 primaryClass = {astro-ph.GA},
       adsurl = {https://ui.adsabs.harvard.edu/abs/2023MNRAS.522.2369S}
}

@ARTICLE{2008ApJ...679..481P,
       author = {{Pineda}, Jaime E. and {Caselli}, Paola and {Goodman}, Alyssa A.},
        title = "{CO Isotopologues in the Perseus Molecular Cloud Complex: the X-factor and Regional Variations}",
      journal = {\apj},
         year = 2008,
        month = may,
       volume = {679},
       number = {1},
        pages = {481-496},
          doi = {10.1086/586883},
archivePrefix = {arXiv},
       eprint = {0802.0708},
 primaryClass = {astro-ph},
       adsurl = {https://ui.adsabs.harvard.edu/abs/2008ApJ...679..481P}
}

@ARTICLE{1982ApJ...262..590F,
       author = {{Frerking}, M.~A. and {Langer}, W.~D. and {Wilson}, R.~W.},
        title = "{The relationship between carbon monoxide abundance and visual extinction in interstellar clouds.}",
      journal = {\apj},
         year = 1982,
        month = nov,
       volume = {262},
        pages = {590-605},
          doi = {10.1086/160451},
       adsurl = {https://ui.adsabs.harvard.edu/abs/1982ApJ...262..590F}
}

@ARTICLE{2023MNRAS.520.1390S,
       author = {{Szil{\'a}gyi}, M{\'a}t{\'e} and {Kun}, M{\'a}ria and {{\'A}brah{\'a}m}, P{\'e}ter and {Marton}, G{\'a}bor},
        title = "{The Gaia view of the Cepheus OB2 association}",
      journal = {\mnras},
         year = 2023,
        month = mar,
       volume = {520},
       number = {1},
        pages = {1390-1410},
          doi = {10.1093/mnras/stad027},
archivePrefix = {arXiv},
       eprint = {2301.02346},
 primaryClass = {astro-ph.SR},
       adsurl = {https://ui.adsabs.harvard.edu/abs/2023MNRAS.520.1390S}
}

@ARTICLE{2015AJ....150..147F,
       author = {{Foster}, T. and {Brunt}, C.~M.},
        title = "{A CGPS Look at the Spiral Structure of the Outer Milky Way. I. Distances and Velocities to Star-forming Regions}",
      journal = {\aj},
         year = 2015,
        month = nov,
       volume = {150},
       number = {5},
          eid = {147},
        pages = {147},
          doi = {10.1088/0004-6256/150/5/147},
archivePrefix = {arXiv},
       eprint = {1405.7003},
 primaryClass = {astro-ph.GA},
       adsurl = {https://ui.adsabs.harvard.edu/abs/2015AJ....150..147F}
}

@dataset{2020yCat.1350....0G,
       author = {{Gaia Collaboration}},
        title = "{VizieR Online Data Catalog: Gaia EDR3 (Gaia Collaboration, 2020)}",
 howpublished = {VizieR On-line Data Catalog: I/350.  Originally published in: 2021A\&A...649A...1G},
         year = 2020,
        month = nov,
          eid = {I/350},
          doi = {10.26093/cds/vizier.1350},
       adsurl = {https://ui.adsabs.harvard.edu/abs/2020yCat.1350....0G}
}

@ARTICLE{2025ApJ...994..222C,
       author = {{C}, Rishi and {Panwar}, Neelam and {Haworth}, Thomas J. and {Sun}, Yan and {Sharma}, Saurabh and {Yadav}, R.~K. and {Ojha}, D.~K. and {Singh}, H.~P. and {Jose}, Jessy and {Singh}, Ajay Kumar and {Jose}, Jincen and {Yadav}, Shubham},
        title = "{Investigating the Influence of Radiative Feedback in the Bright-rimmed Cloud 44}",
      journal = {\apj},
         year = 2025,
        month = dec,
       volume = {994},
       number = {2},
          eid = {222},
        pages = {222},
          doi = {10.3847/1538-4357/ae0f03},
archivePrefix = {arXiv},
       eprint = {2510.08229},
 primaryClass = {astro-ph.SR},
       adsurl = {https://ui.adsabs.harvard.edu/abs/2025ApJ...994..222C}
}

@ARTICLE{2026AJ....171...69D,
       author = {{Dewangan}, L.~K. and {Yadav}, Ram K. and {Sharma}, Saurabh and {Jadhav}, O.~R. and {Maity}, A.~K. and {Goldsmith}, Paul F. and {Panchal}, G.},
        title = "{Discovery of a Rich Population of Compact Hub-filament Systems in a Single Star-forming Complex}",
      journal = {\aj},
         year = 2026,
        month = feb,
       volume = {171},
       number = {2},
          eid = {69},
        pages = {69},
          doi = {10.3847/1538-3881/ae25e7},
archivePrefix = {arXiv},
       eprint = {2512.01455},
 primaryClass = {astro-ph.GA},
       adsurl = {https://ui.adsabs.harvard.edu/abs/2026AJ....171...69D}
}

@ARTICLE{2020A&A...638A...7Z,
       author = {{Zavagno}, A. and {Andr{\'e}}, Ph. and {Schuller}, F. and {Peretto}, N. and {Shimajiri}, Y. and {Arzoumanian}, D. and {Csengeri}, T. and {Figueira}, M. and {Fuller}, G.~A. and {K{\"o}nyves}, V. and {Men'shchikov}, A. and {Palmeirim}, P. and {Roussel}, H. and {Russeil}, D. and {Schneider}, N. and {Zhang}, S.},
        title = "{The role of Galactic H II regions in the formation of filaments. High-resolution submilimeter imaging of RCW 120 with ArT{\'e}MiS}",
      journal = {\aap},
         year = 2020,
        month = jun,
       volume = {638},
          eid = {A7},
        pages = {A7},
          doi = {10.1051/0004-6361/202037815},
archivePrefix = {arXiv},
       eprint = {2004.05604},
 primaryClass = {astro-ph.GA},
       adsurl = {https://ui.adsabs.harvard.edu/abs/2020A&A...638A...7Z}
}

@ARTICLE{2017A&A...599A..94N,
       author = {{Ntormousi}, Evangelia and {Dawson}, Joanne R. and {Hennebelle}, Patrick and {Fierlinger}, Katharina},
        title = "{The role of magnetic fields in the structure and interaction of supershells}",
      journal = {\aap},
         year = 2017,
        month = mar,
       volume = {599},
          eid = {A94},
        pages = {A94},
          doi = {10.1051/0004-6361/201629268},
archivePrefix = {arXiv},
       eprint = {1701.03696},
 primaryClass = {astro-ph.GA},
       adsurl = {https://ui.adsabs.harvard.edu/abs/2017A&A...599A..94N}
}

@ARTICLE{2015A&A...580A..49I,
       author = {{Inutsuka}, Shu-ichiro and {Inoue}, Tsuyoshi and {Iwasaki}, Kazunari and {Hosokawa}, Takashi},
        title = "{The formation and destruction of molecular clouds and galactic star formation. An origin for the cloud mass function and star formation efficiency}",
      journal = {\aap},
         year = 2015,
        month = aug,
       volume = {580},
          eid = {A49},
        pages = {A49},
          doi = {10.1051/0004-6361/201425584},
archivePrefix = {arXiv},
       eprint = {1505.04696},
 primaryClass = {astro-ph.GA},
       adsurl = {https://ui.adsabs.harvard.edu/abs/2015A&A...580A..49I}
}

@ARTICLE{2015ApJ...799...64D,
       author = {{Dawson}, J.~R. and {Ntormousi}, E. and {Fukui}, Y. and {Hayakawa}, T. and {Fierlinger}, K.},
        title = "{A Young Giant Molecular Cloud Formed at the Interface of Two Colliding Supershells: Observations Meet Simulations}",
      journal = {\apj},
         year = 2015,
        month = jan,
       volume = {799},
       number = {1},
          eid = {64},
        pages = {64},
          doi = {10.1088/0004-637X/799/1/64},
archivePrefix = {arXiv},
       eprint = {1411.2708},
 primaryClass = {astro-ph.GA},
       adsurl = {https://ui.adsabs.harvard.edu/abs/2015ApJ...799...64D}
}

@ARTICLE{2021MNRAS.505..459F,
       author = {{Fujii}, Kosuke and {Mizuno}, Norikazu and {Dawson}, J.~R. and {Inoue}, Tsuyoshi and {Torii}, Kazufumi and {Onishi}, Toshikazu and {Kawamura}, Akiko and {Muller}, Erik and {Minamidani}, Tetsuhiro and {Tsuge}, Kisetsu and {Fukui}, Yasuo},
        title = "{Giant molecular cloud formation at the interface of colliding supershells in the large magellanic cloud}",
      journal = {\mnras},
         year = 2021,
        month = jul,
       volume = {505},
       number = {1},
        pages = {459-479},
          doi = {10.1093/mnras/stab1202},
archivePrefix = {arXiv},
       eprint = {2104.14191},
 primaryClass = {astro-ph.GA},
       adsurl = {https://ui.adsabs.harvard.edu/abs/2021MNRAS.505..459F}
}

@ARTICLE{2010A&A...518L.102A,
       author = {{Andr{\'e}}, Ph. and {Men'shchikov}, A. and {Bontemps}, S. and {K{\"o}nyves}, V. and {Motte}, F. and {Schneider}, N. and {Didelon}, P. and {Minier}, V. and {Saraceno}, P. and {Ward-Thompson}, D. and {di Francesco}, J. and {White}, G. and {Molinari}, S. and {Testi}, L. and {Abergel}, A. and {Griffin}, M. and {Henning}, Th. and {Royer}, P. and {Mer{\'\i}n}, B. and {Vavrek}, R. and {Attard}, M. and {Arzoumanian}, D. and {Wilson}, C.~D. and {Ade}, P. and {Aussel}, H. and {Baluteau}, J.-P. and {Benedettini}, M. and {Bernard}, J.-Ph. and {Blommaert}, J.~A.~D.~L. and {Cambr{\'e}sy}, L. and {Cox}, P. and {di Giorgio}, A. and {Hargrave}, P. and {Hennemann}, M. and {Huang}, M. and {Kirk}, J. and {Krause}, O. and {Launhardt}, R. and {Leeks}, S. and {Le Pennec}, J. and {Li}, J.~Z. and {Martin}, P.~G. and {Maury}, A. and {Olofsson}, G. and {Omont}, A. and {Peretto}, N. and {Pezzuto}, S. and {Prusti}, T. and {Roussel}, H. and {Russeil}, D. and {Sauvage}, M. and {Sibthorpe}, B. and {Sicilia-Aguilar}, A. and {Spinoglio}, L. and {Waelkens}, C. and {Woodcraft}, A. and {Zavagno}, A.},
        title = "{From filamentary clouds to prestellar cores to the stellar IMF: Initial highlights from the Herschel Gould Belt Survey}",
      journal = {\aap},
         year = 2010,
        month = jul,
       volume = {518},
          eid = {L102},
        pages = {L102},
          doi = {10.1051/0004-6361/201014666},
archivePrefix = {arXiv},
       eprint = {1005.2618},
 primaryClass = {astro-ph.GA},
       adsurl = {https://ui.adsabs.harvard.edu/abs/2010A&A...518L.102A}
}

@INPROCEEDINGS{2014prpl.conf...27A,
       author = {{Andr{\'e}}, P. and {Di Francesco}, J. and {Ward-Thompson}, D. and {Inutsuka}, S.-I. and {Pudritz}, R.~E. and {Pineda}, J.~E.},
        title = "{From Filamentary Networks to Dense Cores in Molecular Clouds: Toward a New Paradigm for Star Formation}",
    booktitle = {Protostars and Planets VI},
         year = 2014,
       editor = {{Beuther}, Henrik and {Klessen}, Ralf S. and {Dullemond}, Cornelis P. and {Henning}, Thomas},
        month = jan,
        pages = {27-51},
          doi = {10.2458/azu_uapress_9780816531240-ch002},
archivePrefix = {arXiv},
       eprint = {1312.6232},
 primaryClass = {astro-ph.GA},
       adsurl = {https://ui.adsabs.harvard.edu/abs/2014prpl.conf...27A}
}

@ARTICLE{2009ApJ...700.1609M,
       author = {{Myers}, Philip C.},
        title = "{Filamentary Structure of Star-forming Complexes}",
      journal = {\apj},
         year = 2009,
        month = aug,
       volume = {700},
       number = {2},
        pages = {1609-1625},
          doi = {10.1088/0004-637X/700/2/1609},
archivePrefix = {arXiv},
       eprint = {0906.2005},
 primaryClass = {astro-ph.GA},
       adsurl = {https://ui.adsabs.harvard.edu/abs/2009ApJ...700.1609M}
}

@ARTICLE{2012A&A...540L..11S,
       author = {{Schneider}, N. and {Csengeri}, T. and {Hennemann}, M. and {Motte}, F. and {Didelon}, P. and {Federrath}, C. and {Bontemps}, S. and {Di Francesco}, J. and {Arzoumanian}, D. and {Minier}, V. and {Andr{\'e}}, Ph. and {Hill}, T. and {Zavagno}, A. and {Nguyen-Luong}, Q. and {Attard}, M. and {Bernard}, J.-Ph. and {Elia}, D. and {Fallscheer}, C. and {Griffin}, M. and {Kirk}, J. and {Klessen}, R. and {K{\"o}nyves}, V. and {Martin}, P. and {Men'shchikov}, A. and {Palmeirim}, P. and {Peretto}, N. and {Pestalozzi}, M. and {Russeil}, D. and {Sadavoy}, S. and {Sousbie}, T. and {Testi}, L. and {Tremblin}, P. and {Ward-Thompson}, D. and {White}, G.},
        title = "{Cluster-formation in the Rosette molecular cloud at the junctions of filaments}",
      journal = {\aap},
         year = 2012,
        month = apr,
       volume = {540},
          eid = {L11},
        pages = {L11},
          doi = {10.1051/0004-6361/201118566},
archivePrefix = {arXiv},
       eprint = {1203.6472},
 primaryClass = {astro-ph.GA},
       adsurl = {https://ui.adsabs.harvard.edu/abs/2012A&A...540L..11S}
}

@ARTICLE{2022A&A...658A.114K,
       author = {{Kumar}, M.~S.~N. and {Arzoumanian}, D. and {Men'shchikov}, A. and {Palmeirim}, P. and {Matsumura}, M. and {Inutsuka}, S.},
        title = "{Filament coalescence and hub structure in Mon R2. Implications for massive star and cluster formation}",
      journal = {\aap},
         year = 2022,
        month = feb,
       volume = {658},
          eid = {A114},
        pages = {A114},
          doi = {10.1051/0004-6361/202140363},
archivePrefix = {arXiv},
       eprint = {2112.06803},
 primaryClass = {astro-ph.GA},
       adsurl = {https://ui.adsabs.harvard.edu/abs/2022A&A...658A.114K}
}

@ARTICLE{2018ApJ...854..106D,
       author = {{Dewangan}, L.~K. and {Devaraj}, R. and {Ojha}, D.~K.},
        title = "{The Embedded Ring-like Feature and Star Formation Activities in G35.673-00.847}",
      journal = {\apj},
         year = 2018,
        month = feb,
       volume = {854},
       number = {2},
          eid = {106},
        pages = {106},
          doi = {10.3847/1538-4357/aaaa6f},
archivePrefix = {arXiv},
       eprint = {1801.07364},
 primaryClass = {astro-ph.GA},
       adsurl = {https://ui.adsabs.harvard.edu/abs/2018ApJ...854..106D}
}

@ARTICLE{1989ApJ...346..735B,
       author = {{Bertoldi}, Frank},
        title = "{The Photoevaporation of Interstellar Clouds. I. Radiation-driven Implosion}",
      journal = {\apj},
         year = 1989,
        month = nov,
       volume = {346},
        pages = {735},
          doi = {10.1086/168055},
       adsurl = {https://ui.adsabs.harvard.edu/abs/1989ApJ...346..735B}
}

@ARTICLE{1994A&A...289..559L,
       author = {{Lefloch}, B. and {Lazareff}, B.},
        title = "{Cometary globules I. Formation, evolution and morphology.}",
      journal = {\aap},
         year = 1994,
        month = sep,
       volume = {289},
        pages = {559-578},
       adsurl = {https://ui.adsabs.harvard.edu/abs/1994A&A...289..559L}
}

@ARTICLE{2015MNRAS.450.1199D,
       author = {{Dale}, J.~E. and {Haworth}, T.~J. and {Bressert}, E.},
        title = "{The dangers of being trigger-happy}",
      journal = {\mnras},
         year = 2015,
        month = jun,
       volume = {450},
       number = {2},
        pages = {1199-1211},
          doi = {10.1093/mnras/stv396},
archivePrefix = {arXiv},
       eprint = {1502.05865},
 primaryClass = {astro-ph.GA},
       adsurl = {https://ui.adsabs.harvard.edu/abs/2015MNRAS.450.1199D}
}

@ARTICLE{2012ApJ...750...64S,
       author = {{Smith}, Rowan J. and {Shetty}, Rahul and {Stutz}, Amelia M. and {Klessen}, Ralf S.},
        title = "{Line Profiles of Cores within Clusters. I. The Anatomy of a Filament}",
      journal = {\apj},
         year = 2012,
        month = may,
       volume = {750},
       number = {1},
          eid = {64},
        pages = {64},
          doi = {10.1088/0004-637X/750/1/64},
archivePrefix = {arXiv},
       eprint = {1201.6275},
 primaryClass = {astro-ph.GA},
       adsurl = {https://ui.adsabs.harvard.edu/abs/2012ApJ...750...64S}
}

\end{document}